\documentstyle[amstex,prl,aps,twocolumn]{revtex}
\input epsf
\begin{document}

\title{Epitaxial mounding in 
limited mobility models of surface growth}
\author{P. Punyindu$^{(1)}$, Z Toroczkai$^{(1,2)}$, 
and S. Das Sarma$^{(1)}$ }
\address{$^{(1)}$Department of Physics,
University of Maryland, College Park, MD 20742, USA\\
$^{(2)}$Theoretical  Division and Center for Nonlinear Studies, 
Los Alamos National Laboratory, \\
Los Alamos,  New Mexico 87545, USA}
\maketitle

\begin{abstract}{ \footnotesize \protect\bf 
We study, through large scale stochastic simulations
using the noise reduction technique,
surface growth via vapor deposition e.g. molecular
beam epitaxy (MBE), for simple nonequilibrium limited
mobility solid-on-solid growth models, such as the
Family (F) model, the Das Sarma-Tamborenea (DT) model,
the Wolf-Villain (WV) model, the Larger Curvature (LC)
model, and other related models.
We find that d=2+1 dimensional surface growth in
several noise reduced
models (most notably the WV and the LC model)
exhibits spectacular quasi-regular mound formation 
with slope selection
in their dynamical surface morphology in contrast
to the standard statistically scale invariant
kinetically rough surface growth expected
(and earlier reported in the literature) for
such growth models.
The mounding instability in these epitaxial growth models
does not involve the Ehrlich-Schwoebel 
step edge diffusion barrier. The mounded morphology in these 
growth models arises from the interplay between the 
line tension along 
step edges in the plane
parallel to the average surface 
and the suppression  of 
noise and island nucleation. The line 
tension tends to stabilize 
some  of the step orientations that 
coincide with in-plane
high symmetry crystalline directions, 
and thus the mounds 
that are formed  assume quasi-regular structures. 
The noise reduction technique 
developed originally for 
Eden type models can be used to control 
the stochastic noise
and enhance diffusion along the step edge, 
which ultimately leads
to the formation of quasi-regular mounds during growth. 
We show that by increasing the 
diffusion surface length 
together with supression of 
nucleation and deposition noise,
one can obtain a self-organization of the 
pyramids in quasi-regular patterns.
The mounding instability in these simple epitaxial
growth models is closely related to the cluster-edge
diffusion (as opposed to step edge barrier) driven
mounding in MBE growth, which has been recently
discussed in the literature.
The epitaxial mound formation studied here is a
kinetic-topological instability (which can happen
only in d=2+1 dimensional, or higher dimensional,
growth, but {\it not} in d=1+1 dimensional growth
because no cluster diffusion around a closed surface
loop is possible in ``one dimensional'' surfaces),
which is likely to be quite generic in real MBE-type
surface growth. Our extensive numerical simulations
produce mounded (and slope-selected)
surface growth morphologies which 
are strikingly visually similar to many recently reported
experimental MBE growth morphologies.}
\end{abstract} 
\pacs{PACS numbers: 68.35.Ct, 68.36.Bs, 68.55.Jk, 81.10.Aj} 

\makeatother

\section{Introduction}

Crystal growth, particularly high-quality epitaxial thin film growth,  
is one of the most fundamental
processes impacting today's technology \cite{PT}. 
A major issue in interface growth experiments
is to have continuous dynamical control over the deposition
process, such that interfaces with certain desired patterns can
be obtained. For example, while in thin film epitaxy it is
desirable to obtain smooth surfaces,  in nanotechnology
it is also important to be able to create regular, nanoscale
structures with well defined geometry, such as quantum dots,
quantum wires, etc. Growth is usually achieved
by vapor deposition of atoms from
a molecular beam (molecular beam epitaxy, or MBE). 
Thus, in order to be able to
design a controlled deposition process, it is of crucial importance
to understand {\it all} the instability types (which destroy 
controllability) that may occur during growth.
In the present paper we concentrate on MBE growth.
There are several types of instabilities in MBE
among which we mention the Ehrlich-Schwoebel (ES) \cite{ES} 
instability which is  of kinetic-energetic type.
Ehrlich-Schwoebel (ES) barriers
in the lattice potential induce an instability 
by hindering step-edge atoms on upper terraces
from going down to lower terraces
\cite{ES}
which in turn can generate mounded structures during growth. 
The ES instability is thought to be an
ubiquitous phenomenon in real surface deposition processes, and
as such, is widely considered to be the only mechanism
for formation of mounds in surface growth. 
Another kinetic-energetic instability, 
which does not
involve any explicit ES barriers, was discussed by
Amar and Family \cite{AF}.
This instability involves
a negative barrier at the base of a step,  and thus is 
due to a  short range
{\it attraction} between adatoms and islands.
Short range attractions generically lead to clustering
in multiparticle systems, a property which in the language of
MBE translates into formation of mounds.
Note that this attraction-induced instability and ES instability
are essentially equivalent, and both could occur in
d=1+1 or 2+1 dimensions.
Both the attractive instability and the ES barrier instability
cause mounding because atoms on terraces preferentially
collect at up-steps rather than down-steps leading to the
mounded morphology.
In both cases, there is no explicit stabilizing mechanism,
and the mounds should progressively stiffen with time
leading to mounded growth morphology with no slope selection,
where the mound slopes continue to increase as growth progresses.
It is, of course, possible to stop the monotonic slope increase
by incorporating some additional mechanisms
(e.g. the so-called ``downward funneling'' where deposited 
atoms funnel downwards before incorporation)
which produce a downhill mass current on the surface
and thereby opposes the uphill current created by the 
ES barrier.
There is, however, no intrinsic slope selection process
built in the ES barrier mechanism itself.

\begin{figure}[htbp] 
\hspace*{-0.6cm} 
\epsfxsize=3.6 in   
\epsfbox{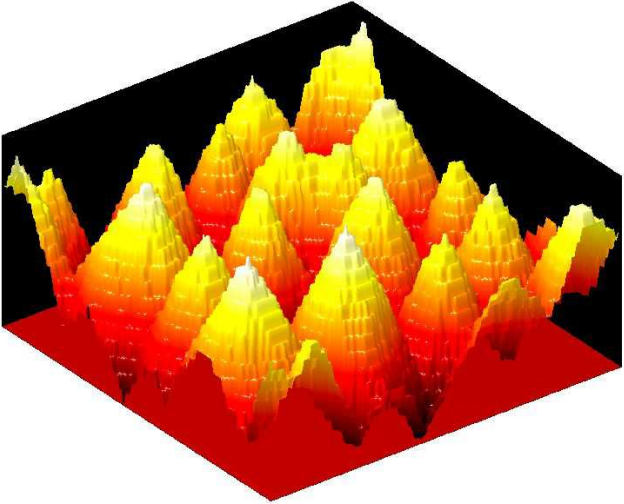}  
\vspace{0.0cm}  
\caption{
Mounded morphology created by SED.
This is a simulation of the LC model as described in the text.
\label{color}} \end{figure}

The purpose of the current article is to discuss a completely
different {\it kinetic-topological} mechanism,
qualitatively distinct from the ES barrier induced mounding,
which leads to spectacular mound formation
(see Fig. \ref{color} for a colorful example)
in MBE growth without involving any ES barriers
(or the closely related island-adatom attraction mechanism)
whatsoever. Based on the direct stochastic numerical
simulations of a large number of MBE-related
nonequilibrium limited mobility solid-on-solid 
epitaxial growth models studied in this paper
we believe that epitaxial mounding of the type 
discussed here is quite generic, and
may actually apply to a variety of experimental 
situations where mounded growth morphologies 
(with slope selection) have been
observed. The mechanism underlying the mounding instability
discussed in this paper is closely related to that proposed in
two recent publications \cite{edge,ecu}
although its widespread applicability to well-known
limited mobility MBE growth models, as described here,
has not earlier been appreciated in the literature.
The mounding instability discussed in this paper,
being topological in nature, can happen only on
physical two-dimensional surfaces (d=2+1 dimensions growth)
or in (unphysical) higher (d$>$2+1) dimensional systems, 
in contrast to
ES type instabilities which,
being of energetic origin, are allowed in all dimensions. 

The instability most recently discovered by Pierre-Louis,
D'Orsogna, and Einstein \cite{edge},  and independently by
Ramana Murty and Copper \cite{ecu} is  entirely different 
from ES instabilities, because it is {\it not}
of kinetic-energetic nature (involving potential barriers)
but rather of a kinetic-{\it topologic} nature.
This instability is generated by strong diffusion along the
edges of monoatomic steps, and its net effect in two or higher
dimensions is to create quasi-regular shaped mounds on the substrate
(see Fig. \ref{color}).
For simplicity of formulation we will call this instability
the step-edge diffusion, or SED instability.
What comes as a bit of a surprise is that in spite of its simplicity
as a mechanism, it has not apparently been recognized
that the SED mechanism may actually be playing a dominant role
in many epitaxial mounding instabilities. 
There are two reasons for this, one is
probably because the ES instability has inadvertently
and tacitly been accepted as the only mechanism for mounding,
and therefore the mounded
surface morphologies coming from experiments
have been {\it apriori } analyzed with the ES mechanism
in mind, and the second
reason is that SED is not easy to see 
in simulations unless other conditions
(related to reduced noise) are met in addition to 
the existence of edge-current, which we
will discuss in details in the present paper.

In this paper we discuss the SED instability, and analyze
its effects on various growth models, giving both a simple continuum
and discrete description. The SED is shown to exist in a large
number of discrete, limited mobility growth models, and we
identify the conditions under which SED-induced
epitaxial mounding is manifested.
We study the instability by producing dynamical growth morphologies
through direct numerical simulations and
by analyzing 
the measured height-height correlation
functions \cite{short} of the simulated morphologies.
This paper is organized as follows: in Section II we define
a number of limited mobility growth models (Subsection II.A)
studied here,
we briefly review the noise reduction technique (Subsection II.B),
then present a series of simulated growth
morphologies, both from early and late
times, with and without noise reduction (Subsection II.C); 
in Section III we 
describe and discuss the SED mechanism,
by first presenting a continuum description in terms
of local currents (Subsection III.A) and its discrete counterpart
with particular emphasis on the Wolf-Villain and Das Sarma - Tamborenea
models (Subsection III.B); in Subsection III.C we 
introduce the notion of conditional site occupation rates to      
analyze the effects of noise reduction;
Section IV is devoted to the properties of the mounded morphologies
with the emphasis on the relevance of the height-height correlation
function (Subsection IV.A), where we note possible comparison with
experiments and discuss the relevance of the global diffusion current
in the simulations for the mounding instability.

\section{Morphologies of discrete limited mobility growth models}

\subsection{Models}

All the models  studied here are dynamical limited mobility
growth models in which an adatom is allowed to diffuse
(according to specific sets of diffusion rules for each model)
within a finite diffusion length of $l$ sites. 
The diffusion process is
instantaneous and once the adatom has found its final site, 
it is incorporated permanently into the substrate
and can no longer move.
These models are also based on the solid-on-solid (SOS) constraint
where bulk vacancies and overhangs are not allowed.
Desorption from the growth front is neglected.

\begin{figure}[htbp]\noindent 
\begin{minipage}{2.4 in} 
\hspace*{1.3 cm}\epsfxsize=2.4 in  \epsfbox{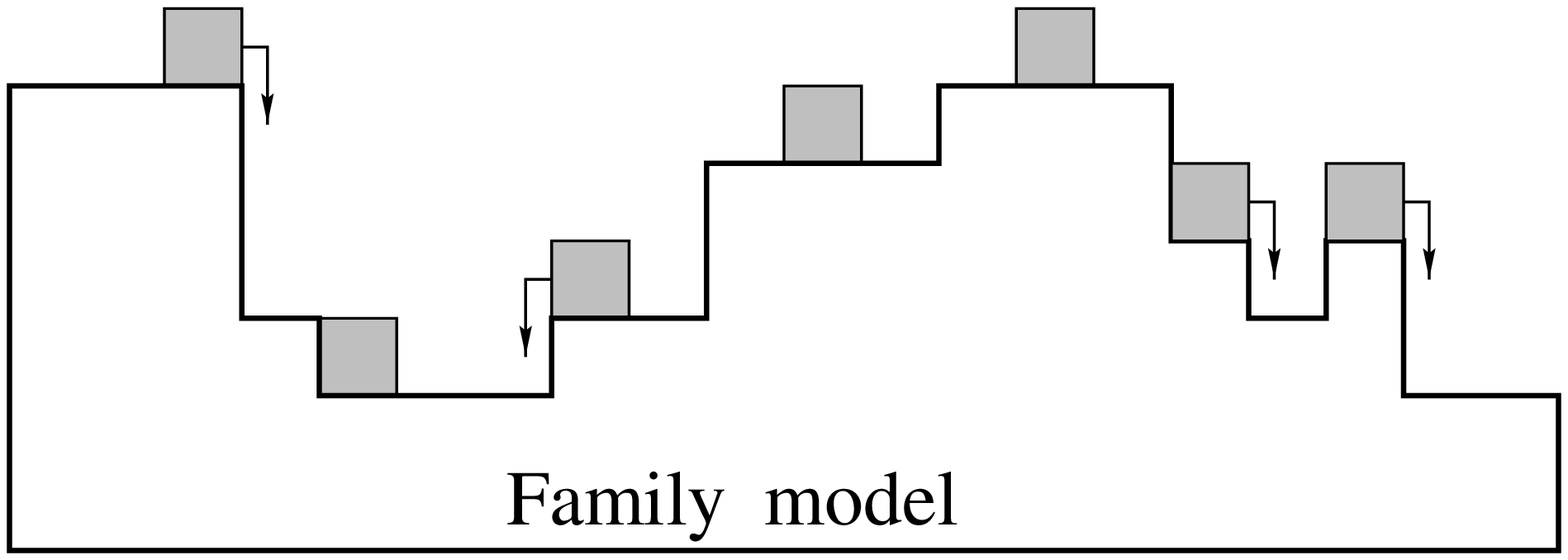} 
\end{minipage}\\ 
\noindent\begin{minipage}{2.4 in} 
\hspace*{1.3 cm}\epsfxsize=2.4 in  \epsfbox{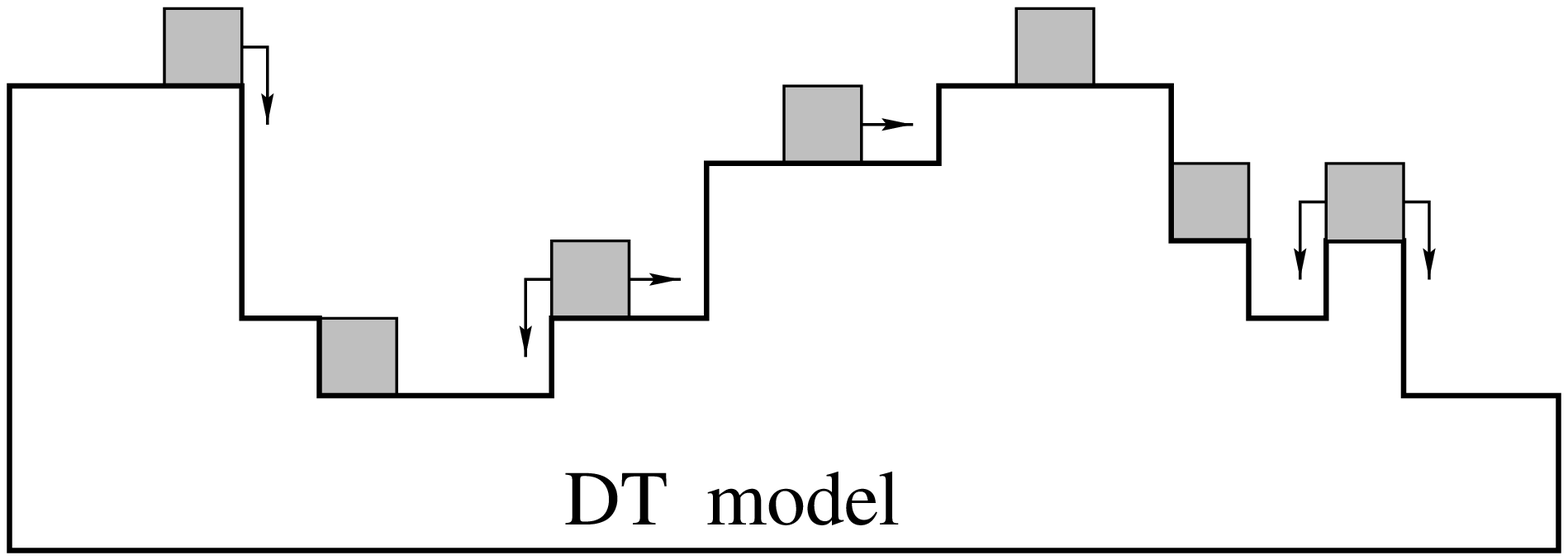} 
\end{minipage}\\ 
\noindent\begin{minipage}{2.4 in} 
\hspace*{1.3 cm}\epsfxsize=2.4 in  \epsfbox{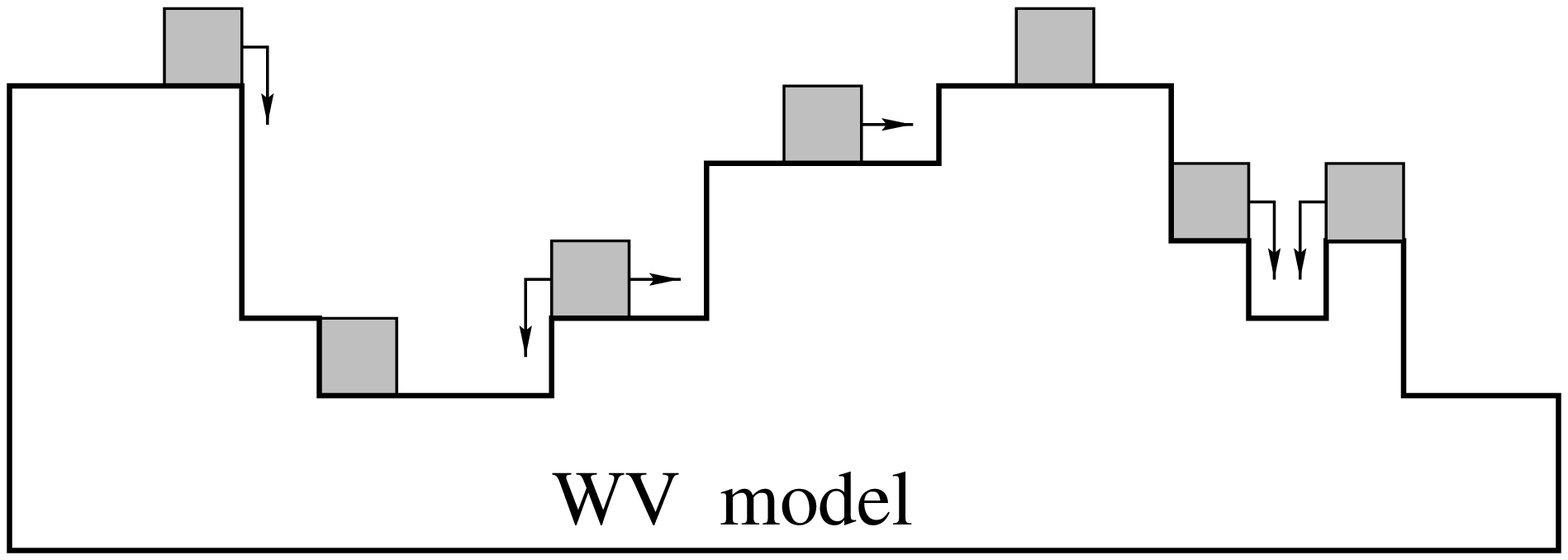} 
\end{minipage}\\ 
\noindent\begin{minipage}{2.4 in} 
\hspace*{1.3 cm}\epsfxsize=2.4 in  \epsfbox{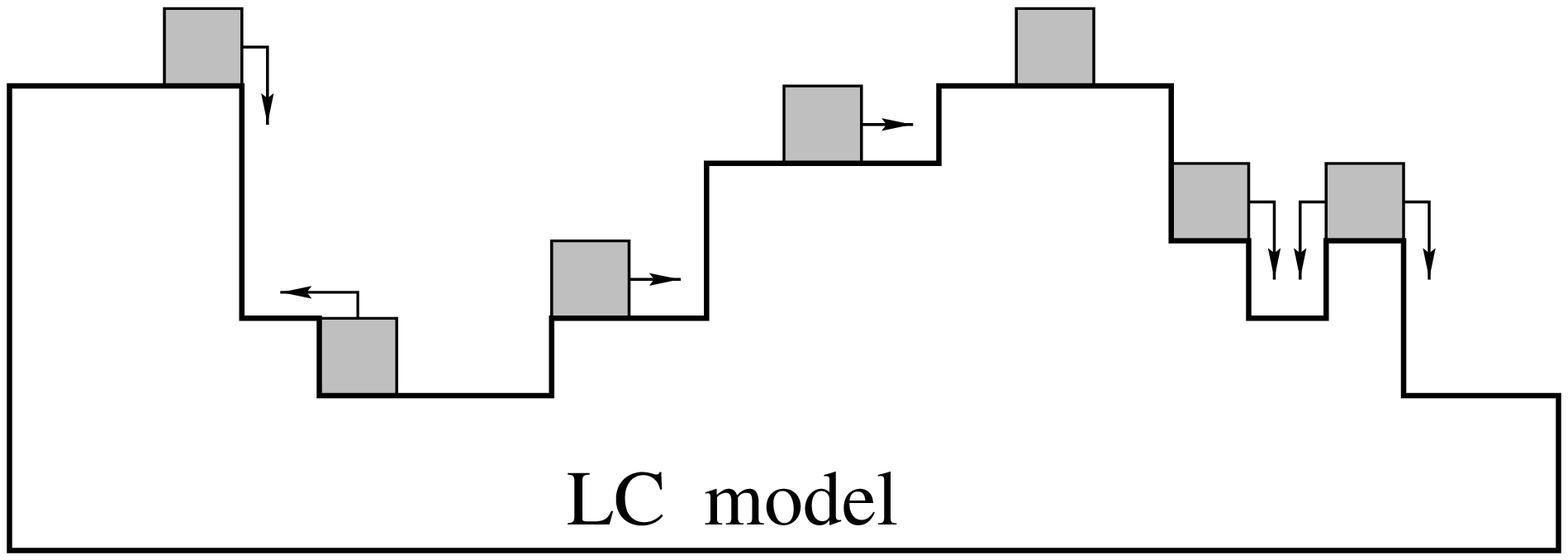} 
\end{minipage}\\ \vspace{0.5cm} 
\caption{
Schematic configurations defining the diffusion growth rules
in one dimension for various dynamical growth
models discussed in this paper. Here $l=1$.
\label{rule}} 
\end{figure}

With this SOS constraint,
growth can be described on a coarse-grained level 
by the continuity equation
\begin{equation}
\frac{\partial h}{\partial t}+\nabla \cdot {\bf j} =
F \Omega + \eta ({\bf x},t), \label{cont}
\end{equation}
where $h({\bf x},t)$ is the surface height measured in the
growth direction, ${\bf j}$ is the local (coarse-grained)
surface diffusion current, $F$ is the incoming particle flux,
$\Omega = a_{\perp}a_{\|}^d$ is the surface cell volume
with $a_{\perp}$ and $a_{\|}$ being the lattice constants
in the growth direction and in the $d$-dimensional
substrate, respectively. The shot noise is Gaussian 
uncorrelated white noise with correlator:
\begin{equation}
\langle \eta({\bf x},t)\eta({\bf x'},t') \rangle
=D \delta({\bf x}-{\bf x'})\delta(t-t') \label{noise}
\end{equation}
where the amplitude of the fluctuations $D$ is directly
proportional to the average incoming particle flux through the
relation: $D=F\Omega^2$.

The most general form of this continuity equation \cite{ld}
which preserves all the symmetries of the problem can be
written as:
\begin{equation}
\frac{\partial h}{\partial t} = \nu_2 \nabla^2h - \lambda_4
\nabla^4h + \lambda_{22} \nabla^2 
( \mbox{\boldmath $\nabla$} h )^2 + ... + \eta,
\label{full4theq}
\end{equation}
where $h$ from now on indicates the height fluctuation around 
the average interface height $\langle h \rangle$.
Note that the unit of time in 
our simulations is defined through $\langle h \rangle$;
we arbitrarily choose the
deposition rate of 1 monolayer (ML) per second, and all
simulations are done with periodic boundary conditions
along the substrate.

\subsubsection{Family model}

This is a model introduced by Family (F) \cite{fam} with the 
simplest diffusion rule: adatoms diffuse to the local height
minima sites within range $l$ (see the one dimensional
version in Fig. 2).
This is a well studied model and it is described on a coarse-grained
level by
the continuum Edwards-Wilkinson \cite{R1} equation 
$\frac{\partial h}{\partial t} = \nu \nabla^2 h + \eta$
and hence the model asymptotically
belongs to the Edward-Wilkinson (EW) universality class.

\subsubsection{Das Sarma-Tamborenea model}

The Das Sarma-Tamborenea (DT) model was proposed \cite{dt1}
as a simple limited mobility nonequilibrium model for 
MBE growth. 
In this model, an adatom is randomly dropped on an 
initially flat substrate.
If the adatom already has a coordination number of two or more 
(i.e. one from the neighbor beneath it to satisfy the SOS condition,
and at least one lateral neighbor)
at the original deposition site, it is incorporated 
at that site. 
If the adatom does not have a lateral nearest neighbor at the 
deposition site, it is allowed to search, within a finite diffusion
length, and diffuse to a final site with higher coordination number
than at the original deposition site. In the case where there is no
neighbor with higher coordination number, the adatom will remain
at the deposition site. 
At the end of the diffusion process, the
adatom becomes part of the substrate and the next adatom is 
deposited.

It is important to note that adatoms in the DT model search for
final sites with higher coordination numbers compared to the deposition
site. The final sites are not necessarily the local sites with
maximum coordination numbers. In other words, in the DT model
adatoms try to {\it increase}, but not necessarily {\it maximize},
the local coordination number.
Also, deposited adatoms with more than one nearest-neighbor bond
do not move at all in the DT model.

\begin{figure}[htbp]
\begin{minipage}{2.3 in} 
\hspace*{1.3cm}\epsfxsize=2.3 in  \epsfbox{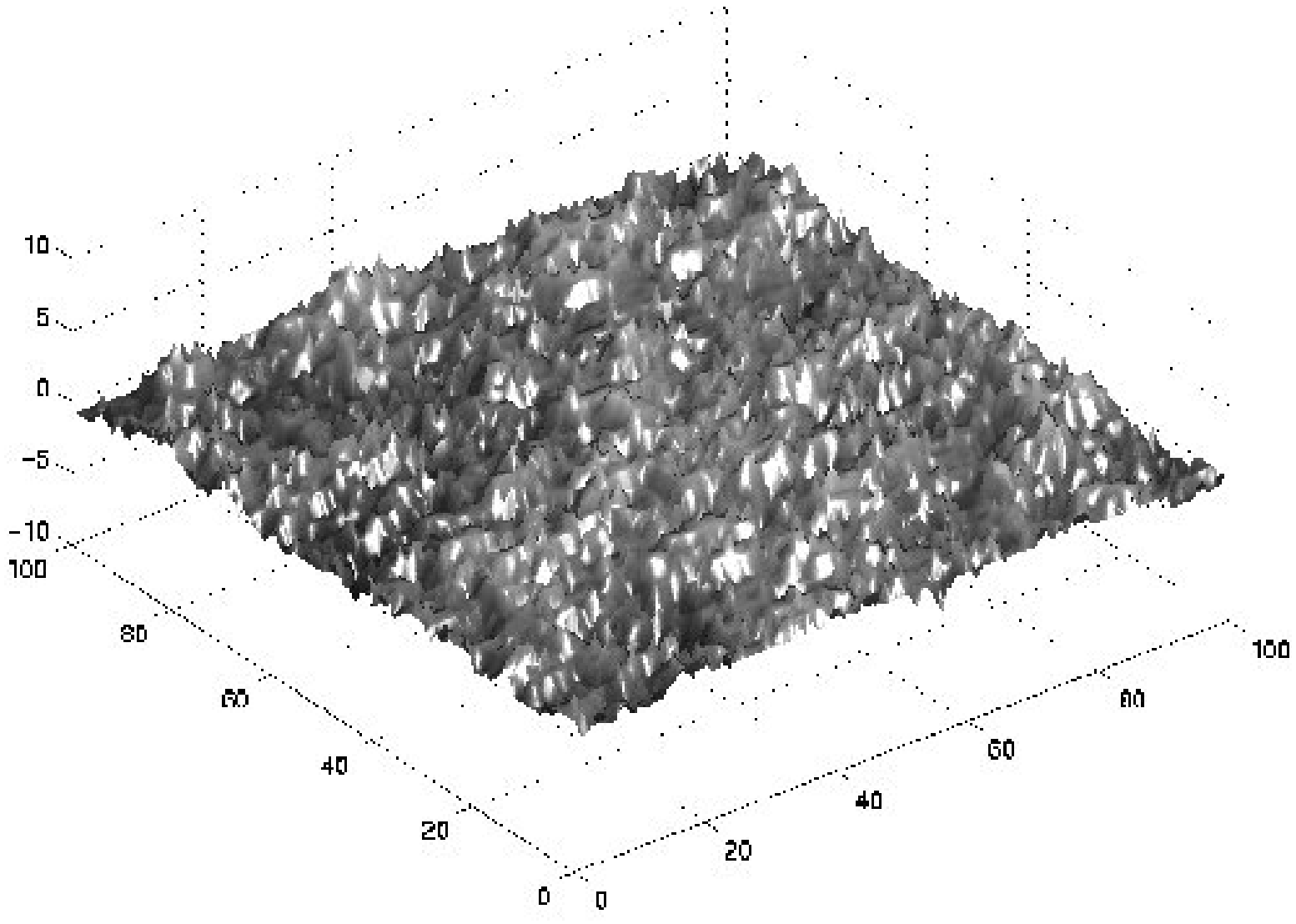} 
\end{minipage}\\ 
\noindent\begin{minipage}{2.3 in} 
\hspace*{1.3cm}\epsfxsize=2.3 in  \epsfbox{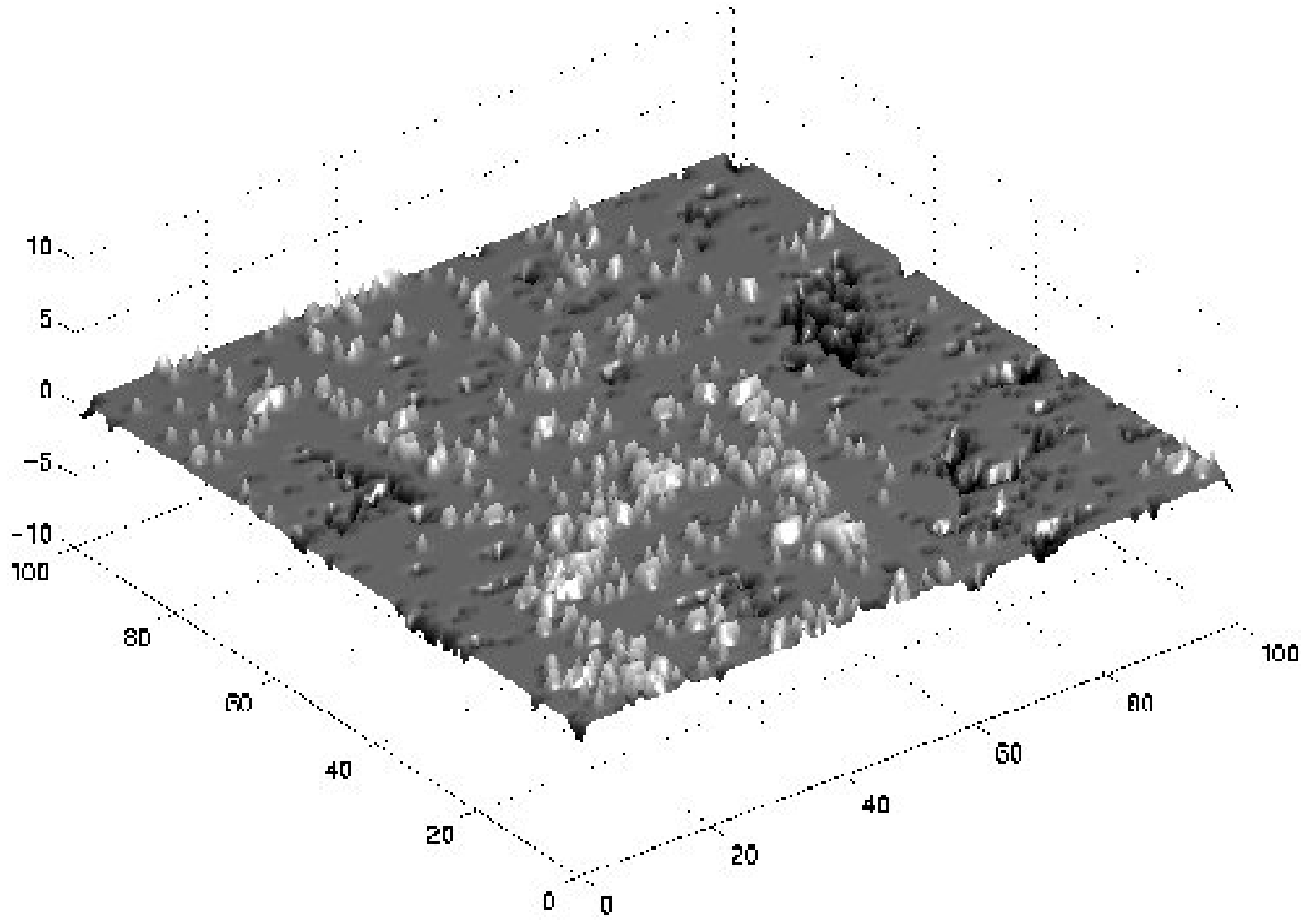} 
\end{minipage}
\caption{
Morphologies from F model with $n_r=1$ (top) and
$n_r=5$ (bottom) from substrates of size $L=100 \times 100$
after $10^6$ ML.
\label{fmsur}} \end{figure}

\subsubsection{Wolf-Villain model}

The Wolf-Villain (WV) model \cite{wv} is very similar to the
DT model in the sense that diffusion is controlled by the
local coordination number. 
There are, however, two important distinctions between the two
models. 
For one thing, in the WV model, an adatom with lateral nearest neighbors
can still diffuse if it can find a final site with higher coordination
number. 
The other difference is that, adatoms in the WV model try to
{\it maximize} the coordination number, and not just to {\it increase}
it as in the DT model (see Fig. \ref{rule}).
The two models have very similar behavior in one dimensional
(d=1+1) simulations, and there have been substantial confusion
regarding these two models in the literature.
But in two dimensional (d=2+1) systems, after an
initial crossover period, the two models behave very differently,
as will be shown in this paper.

\subsubsection{Kim - Das Sarma class of conserved growth models}

A number of discrete growth models can be defined by 
applying the approach
originally developed by Kim and 
Das Sarma \cite{kd} for obtaining precise discrete counterparts of
continuum conserved growth 
equations. 
In this method, the surface current 
${\bf j}$ is expressed as
the gradient of a scalar field 
$K$, which in turn is written as
a combination of $h$ and its 
various differential forms,
$\nabla^2 h$, $(\nabla h)^2$, ... etc., 
where certain combinations
are ruled out by symmetry constraints:
\begin{equation}
{\bf j}({\bf x},t)=-\mbox{\boldmath $\nabla$} K\;,
\end{equation}
where 
\begin{equation}
K = \nu_2 h - \lambda_4 \nabla^2 h
       + \lambda_{22} 
( \mbox{\boldmath $\nabla$} h)^2 + ...
\label{kappa}
\end{equation}

\begin{figure}[htbp]
\begin{minipage}{2.4 in} 
\hspace*{1.3cm}\epsfxsize=2.4 in  \epsfbox{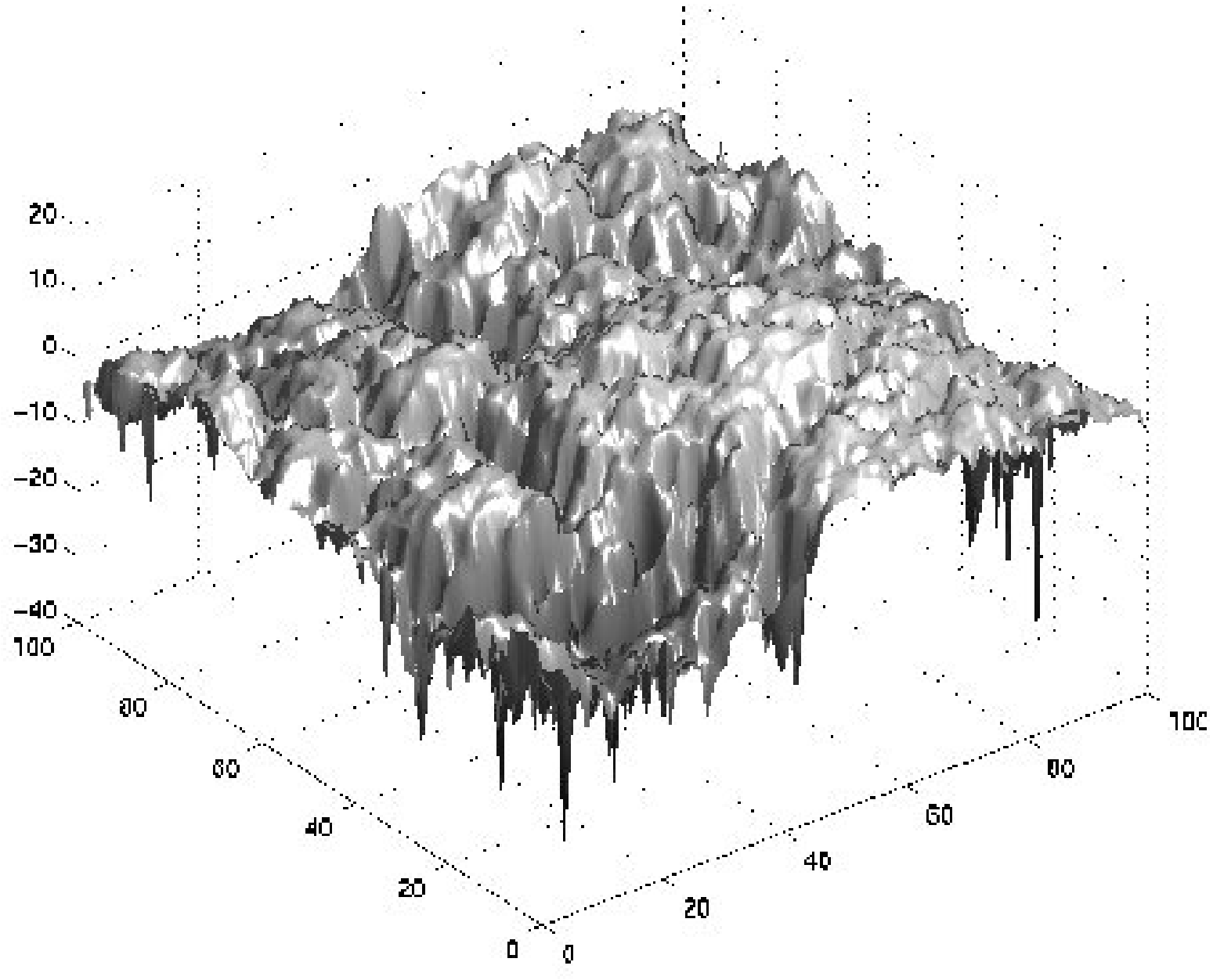} 
\end{minipage}\\ 
\noindent\begin{minipage}{2.4 in} 
\hspace*{1.3cm}\epsfxsize=2.4 in  \epsfbox{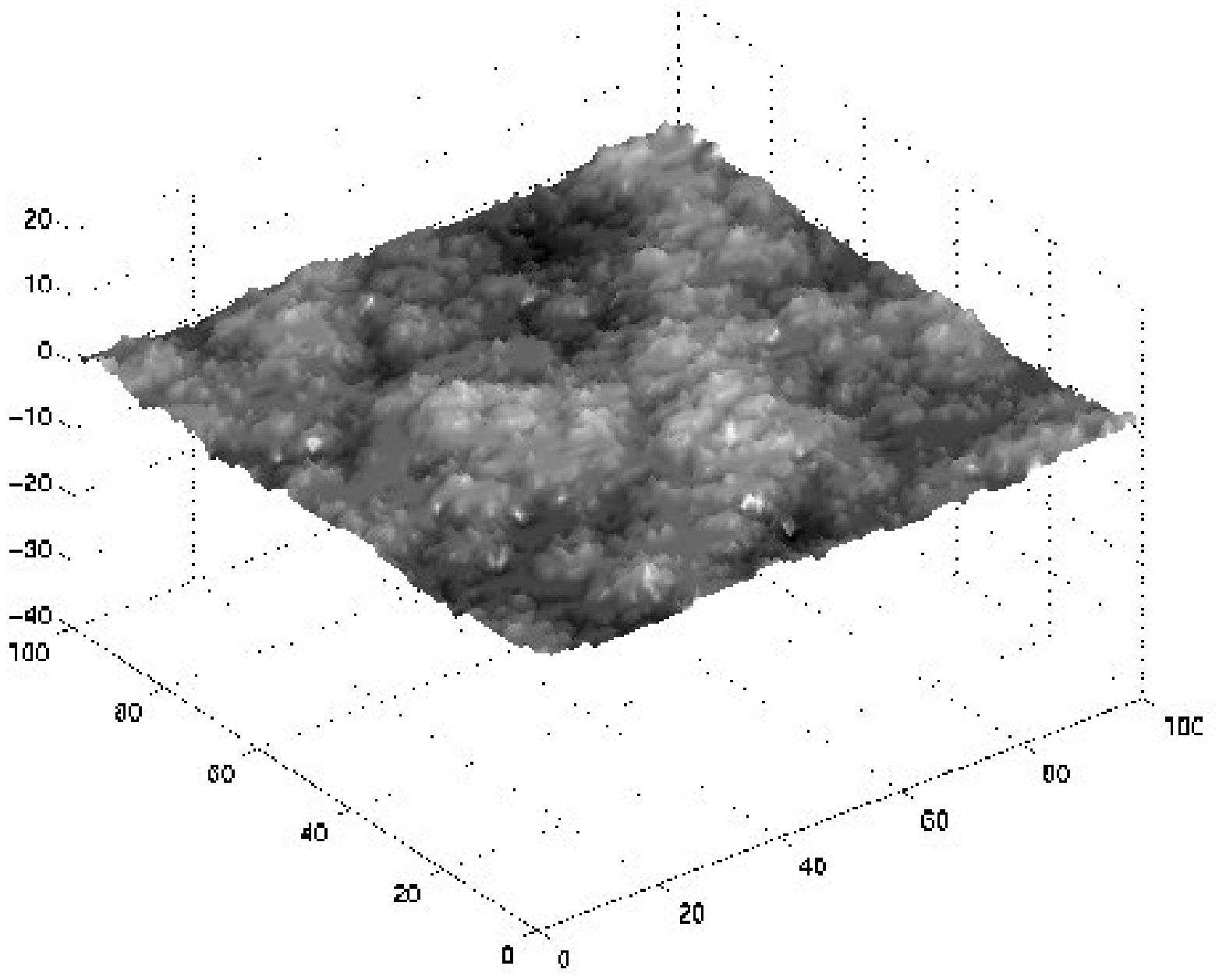} 
\end{minipage}\\ \vspace{0.2cm} 
\caption{
Morphologies from DT model with $n_r=1$ (top) and
$n_r=5$ (bottom) from substrates of size $L=100 \times 100$
after $10^6$ ML.
\label{dtsur}} \end{figure}

Using the above expression for $K$,
the general conserved growth equation, 
Eq. (\ref{full4theq}), is obtained.
Due to the conserved nature of growth,
one concludes that the 
local particle transport
happens in the direction of the largest 
variation of the scalar $-K$,
i.e., in the direction where $K$ has the 
largest decay. This
allows for the definition of the following Kim-Das Sarma discrete 
atomistic rules: 1)
a site $i$ is chosen at random, 2) the 
scalar $K$ is computed on
the lattice  
at site $i$ and for all its
nearest neighbors, 3) then a particle is added to the 
site that has the
smallest value of $K$; if site $i$ is among 
the sites with the
smallest $K$, then the particle is deposited at site $i$, 
otherwise one picks a final site with equal
probability among those sites that have the 
common smallest $K$.

\begin{figure}[htbp]
\begin{minipage}{2.4 in} 
\hspace*{1.3cm}\epsfxsize=2.4 in  \epsfbox{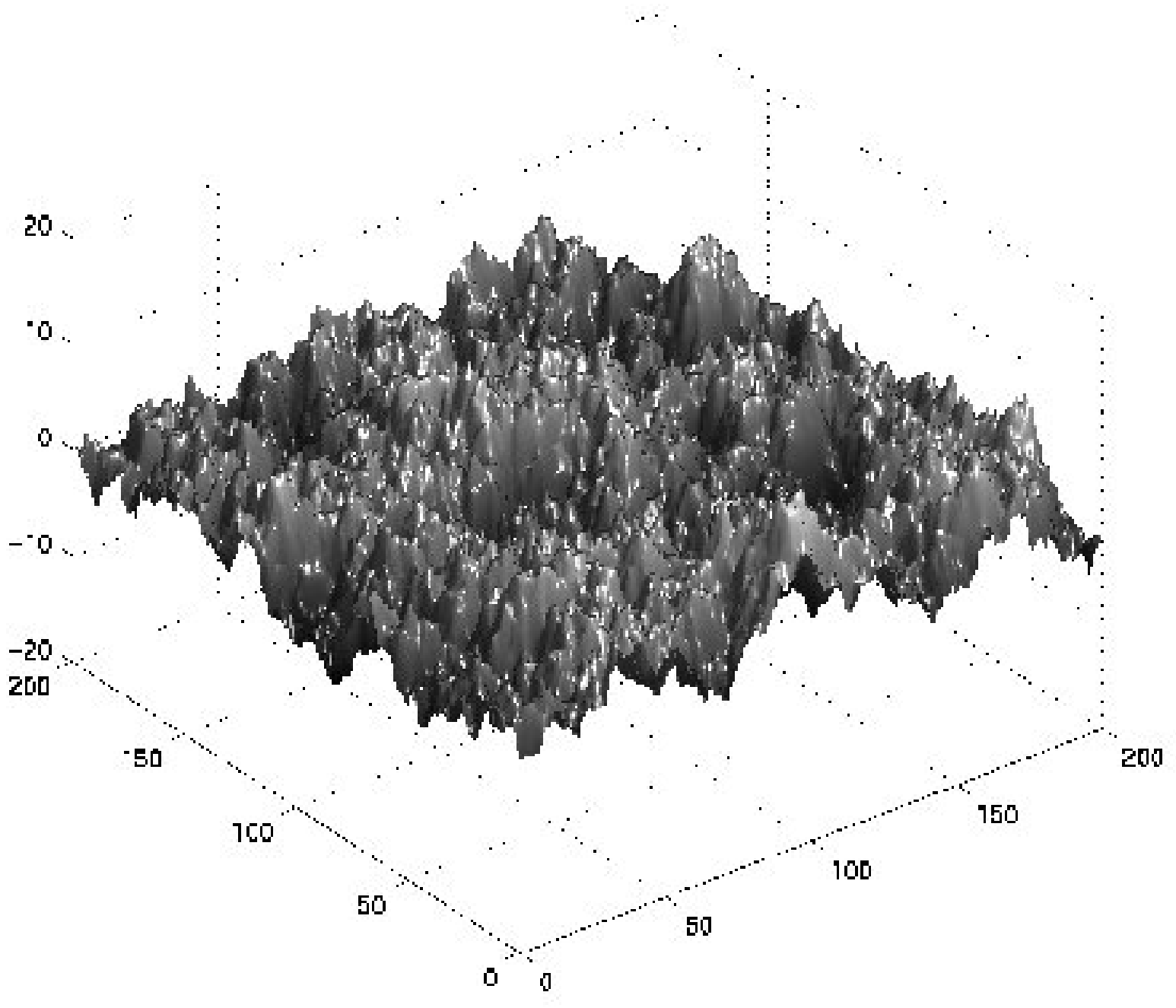}
\end{minipage}\\ 
\noindent\begin{minipage}{2.4 in} 
\hspace*{1.3cm}\epsfxsize=2.4 in  \epsfbox{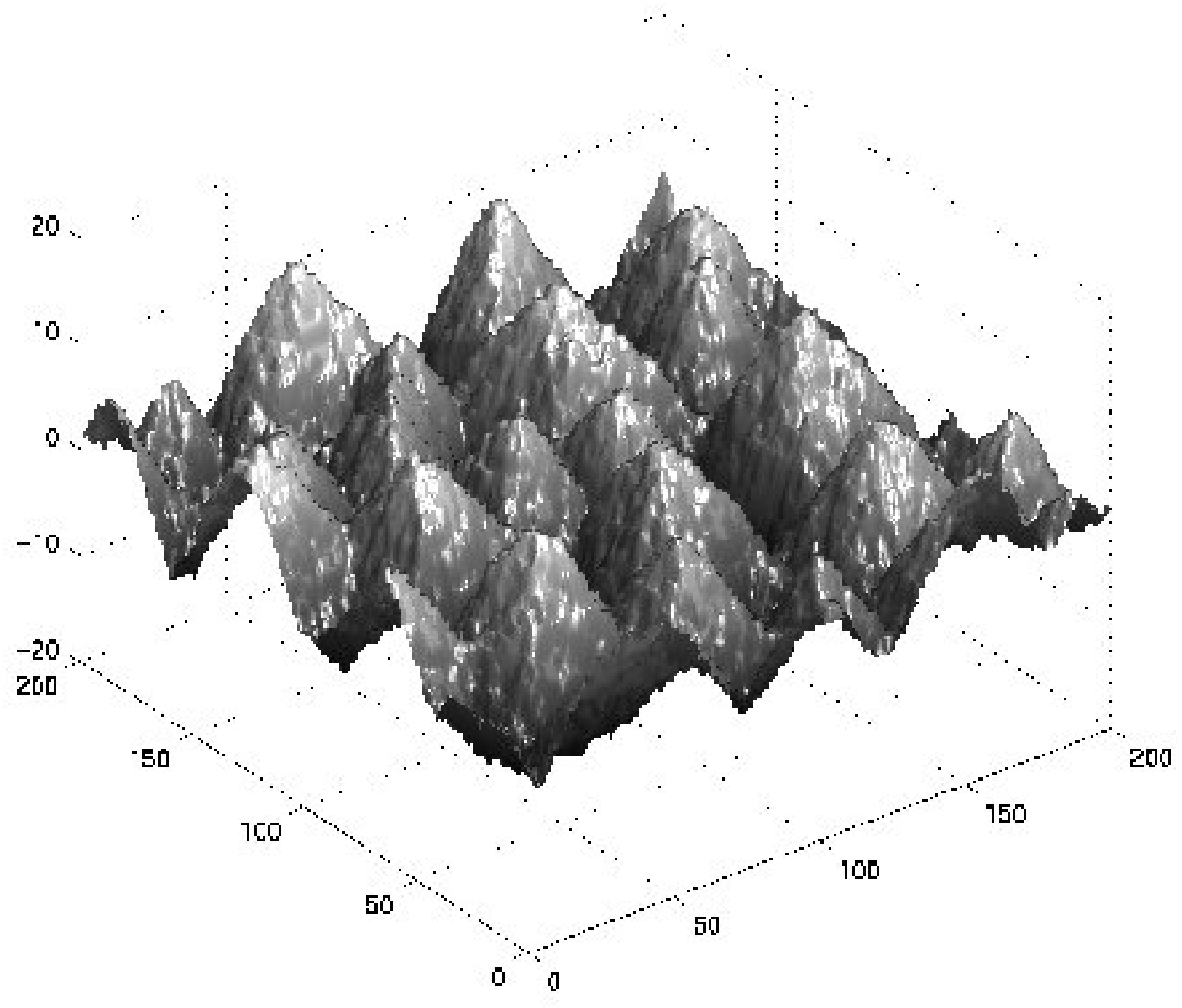} 
\end{minipage}\\\vspace{0.2cm} 
\caption{
Morphologies from WV model with $n_r=1$ (top) and
$n_r=5$ (bottom) from substrates of size $L=500 \times 500$
after $10^4$ ML.
\label{wvsur}} \end{figure}

For example, the F model
(where adatoms relax to the sites with local 
height minima) is described  by the linear second
order growth equation, with $K \sim h$ in Eq. (\ref{kappa}).
Thus the well-known F model obeying the linear EW equation
is the simplest example of the Kim-Das Sarma class of
discrete growth models.
A well-known member of the Kim - Das Sarma class
is the {\it Larger Curvature model} (LC) \cite{kd,KrugPRL}
where $K \sim -\nabla^2 h$ 
(which generates the linear fourth order growth equation).
Adatoms in this model diffuse to sites with minimum $-\nabla^2 h$,
i.e. with maximum curvature, hence the name of the model.

In this paper we also study the nonlinear 
fourth order equation \cite{ld,R2} where:
\begin{equation}
K = -\lambda_4 \nabla^2 h 
+ \lambda_{22}(\mbox {\boldmath $\nabla$}h)^2
\label{k4neq}
\end{equation}
with $\lambda_{22}$ being a small constant controlling the 
instability in the equation.
Recently, a nonlinear fourth order equation with controlled
higher order nonlinearities \cite{ddk,dasgupta} has been proposed
as a continuum description for the DT model. 
The scalar $K$ in this equation is
\begin{equation}
K = -\lambda_4 \nabla^2 h + \frac{\lambda_{22}}{C}
          [1 - e^{-C|\nabla h|^2} ]\;.
\label{4seq} \end{equation} 
The last variant we analyze is the
the linear sixth order
continuum growth equation, with 
$K \sim -\nabla^4 h$.

\begin{figure}[htbp]
\begin{minipage}{2.4 in} 
\hspace*{1.3cm}\epsfxsize=2.4 in  \epsfbox{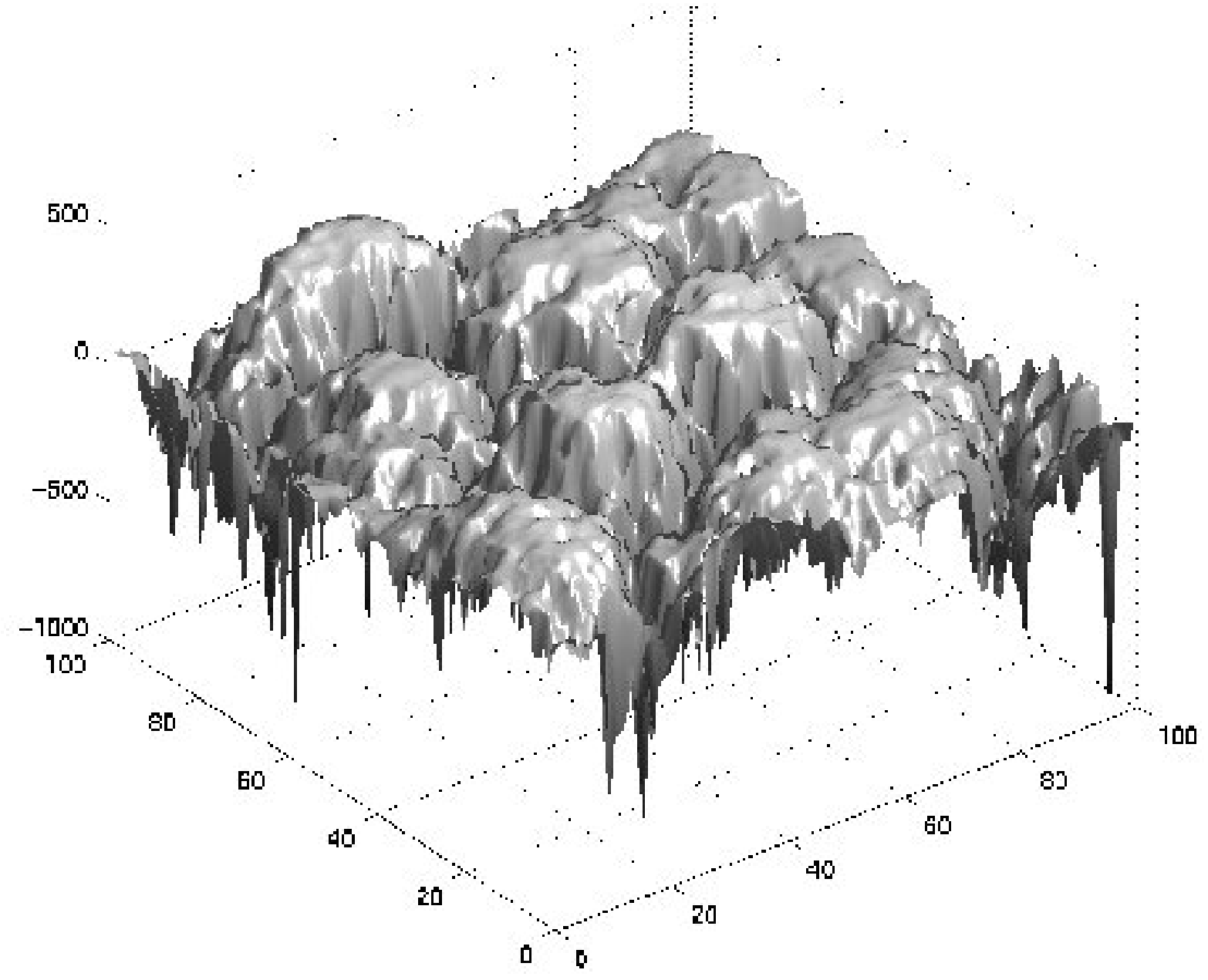} 
\end{minipage}\\ 
\noindent\begin{minipage}{2.4 in} 
\hspace*{1.3cm}\epsfxsize=2.4 in  \epsfbox{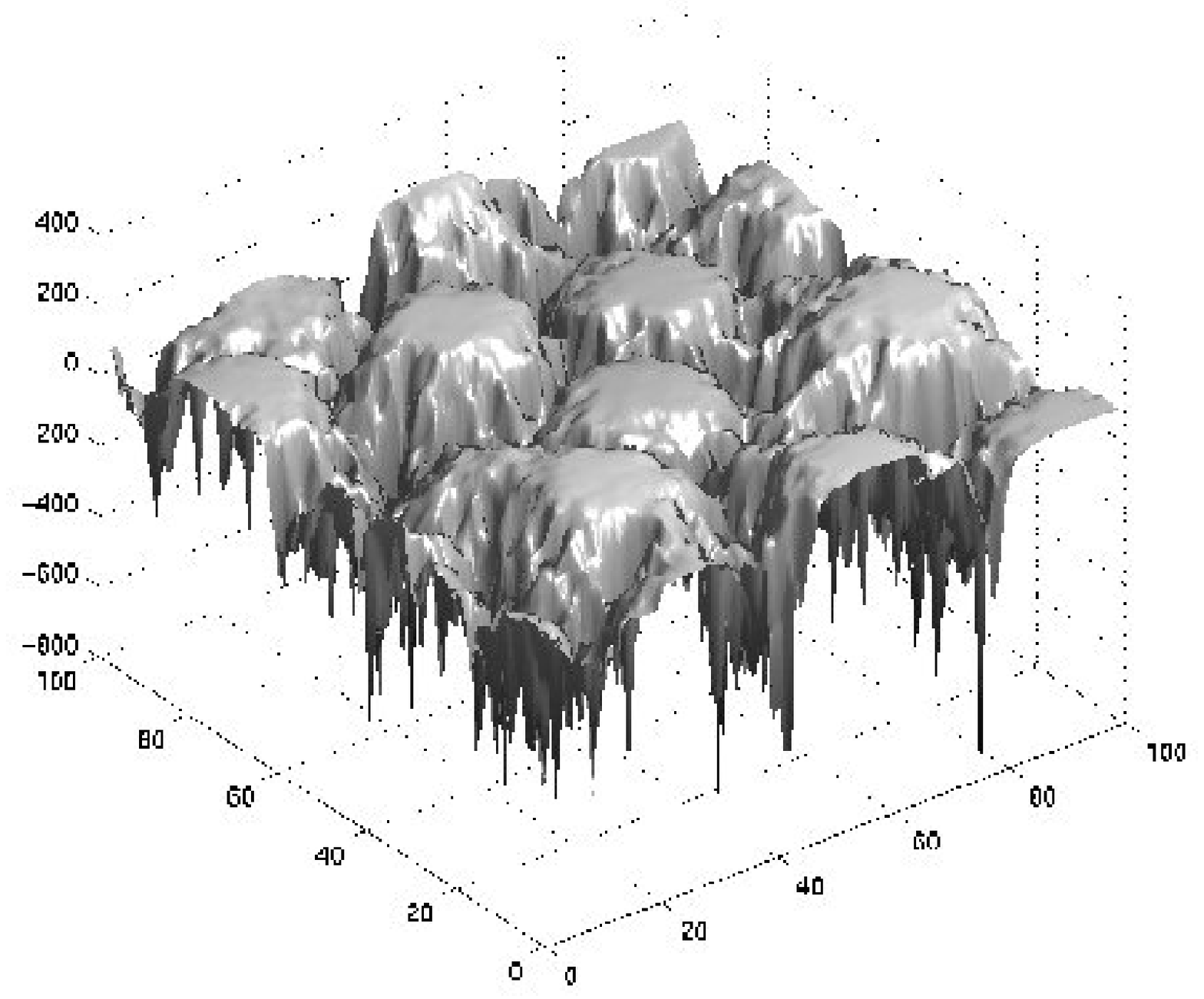} 
\end{minipage}\\\vspace{0.2cm} 
\caption{
Morphologies from DT model in the presence of an ES barrier
with $n_r=1$ (top) and
$n_r=5$ (bottom) from substrates of size $L=100 \times 100$
after $10^6$ ML.
\label{dtessur}} \end{figure}

\subsection{The noise reduction technique}

In this section we briefly discuss a well-known 
theoretical method used to
reduce stochastic noise in the stochastic growth simulations of
surface morphologies, 
in particular its short wavelength
component. The short wavelength noise 
component generates an intrinsic
width ($w_i$) in the surface roughness 
giving rise to strong nontrivial corrections to scaling
\cite{ppsds},
and typically has different
scaling properties than the long wavelength
quantities, such as the dynamical width ($w \sim t^{\beta}$)
and the correlation length parallel to the surface 
($\xi \sim t^{1/z}$). 
Reducing the intrinsic width through the noise reduction technique, 
the scaling behavior
improves dramatically, and the asymptotic critical
properties emerge in the simulations \cite{ppsds,KW,WK}.
In the simplest growth models of MBE,  two main 
competing kinetic processes are kept and
simulated: (1) deposition from an atomic beam, and (2) surface
diffusion of the freshly landed atoms until they incorporate
into the surface according to some diffusion rules.
The shot noise in the beam is simulated in (1) by randomly selecting
the landing site, and the surface diffusion performed by the
freshly landed atoms is simulated in (2) essentially by a random
walk of $l$ steps biased by the local 
coordination (filled and unfilled sites) landscape of the surface.

\begin{figure}[htbp]
\begin{minipage}{2.4 in} 
\hspace*{1.3cm}\epsfxsize=2.4 in  \epsfbox{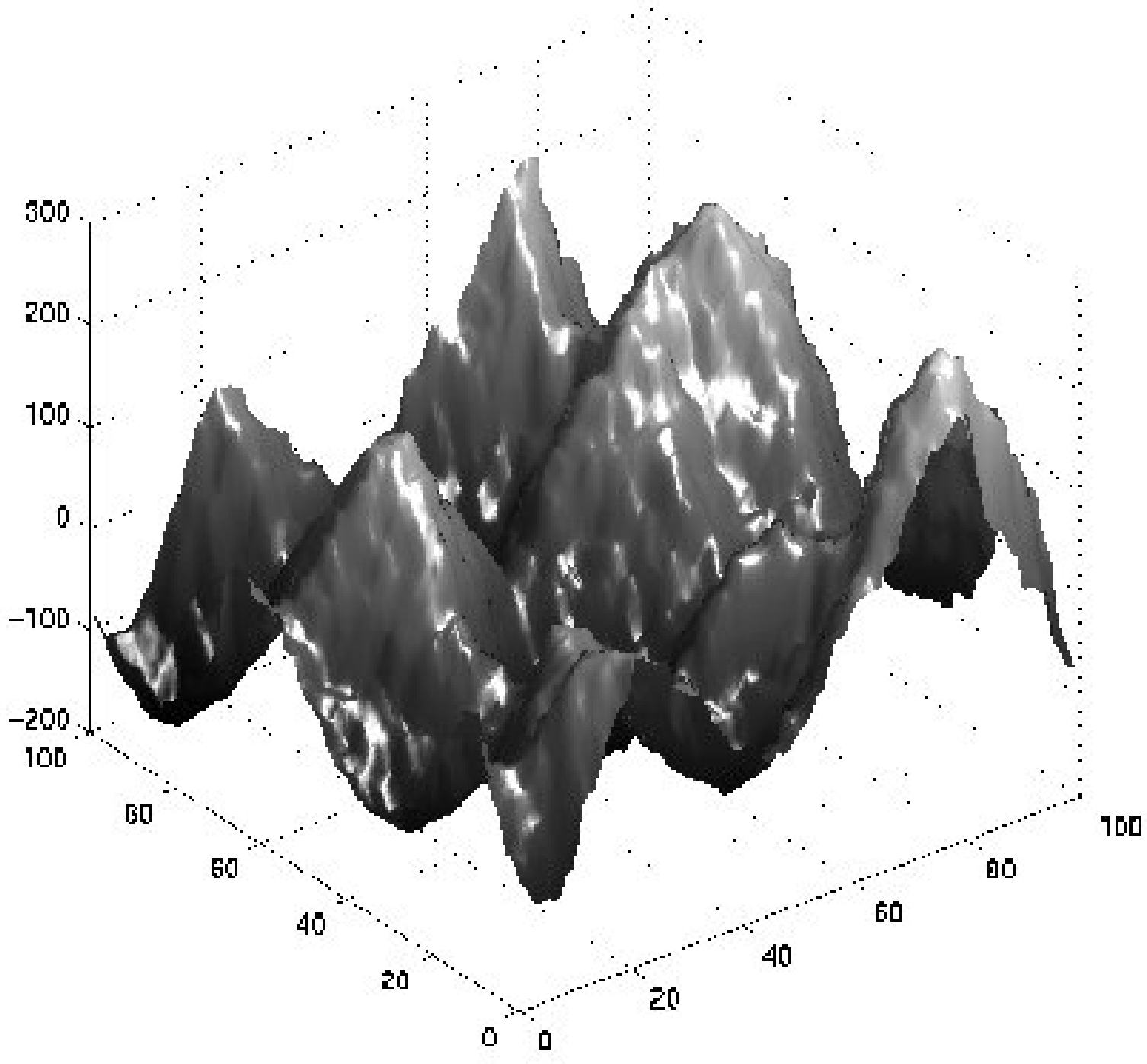} 
\end{minipage}\\ 
\noindent\begin{minipage}{2.4 in} 
\hspace*{1.3cm}\epsfxsize=2.4 in  \epsfbox{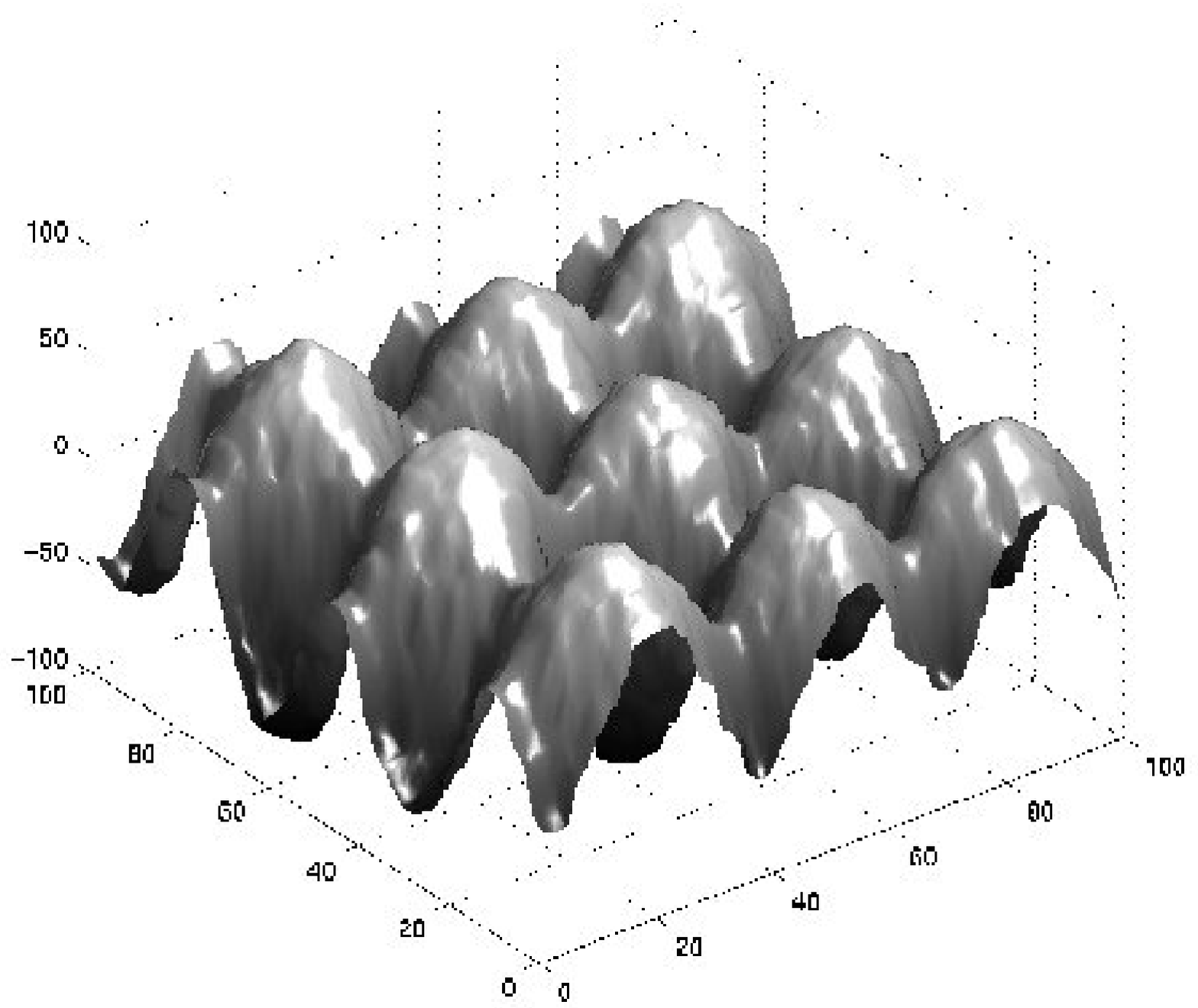} 
\end{minipage}\\\vspace{0.2cm} 
\caption{
Morphologies from WV model in the presence of an ES barrier
with $n_r=1$ (top) and
$n_r=5$ (bottom) from substrates of size $L=100 \times 100$
after $10^6$ ML.
\label{wvessur}} \end{figure}

The noise reduction method,
originally introduced \cite{KW,WK} to study the Eden model, 
proceeds as follows:  
a site is chosen randomly (with uniform probability $L^{-d}$),  
then the $l$-step random walk (``diffusion'') is performed  
according to the local diffusion rules of the model. 
Assume that the site where the atom would 
be incorporated is site $i$. Instead of actually 
depositing the atom at site $i$, first   
a counter $c_i$ associated with site $i$ is increased. The deposition 
attempt becomes a ``true'' deposition process only if the counter at the 
attempt site reaches a predetermined
threshold value $n_r$. After a true 
deposition the counter is reset to zero, and the process is repeated. 
The integer $n_r$ is the {\it noise reduction factor} and it  
quantifies the level of noise suppression, larger values meaning 
a stronger supression. Obviously $n_r =1$ simply means that 
there is no noise reduction. 
The noise reduction technique has been successfully used
in limited mobility MBE growth models
\cite{ppsds,sdsppdtes}.
The effects of noise reduction can also be understood as a coarsening 
process, as shown in Ref. \cite{BKW}: noise reduction performs an effective  
coarsening of the shot noise within regions of 
linear size $\ell$, which is defined 
as the linear distance within which no nucleation of new islands can 
form on the terraces.

An important observation is that when the noise reduction
technique (NRT) is used, the `material parameters' such as
the noise strength, diffusion constant, step stiffness, etc.
in general become dependent on the
level of noise reduction \cite{BKW}, which is quantified by the
noise reduction parameter $n_r$. 

\begin{figure}[htbp]
\begin{minipage}{2.4 in} 
\hspace*{1.3cm}\epsfxsize=2.4 in  \epsfbox{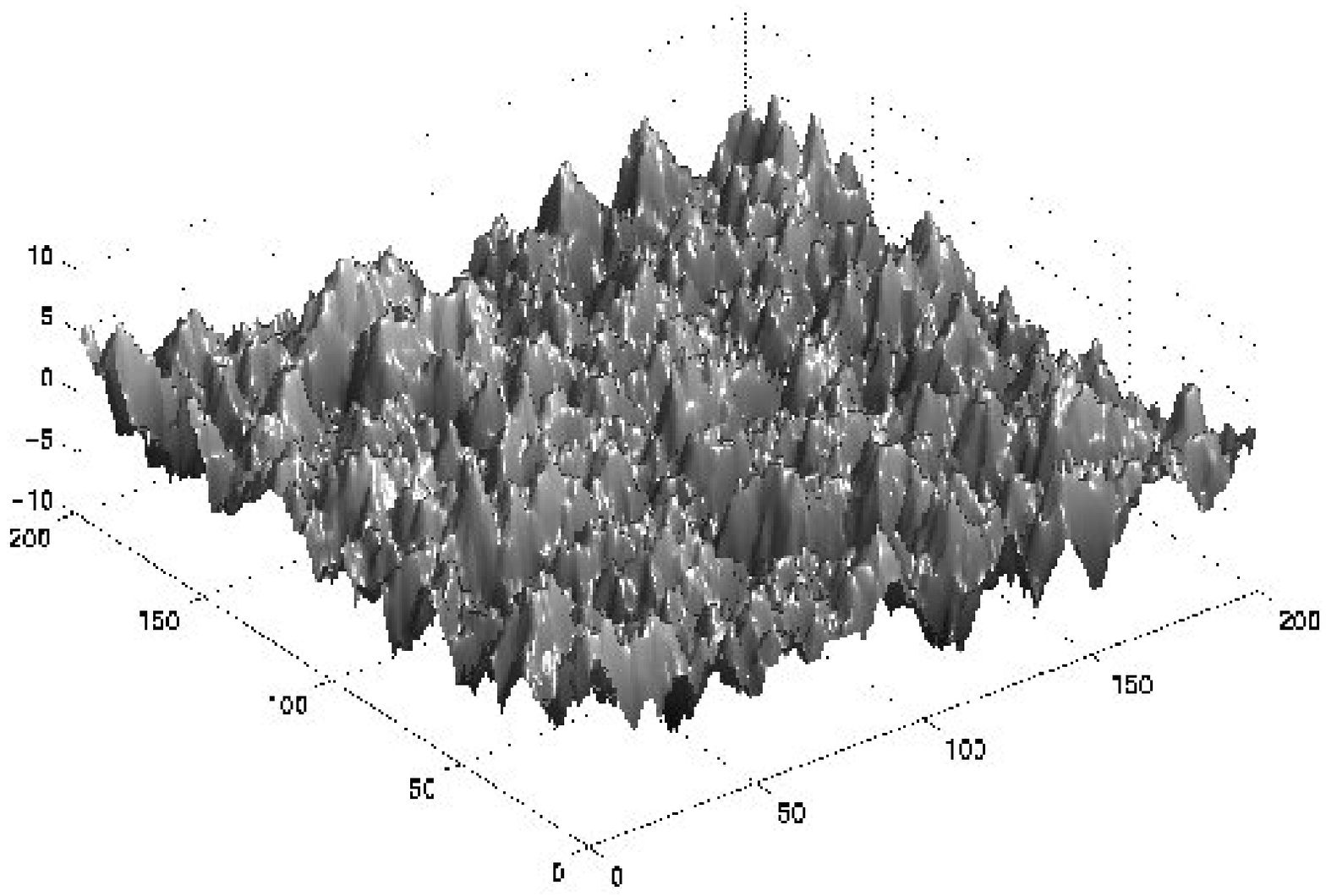} 
\end{minipage}\\ 
\noindent\begin{minipage}{2.4 in} 
\hspace*{1.3cm}\epsfxsize=2.4 in  \epsfbox{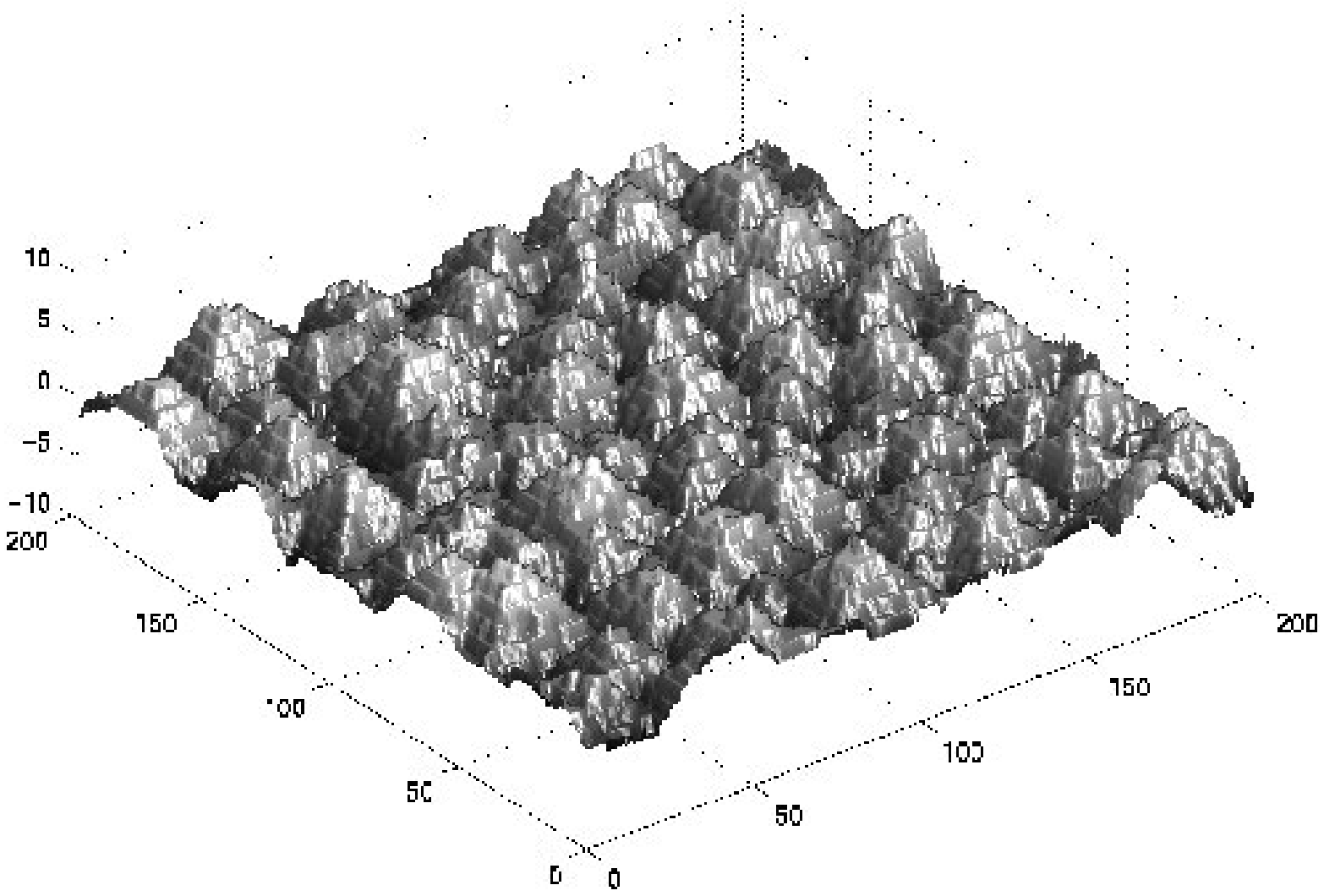} 
\end{minipage}\\\vspace{0.2cm} 
\caption{
Morphologies from LC model with $n_r=1$ (top) and
$n_r=8$ (bottom) from substrates of size $L=200 \times 200$
after $10^3$ ML.
\label{4lsur}} \end{figure}

Brendel, Kallabis and Wolf have
recently identified \cite{BKW} the dependence of 
the material parameters
on the level of noise reduction for kinetic roughening described
on coarse-grained scale by the KPZ equation,
$\frac{\partial h}{\partial t}
= \nu \nabla^2 h + \lambda (\nabla h)^2 + F \Omega
+\eta$ in the context of layer-by-layer growth. In this 
case it turns out that the parameters $\nu$ and 
$\lambda$ assume power law behavior: $\nu \sim (n_r)^{e_{\nu}}$
and $\lambda \sim (n_r)^{e_{\lambda}}$, where the exponents
are not universal and may depend on the details of the
way in which the model is simulated. A simple argument by Kert\'esz
and Wolf \cite{KW} shows that the noise strength $D$ is weakened
by NRT according to:
\begin{equation}
D(n_r) \propto 1/n_r\;.  \label{nm}
\end{equation}

Two of the current authors have earlier studied 
\cite{ppsds} the effect of noise
reduction in the growth simulations of limited mobility 
DT, WV, and F models, finding that NRT is extremely effective
in suppressing crossover and correction to scaling problems
in these growth models, leading effectively to the asymptotic
critical exponents rather compellingly.
An important feature of the current article is to obtain
simulated 2+1-dimensional growth morphologies of the various
growth models under NRT.
We find that NRT is extremely effective in suppressing noise
and giving rise to the epitaxial mounded morphologies in
models where SED mechanism is operational.

\begin{figure}[htbp]
\begin{minipage}{2.4 in} 
\hspace*{1.3cm}\epsfxsize=2.4 in  \epsfbox{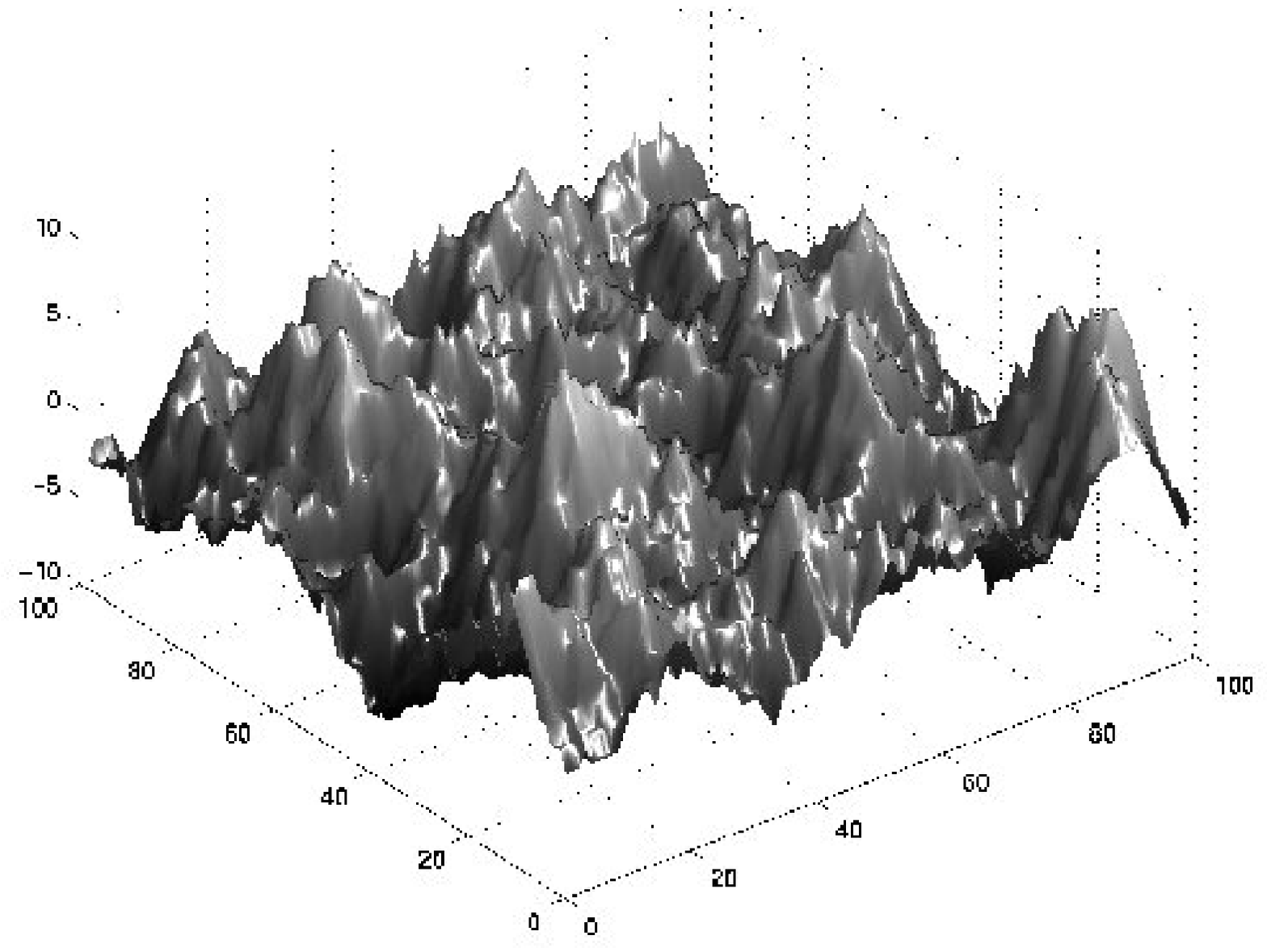} 
\end{minipage}\\ 
\noindent\begin{minipage}{2.4 in} 
\hspace*{1.3cm}\epsfxsize=2.4 in  \epsfbox{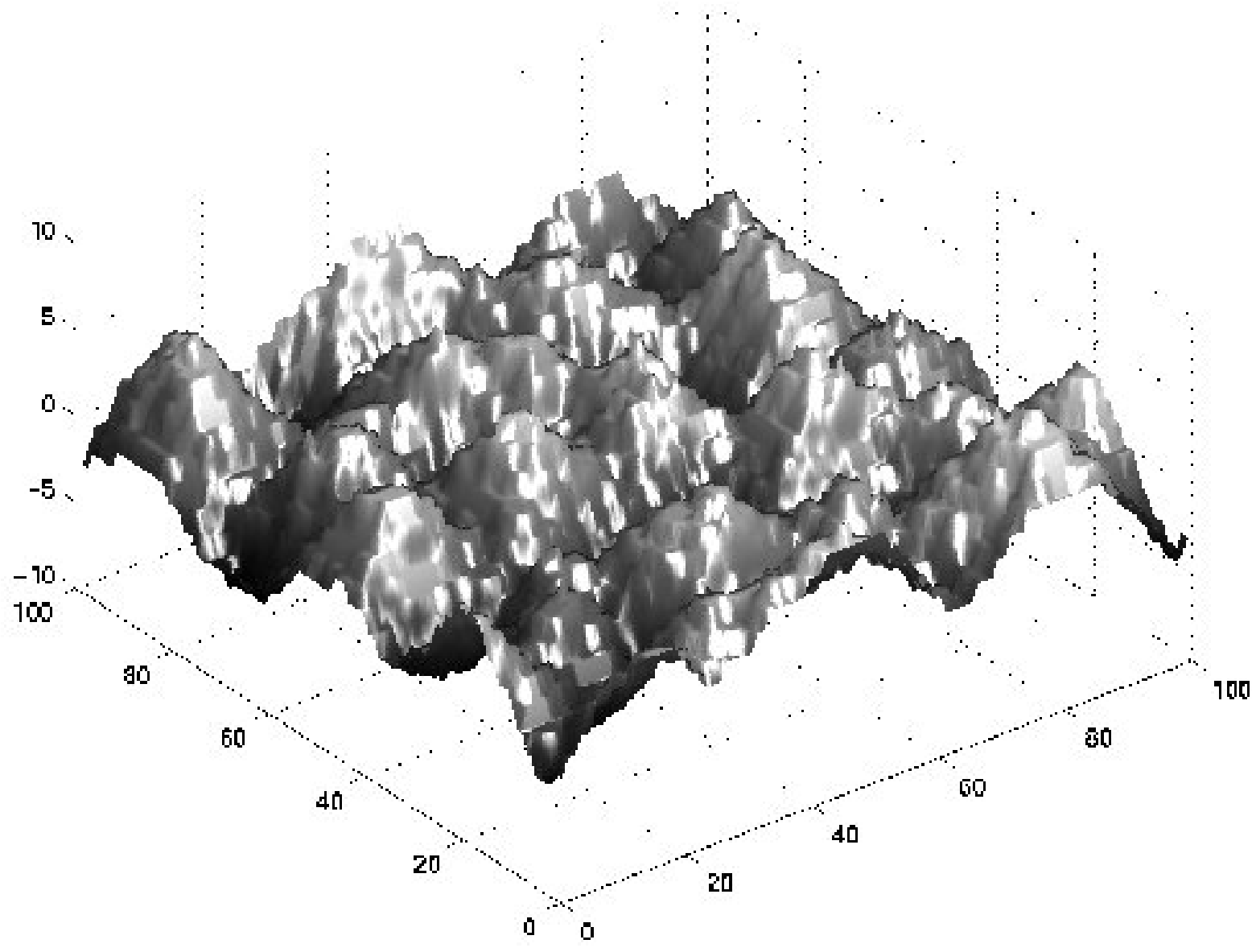} 
\end{minipage}\\\vspace{0.2cm} 
\caption{
Morphologies from nonlinear fourth order
model with $n_r=1$ (top) and
$n_r=5$ (bottom) from substrates of size $L=100 \times 100$
after $10^3$ ML. The coefficient of the nonlinear term is
$\lambda_{22} = 0.5$.
\label{4nsur}} \end{figure}

Often (almost always) the limited mobility models of interest
in our paper are far too noisy (because of the extreme limited
role of diffusion in the models) to exhibit epitaxial mounding
without effective noise reduction, at least in treatable
simulation times.

\subsection{Growth morphologies}

In this section, we show simulated dynamical growth morphologies 
of the discrete SOS models described earlier, with and
without NRT.

The first model is the F model which is known
to be described exactly by the linear second order growth
equation and belong to the well known EW universality class.
As shown in Fig. \ref{fmsur}, F morphology 
with $n_r=1$ (without NRT)
is already quite
smooth. However, when noise is suppressed with $n_r=5$,
the layer-by-layer growth process persists as long as 
$10^6$ monolayers, which is our longest simulation time for
the system. With layer-by-layer growth, the interface is
extremely smooth and flat and this agrees with the EW universality 
definition where the growth ($\beta$) and the roughness
($\alpha$) exponents in 2+1 dimensions is zero
corresponding to smooth growth.

\begin{figure}[htbp]
\begin{minipage}{2.4 in} 
\hspace*{1.3cm}\epsfxsize=2.4 in  \epsfbox{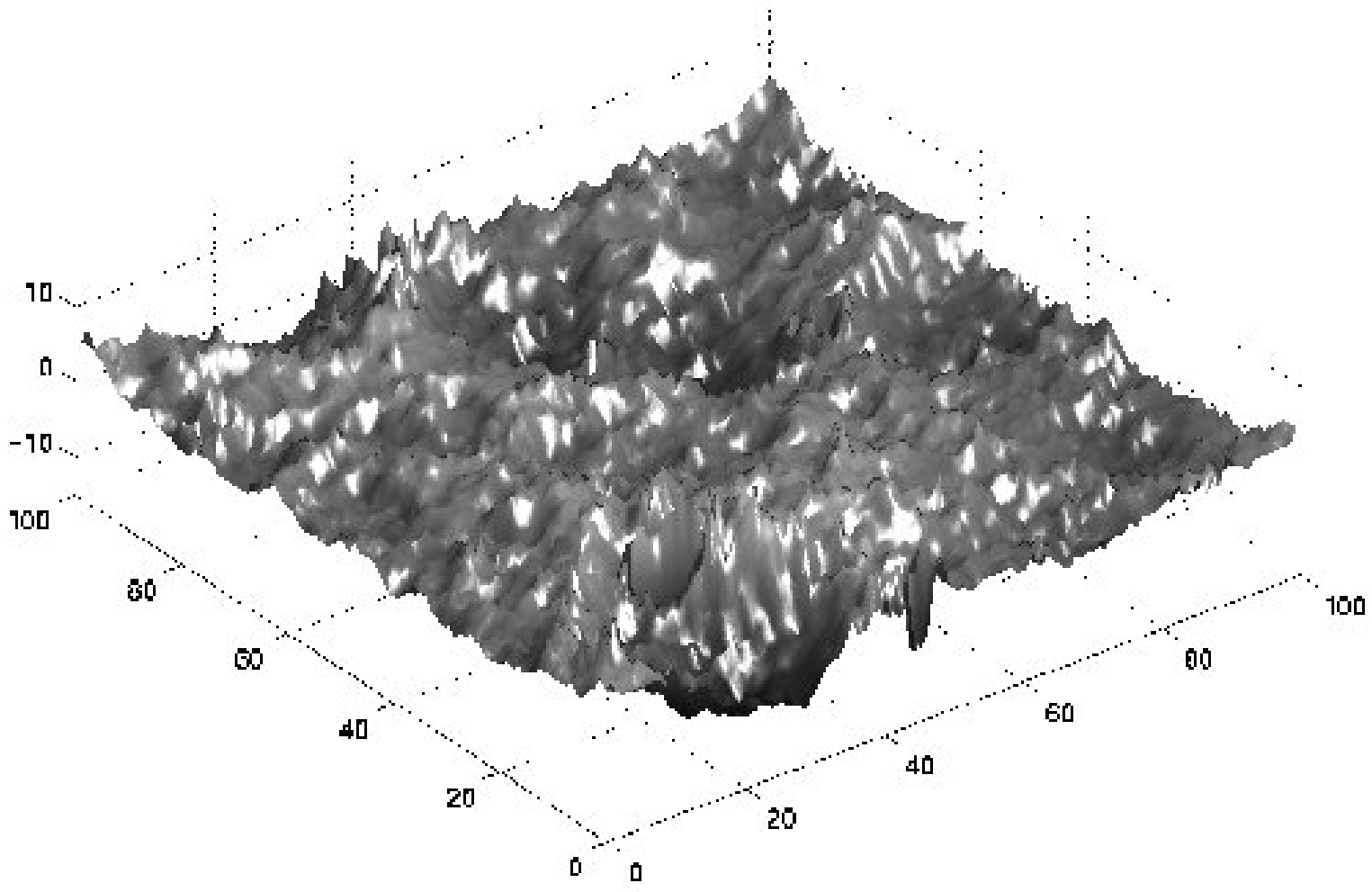} 
\end{minipage}\\ 
\noindent\begin{minipage}{2.4 in} 
\hspace*{1.3cm}\epsfxsize=2.4 in  \epsfbox{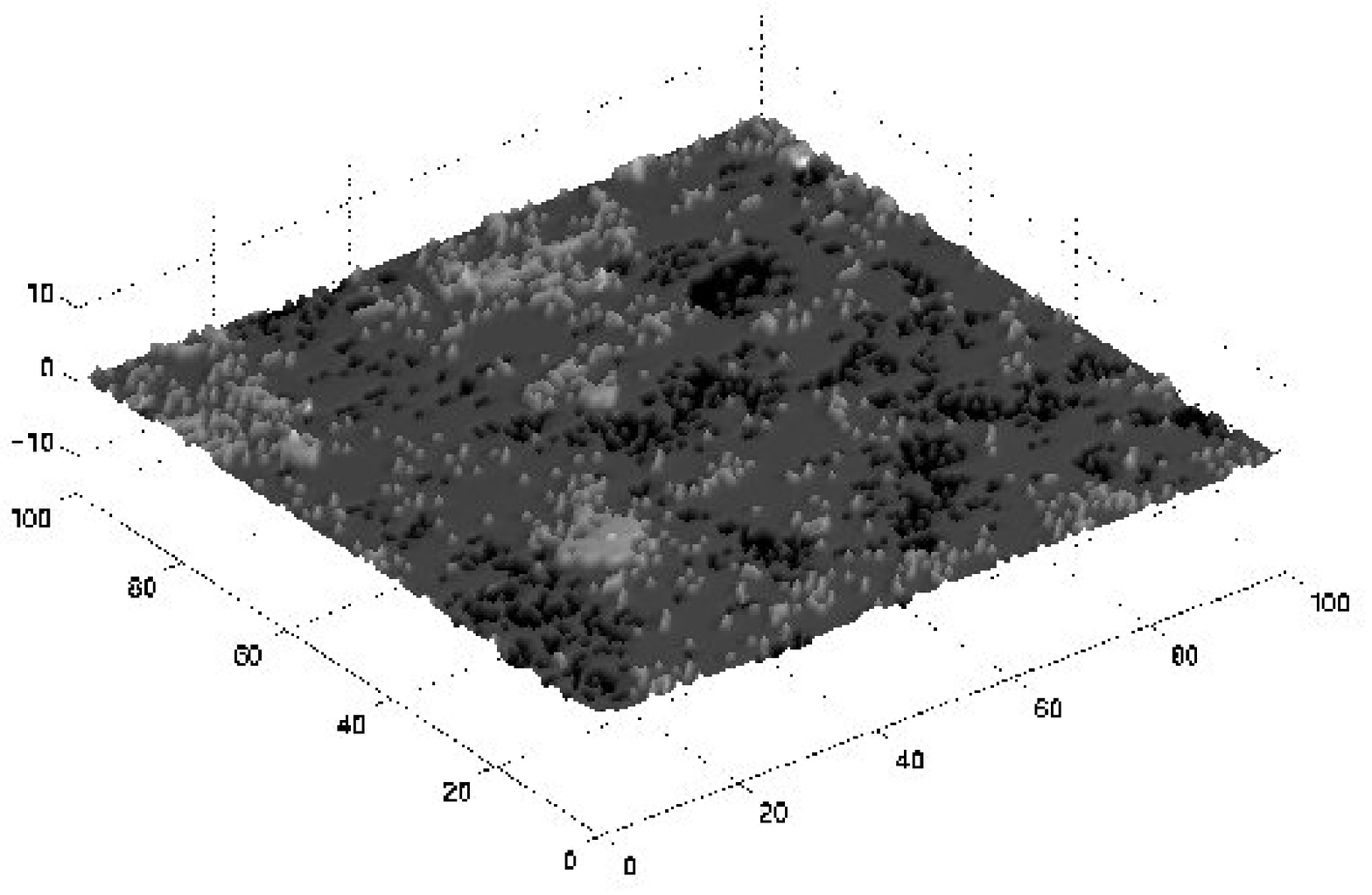} 
\end{minipage}\\\vspace{0.2cm} 
\caption{
Morphologies from nonlinear fourth order model
with an infinite series of higher order nonlinear terms
with $n_r=1$ (top) and
$n_r=5$ (bottom) from substrates of size $L=100 \times 100$
after $10^3$ ML. The coefficients of the nonlinear term are
$\lambda_{22} = 5.0$ and $C=0.085$.
\label{4ssur}} \end{figure}

The DT model, as shown in Fig. \ref{dtsur}, 
shows kinetically rough
interface in the $n_r=1$ simulation. But with $n_r=5$,
the morphology becomes considerably smoother, without 
any specific mounding pattern. 
Similar to the DT model, the WV model 
(Fig. \ref{wvsur}) with $n_r=1$ shows
rough morphology without any special mounding features. 
But when we
suppress the noise, using $n_r>1$, the WV morphology changes
drastically and patterned mound formations can clearly
be seen. 
This mound formation in the WV model is intriguing
because there is no ES barrier involved in the
simulation. As it turns out, the WV diffusion rule 
has an intrinsic machanism to create a local uphill
particle current which is strong under NRT,
and hence a pyramid-like mounded morphology formed.
This topic is discussed in the next section.

We have actually carried out both DT and WV model growth simulations
\cite{sdsppdtes} in the presence of an ES barrier also. For the sake
of comparison with the epitaxial mounding in the WV model
without any ES barrier (but with NRT),
as shown in Fig. \ref{wvsur}, we present in 
Figs. \ref{dtessur} and \ref{wvessur} the observed mounded
morphologies in the DT and WV models in the presence of an
ES barrier (both with and without NRT).
One can clearly see that the ES barrier induced mounding 
instability in Figs. \ref{dtessur} and \ref{wvessur}
is qualitatively different from the epitaxial
mounding in Fig. \ref{wvsur}.
While the ES barrier induced DT/WV mounding has flat tops,
the intrinsic WV mounding of Fig. \ref{wvsur} is of
pyramid shape.

\begin{figure}[htbp]
\begin{minipage}{2.4 in} 
\hspace*{1.3cm}\epsfxsize=2.4 in  \epsfbox{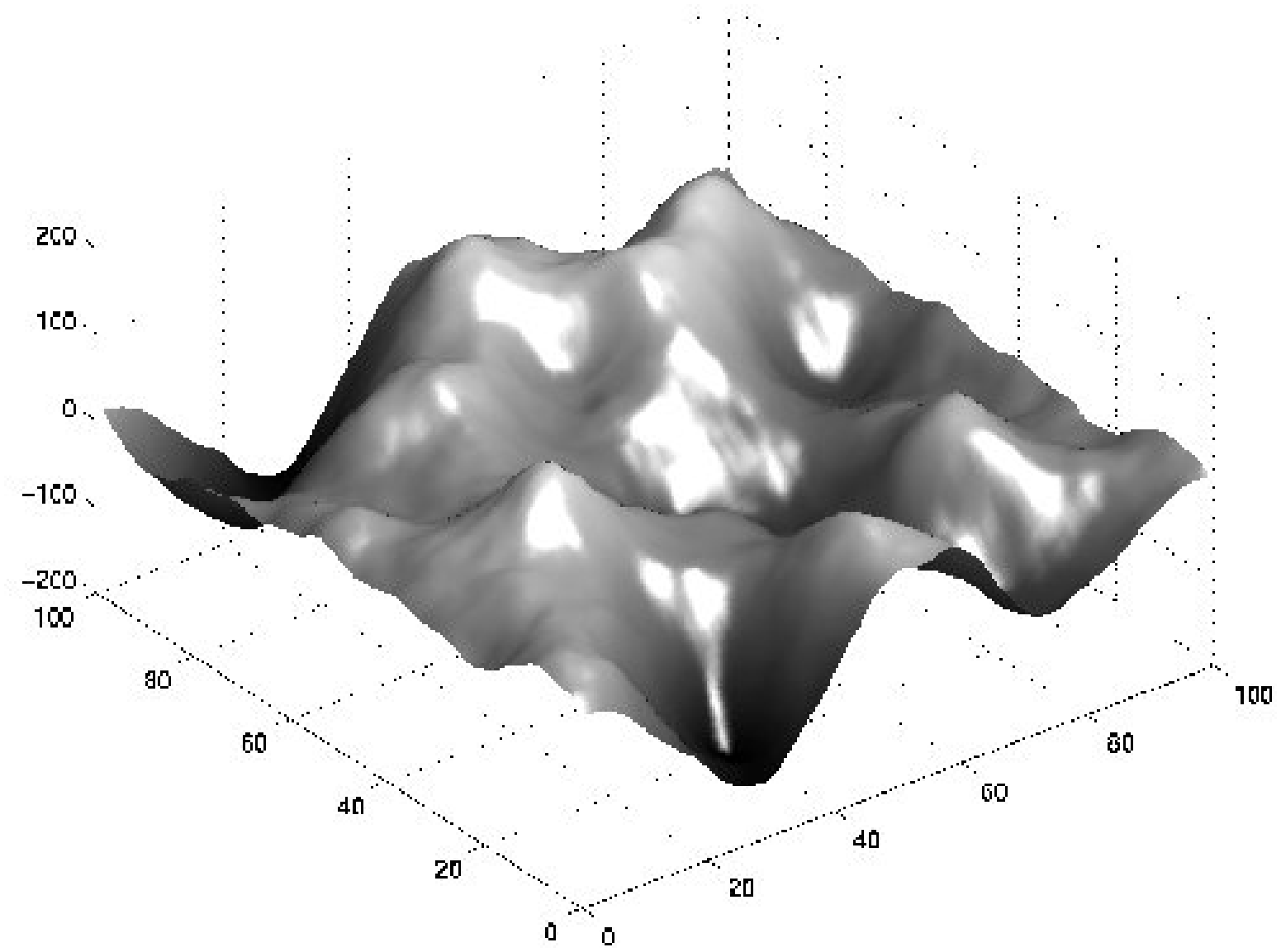} 
\end{minipage}\\ 
\noindent\begin{minipage}{2.4 in} 
\hspace*{1.3cm}\epsfxsize=2.4 in  \epsfbox{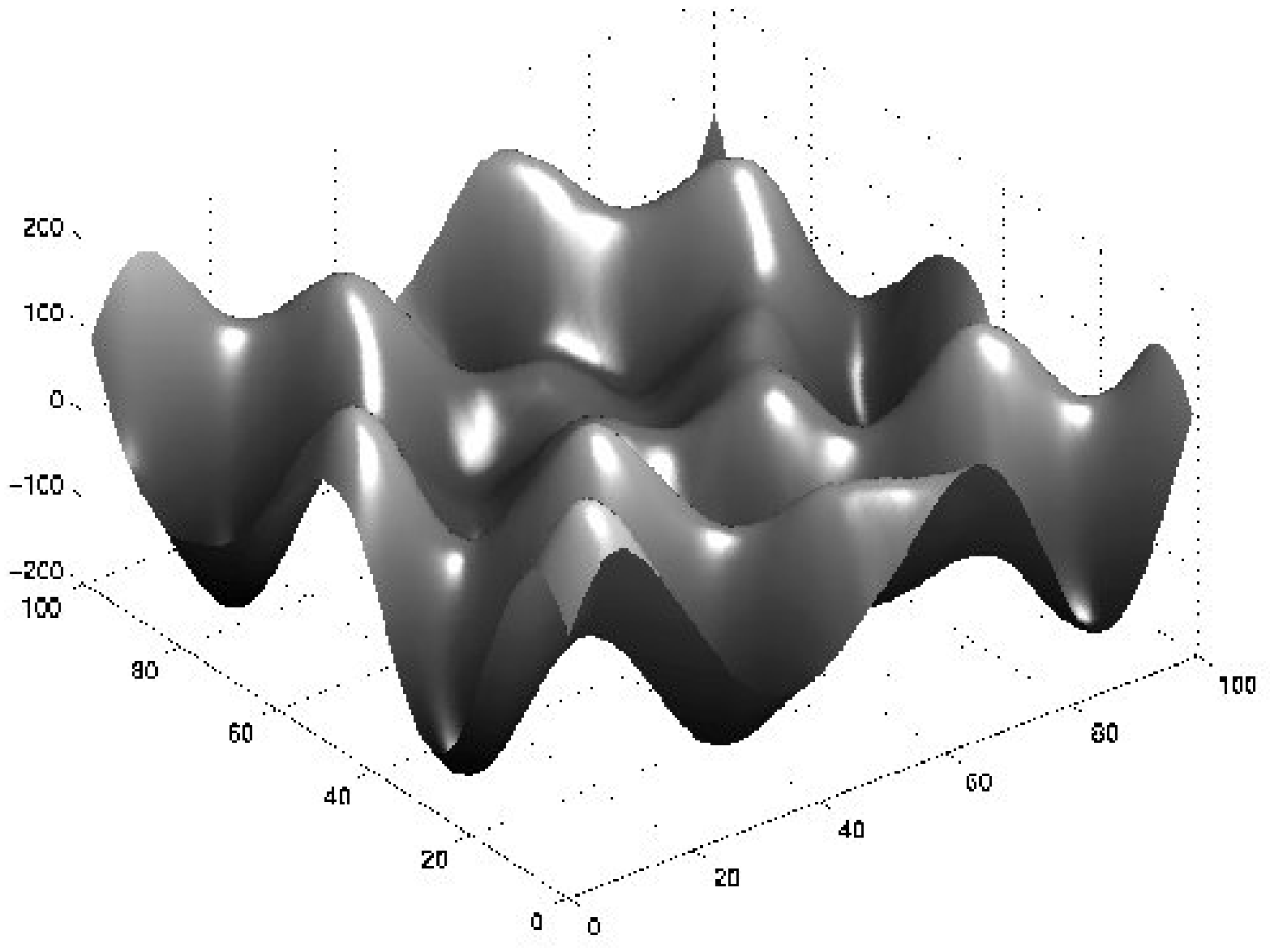} 
\end{minipage}\\\vspace{0.2cm} 
\caption{
Morphologies from linear sixth order model with $n_r=1$ (top) and
$n_r=5$ (bottom) from substrates of size $L=100 \times 100$
after $10^6$ ML.
\label{6lsur}} \end{figure}

The linear fourth order model (the LC model)
also creates mounds without ES barrier in a way similar to
the WV model. This is shown in Fig. \ref{4lsur}
(and also Fig. \ref{color}).

Although the epitaxial mounding in Figs. \ref{wvsur} and 
\ref{4lsur} 
look similar,
the mounded morphologies in LC and WV
models are not identical, however, as the LC is a linear model
and hence the morphology exhibits up-down symmetry, while 
the WV morphology does not look the same when it is turned
upside down,
and therefore does not possess the up-down symmetry.
This lack of up-down symmetry in the WV model arises
from the presence of the nonlinear 
$\nabla^2(\mbox {\boldmath $\nabla$} h)^2$
term in its dynamics.
When we add a small instability into the LC equation,
by adding the nonlinear fourth order term with a small
coefficient, the interface (shown in Fig. \ref{4nsur})
is somewhat mounded
even for $n_r=1$ simulation.
But the surface in this $n_r=1$ model is so rough 
that a specific mound pattern cannot be discerned.
When we reduce the noise, using $n_r=5$, clear
mound formation (without up-down symmetry)
can be seen in the morphology as
one would expect.

In Fig. \ref{4ssur} we present our simulation results for the 
full nonlinear equation (Eq. (\ref{4seq}) with $C=0.085$)
with an infinite number of nonlinear terms.
The rationale for studying this particular continuum
equation with an infinite number of nonlinear terms of the
$\nabla^{2n}(\mbox {\boldmath $\nabla$}h)^{2n}$
form is the fact that recently it has been established
\cite{ddk,dasgupta}
that such as infinite order nonlinear dynamical equation
is the most likely continuum description 
(at least in d=1+1 dimensions, where this issue has so far 
been studied intensively) of the discrete DT model.

\begin{figure}[htbp]\vspace*{-0.0cm} \noindent 
\hspace*{0.5cm}\begin{minipage}{2.8 in} 
\epsfxsize=2.8 in  \epsfbox{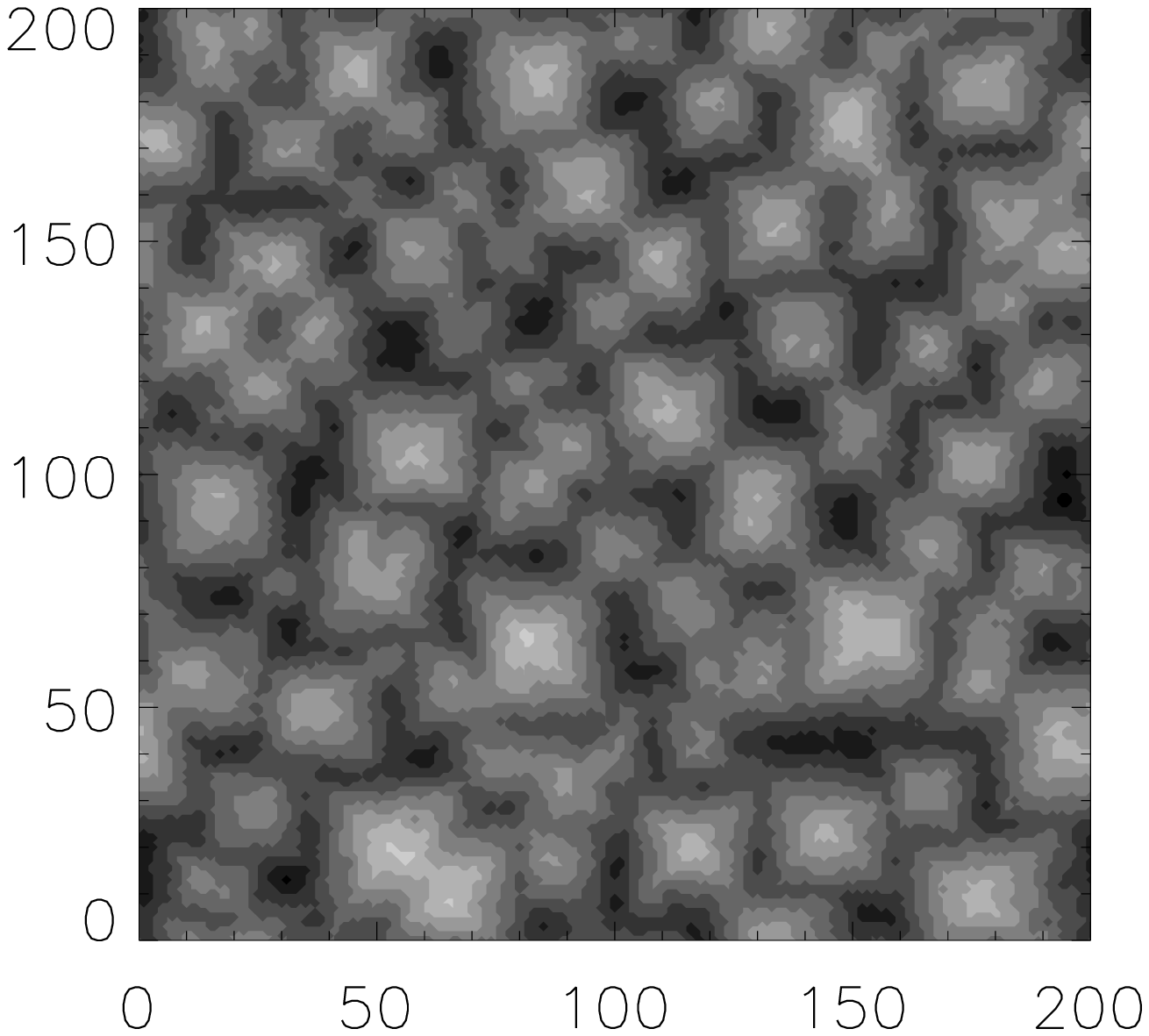} 
\vspace*{-0.3cm} 
\end{minipage}\\ 
\hspace*{0.5cm}\begin{minipage}{2.8 in} 
\epsfxsize=2.8 in  \epsfbox{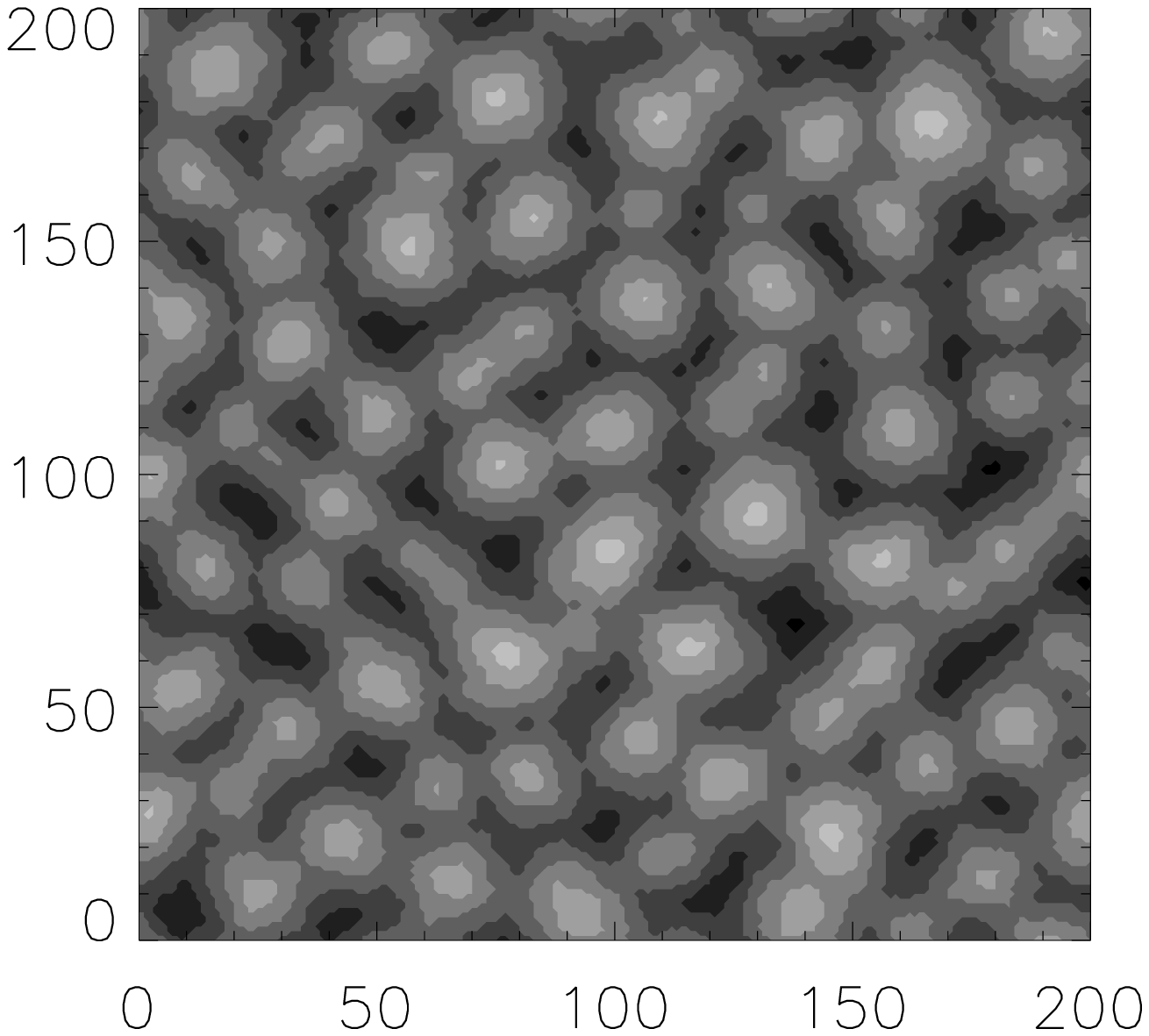} 
\vspace*{-0.1cm} 
\end{minipage}
\caption{
Contour plots of LC morphologies from $L=200 \times 200$
at $10^3$ ML with $n_r = 8$.
Lighter (darker) shade represents higher (lower) points.
Plot (a) corresponds to the  ``higher priority''
version while plot (b) is from the
``equal priority'' version. Here $l=1$.
\label{ehl1}} \end{figure}

Since the DT model does not seem to exhibit
(see Fig. \ref{dtsur}) any epitaxial mounding instability
in our simulations, it is of considerable interest to see
how the infinite order nonlinear continuum equation behaves.
It is, therefore, gratifying to see, as shown in Fig. \ref{4ssur},
that the growth morphology in the infinite order nonlinear
equation does not exhibit any discernible mounding patterns ---
the morphologies depicted in Fig. \ref{4ssur} look vaguely
similar to the kinetically rough growth of the DT model
shown in Fig. \ref{dtsur}.
It is, however, remarkable that the fourth order continuum
equation (either the linear LC model as in Figs. \ref{color}
and \ref{4lsur} or the nonlinear model with a small nonlinearity
as in Fig. \ref{4nsur}) manifests striking mound formation,
but the infinite order nonlinear equation does {\it not}
manifest any epitaxial mounding.
We do not have a precise mathematical understanding of this
dichotomy other than pointing out that the presence of
an infinite series of nonlinear terms clearly overwhelm
the SED instability arising from the $\nabla^4 h$ surface
diffusion term.
The precise understanding of the combined effects of 
noise reduction and growth nonlinearities is not available
at the present time.
Finally, in Fig. \ref{6lsur} we present the surface 
morphologies associated with the linear
sixth order growth equation. The mounding pattern can be
barely seen in the $n_r=1$ simulation. 
But the spectacular mound formation
can be seen clearly when we reduce the noise ($n_r=5$).

\begin{figure}[htbp]\noindent
\hspace*{0.8cm}\begin{minipage}{2.8 in} 
\epsfxsize=2.8 in  \epsfbox{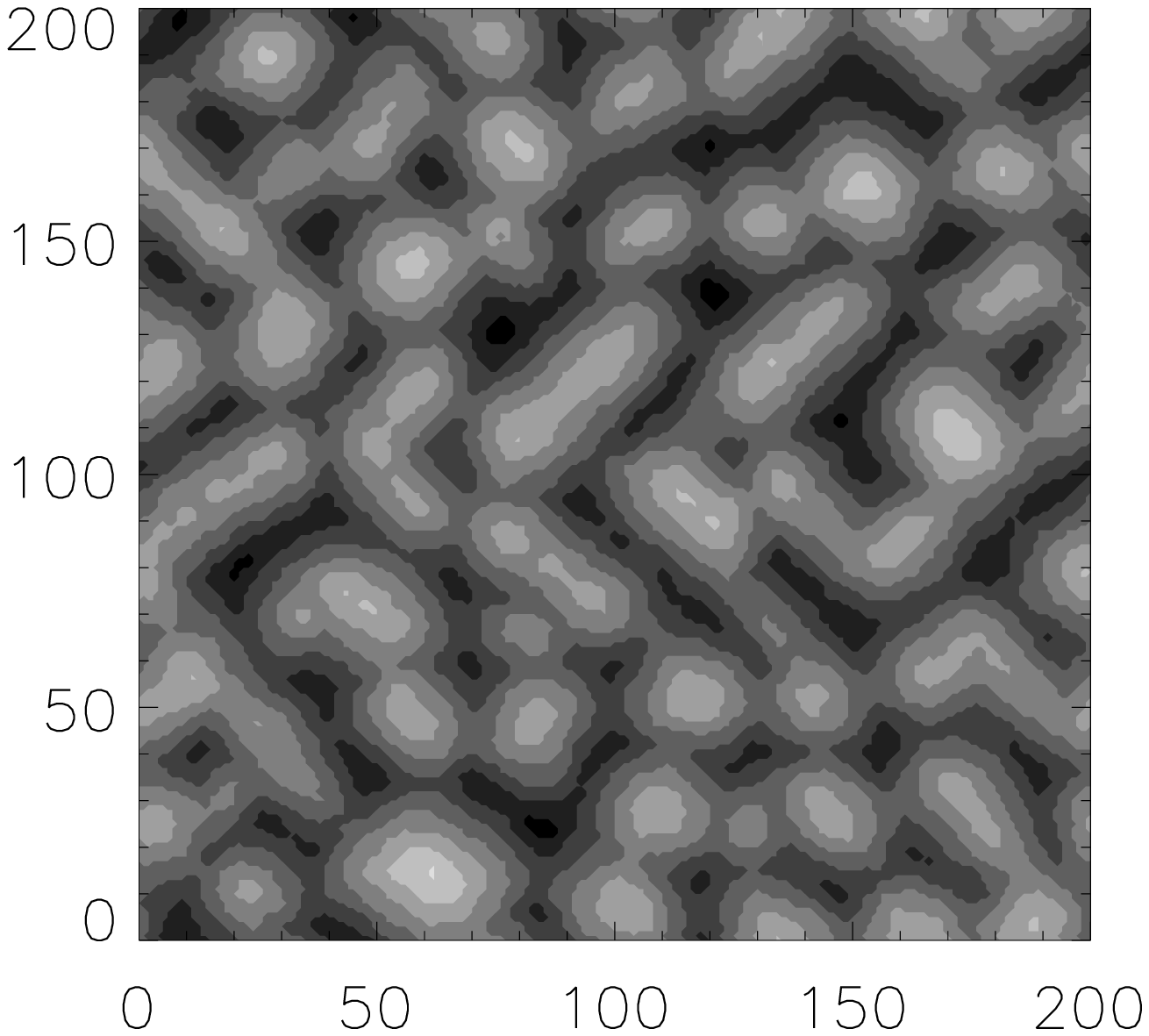} 
\vspace*{-0.1cm} 
\end{minipage}\\ 
\hspace*{0.8cm}\begin{minipage}{2.8 in} 
\epsfxsize=2.8 in  \epsfbox{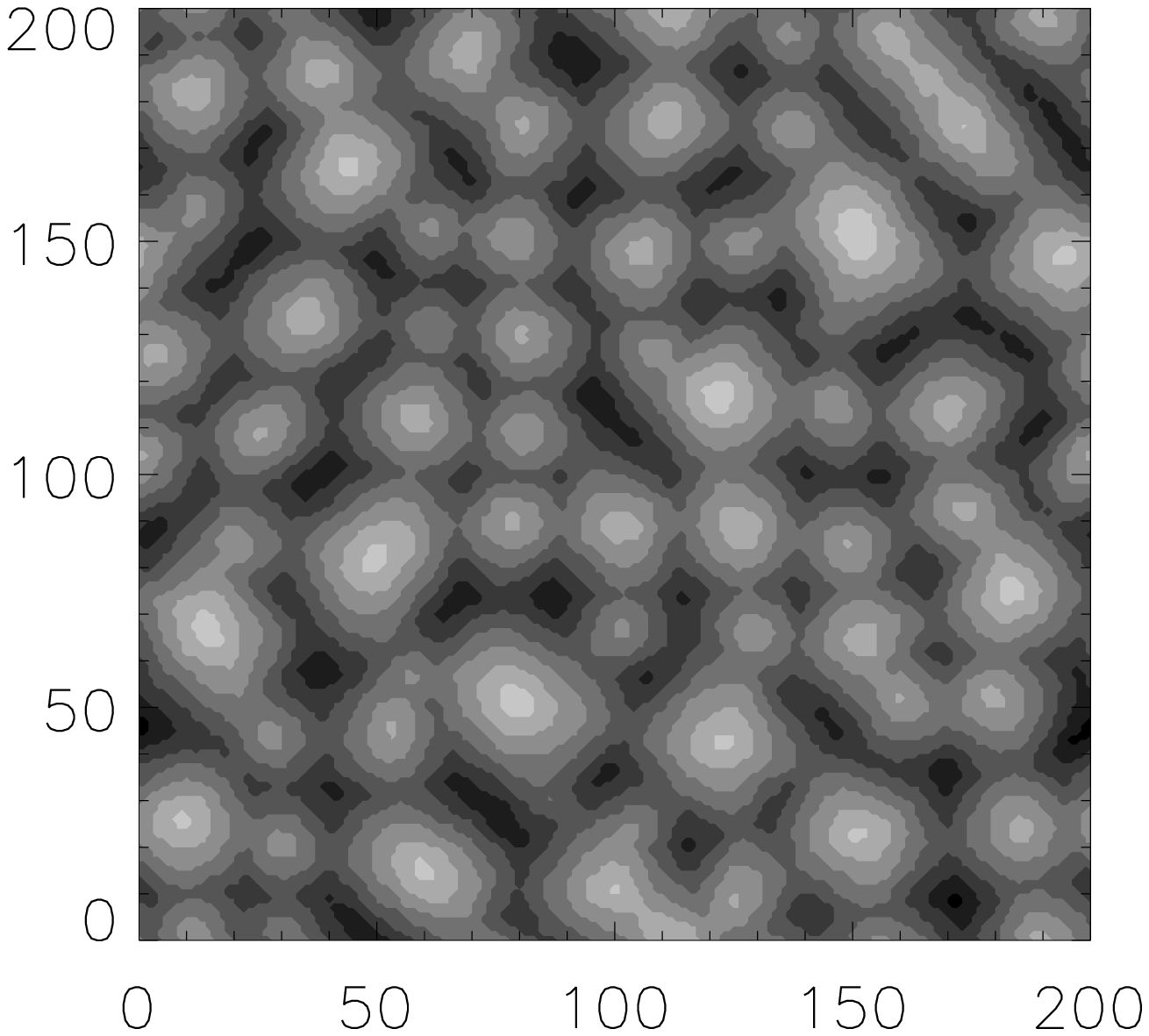} 
\end{minipage} 
\vspace{0.2cm} 
\caption{
Contour plots of LC morphologies from $L=200 \times 200$
at $10^3$ ML with $n_r = 8$.
Lighter (darker) shade represents higher (lower) points.
Plot (a) corresponds to the  ``higher priority''
version while plot (b) is from the
``equal priority'' version. Here $l=2$
\label{ehl2}} \end{figure}

These simulated morphologies are very 
sensitive to the local atomistic
diffusion rules, as can be seen when we make some minor
modifications to the LC rule. 
The first thing we modify is to make the adatoms more
``energetic''. In the original LC model, 
an adatom searches for a site
with the largest curvature, and if the original random deposition 
site is one of the sites with the largest curvature then the
adatom remains at the deposition site. So in the original version
of the model, an adatom will not move if it is already at one
of the preferred sites. 
We call this original model the ``higher priority'' version 
of the LC model, as the deposition sites have {\it higher priority}
than the neighboring sites. 
In the modified version, called ``equal priority'' 
version, the adatom 
chooses its final site among all the sites with largest curvature
with equal probability. So even if the original deposition site
is one of the largest curvature sites, the adatom may still end
up being incorporated at a neighboring site if that neighbor
also has the same curvature as the deposition site.
The second modification we make is simply to increase the
diffusion length $l$ from nearest-neighbor-only diffusion
($l=1$) to $l>1$ where diffusion rules now extend to
$l(>1)$ sites. 

\begin{figure}[htbp]  \noindent
\hspace*{0.8cm}\begin{minipage}{2.8 in} 
\epsfxsize=2.8 in  \epsfbox{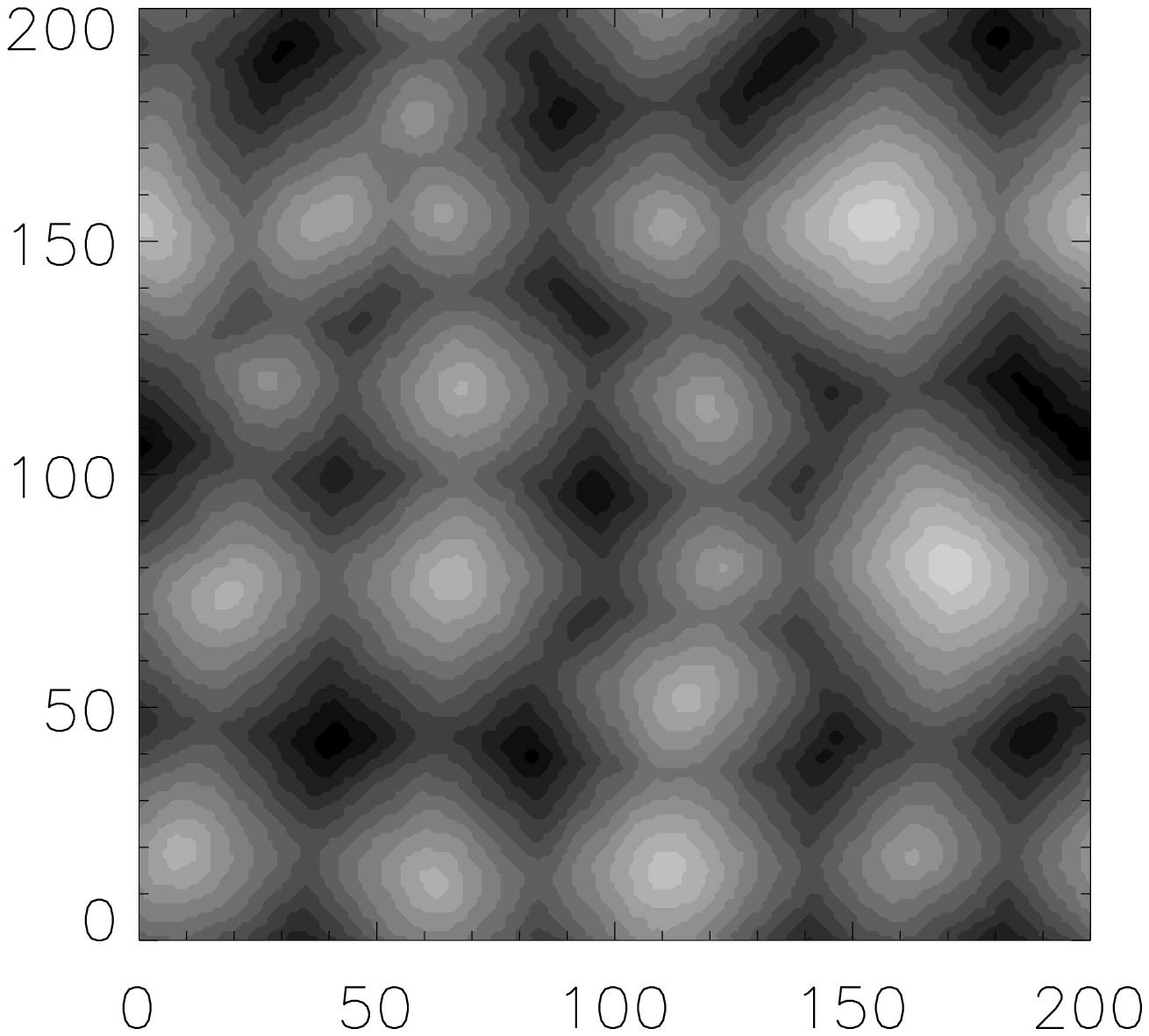} 
\vspace*{-0.1cm} 
\end{minipage}\\ 
\hspace*{0.8cm}\begin{minipage}{2.8 in} 
\epsfxsize=2.8 in  \epsfbox{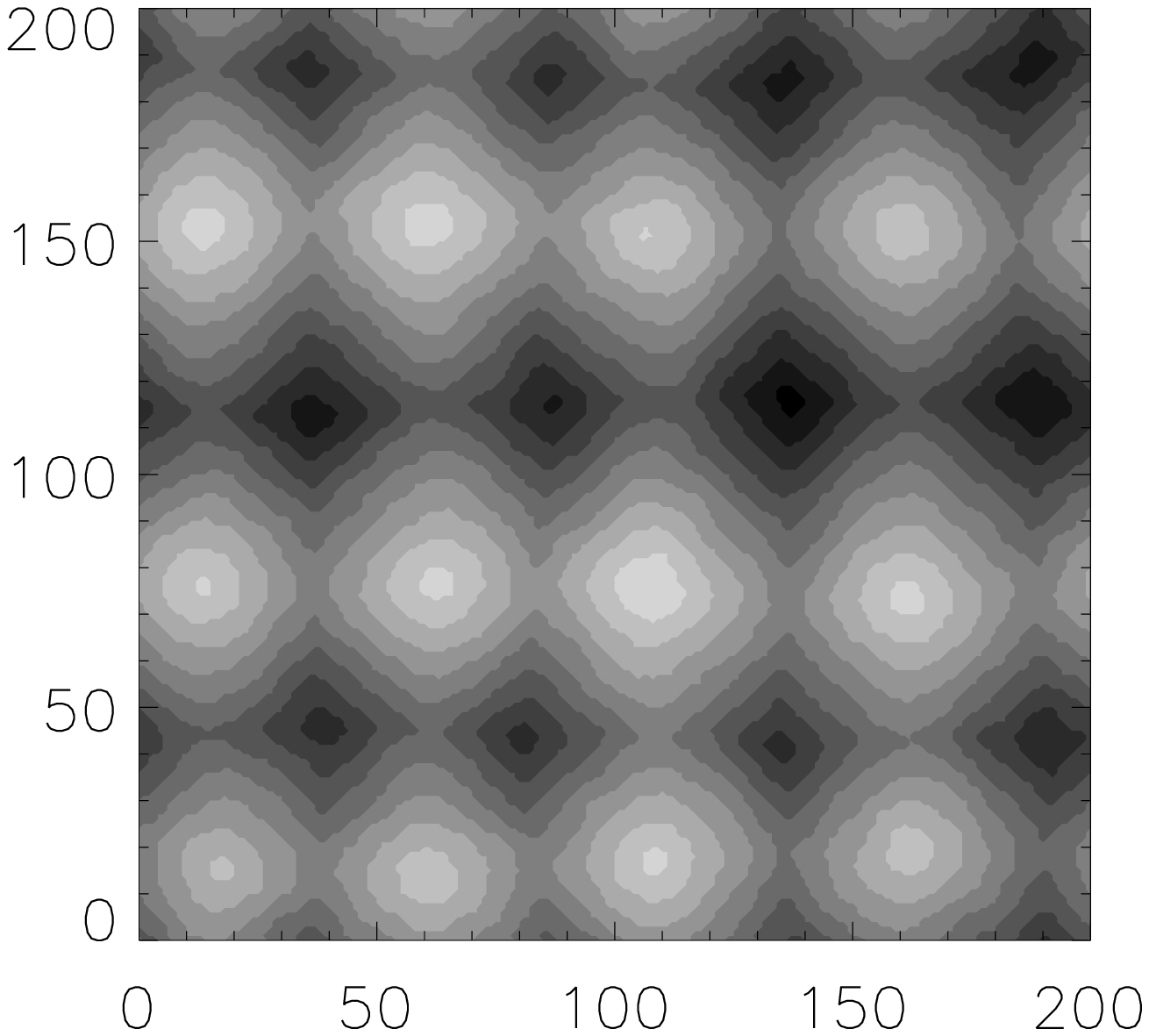} 
\vspace*{-0.1cm} 
\end{minipage} \\
\hspace*{0.8cm}\begin{minipage}{2.8 in} 
\epsfxsize=2.8 in  \epsfbox{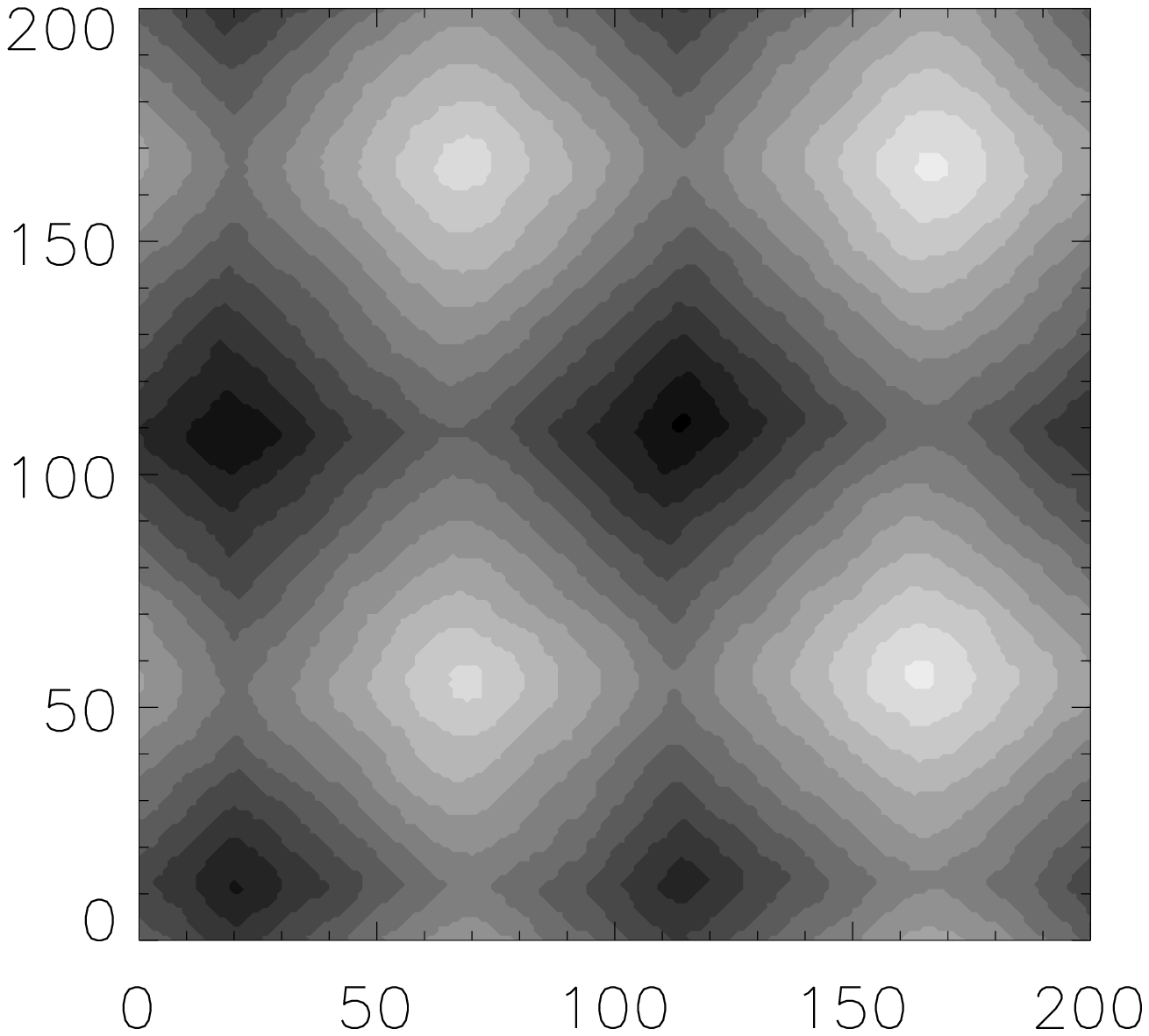} 
\vspace*{-0.1cm} 
\end{minipage}
\vspace{0.2cm} 
\caption{Contour plots of ``equal priority'' LC model
with $l=2$ and $n_r=8$ at (a) $10^4$ ML,
(b) $10^5$ ML and (c) $10^6$ ML.
\label{long}} \end{figure}

It is known \cite{dt1} that 
increasing diffusion length does not change the asymptotic
universality class of limited mobility growth models
as long as the substrate 
size is large enough so that finite size effects remain
neglegible. But increasing diffusion length can most certainly
change the early-time morphology of the system by strongly
affecting finite size effects.

Both versions of LC model show mounded morphologies,
as can be seen in Fig. \ref{ehl1}. But
it seems that the ``equal priority'' version, the one with more
energetic adatoms, has mounds with the bases lined in the
110 direction. In the ``higher priority'' version the
mounds are more stable when the bases are in 100 direction.
Increasing diffusion length in the simulations encourages
faster coarsening process (Fig. \ref{ehl2}), 
as should be expected, and
individual mounds in the $l=2$ simulations ``link''
to each others more than in the $l=1$ simulations.
To further investigate the mound coarsening process,
we study the ``equal priority'' LC model
with $l=2$ in long time limit and find that the
mounds rearrange themselves into a regular pattern
as shown in Fig. \ref{long}.
This regular re-arrangement of the mound morphology, caused by
longer diffusion lengths as well as the different symmetry
properties (110 versus 100) of the equal priority and
higher priority LC models, show the extreme sensitivity of the
epitaxial mound morphology to the details of the diffusion
process controlling growth --- in general, we find the SED
instability in all our simple models to be quite complex in
the sense of the mound patterns (and even whether mounds 
exist or not as in WV and DT models)
seem to be very sensitive to the details of the 
surface diffusion rules.

\section{Kinetic-topological ``instability'': step-edge diffusion}

\subsection{A continuum description}

\begin{figure}[htbp]
\begin{minipage}{1.0 in}\epsfxsize=1.0 in  
\epsfbox{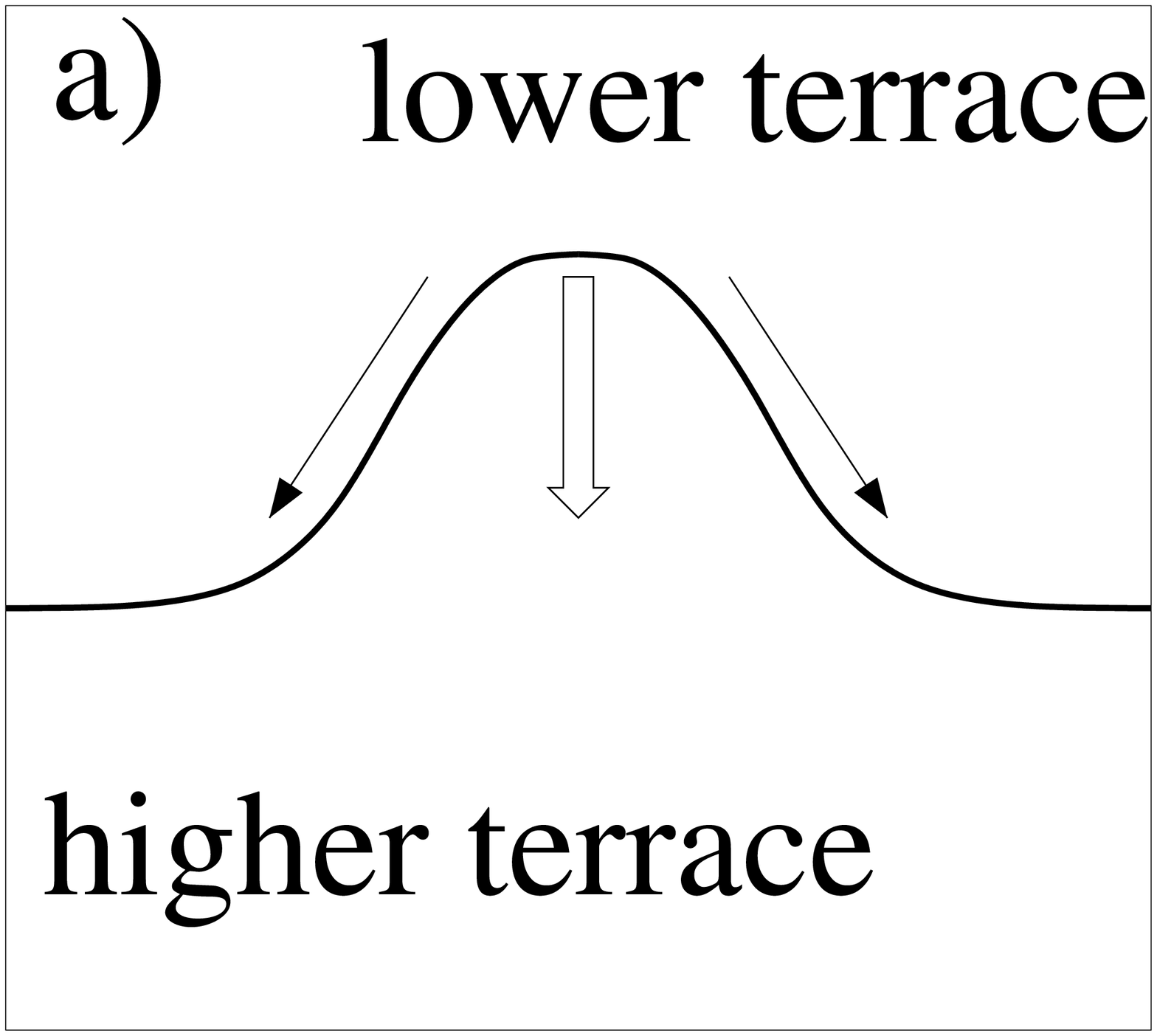} 
\end{minipage} 
\hspace*{0.1cm} 
\begin{minipage}{1.0 in}\epsfxsize=1.0 in  
\epsfbox{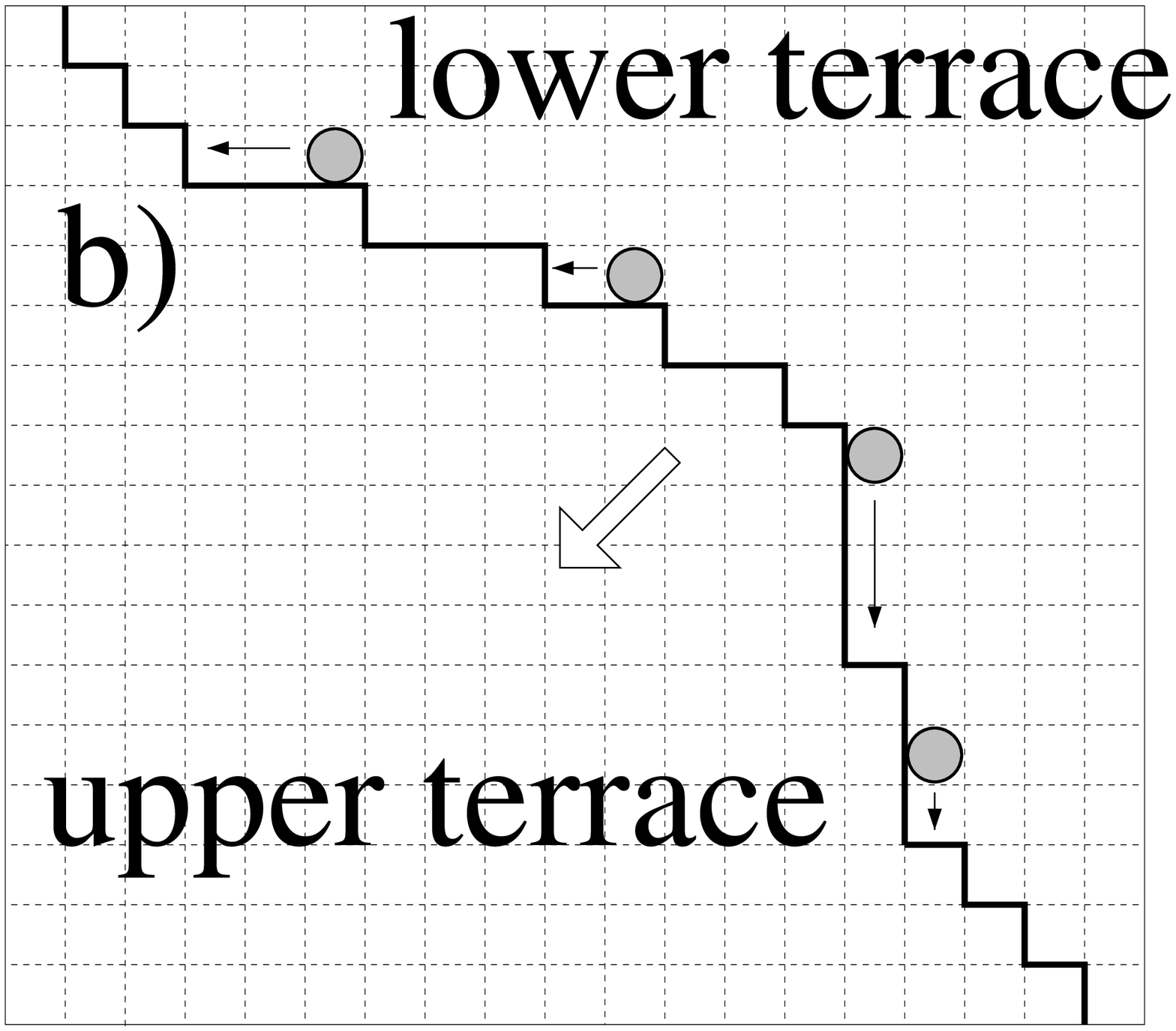} 
\end{minipage} 
\hspace*{0.1cm} 
\begin{minipage}{1.0 in}\epsfxsize=1.0 in  
\epsfbox{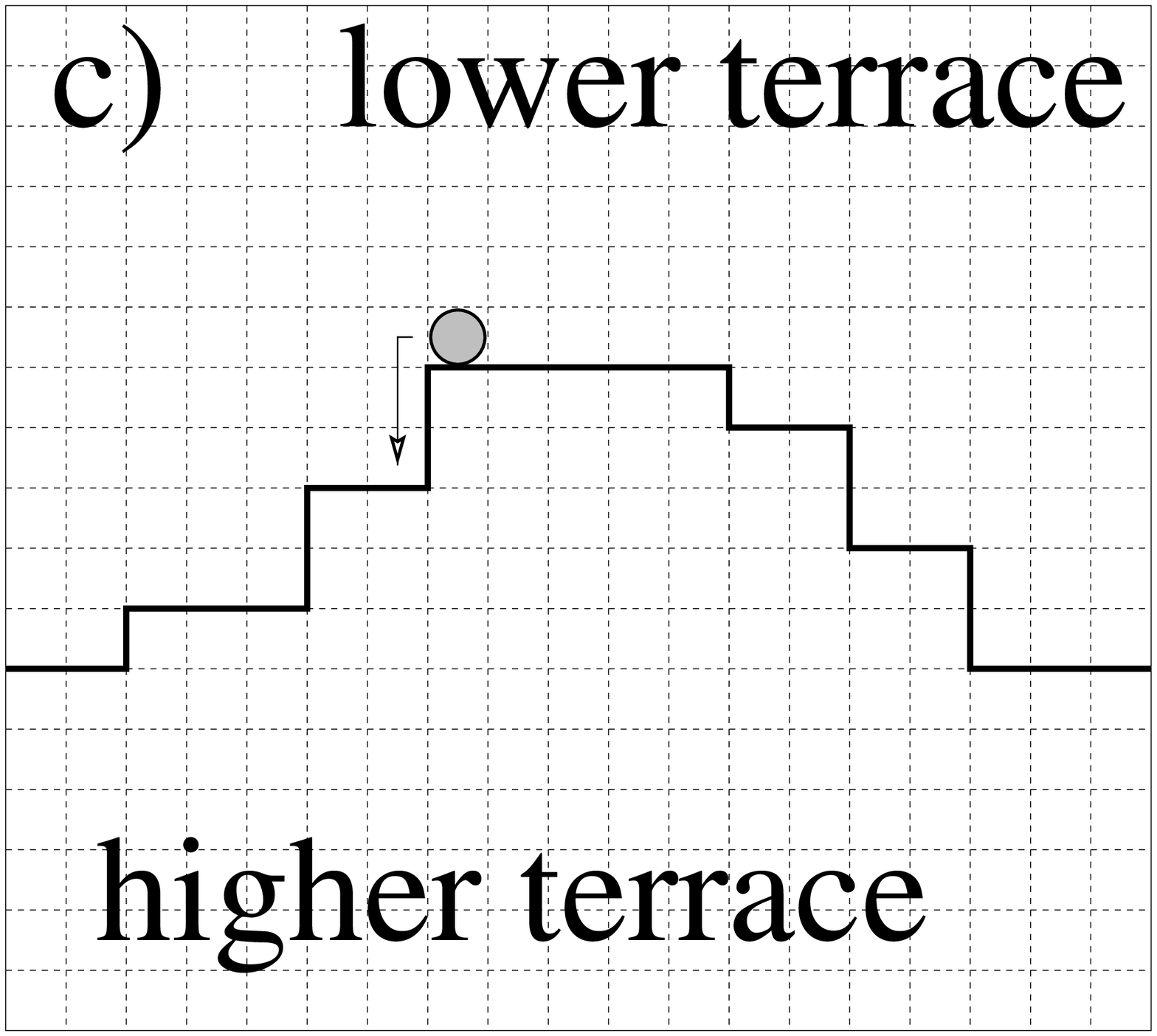} 
\end{minipage} 
\vspace{0.5cm} 
\caption{
Schematic illustration of the instability caused
by the line tension in the step-edge. View from above,
the thick line is the boundary between two terraces whose
difference in elevation is one monoatomic layer. a) shows
a continuum picture while b) and c) represent
the same on a lattice
for different step-boundary orientations.
\label{schematic}} \end{figure}

It has been recently recognized \cite{edge} 
that step-edge diffusion generates a local {\it uphill} 
surface diffusion current during the deposition process, 
leading to a morphological ``instability'',
i.e. the starting substrate is no longer stable after growth.  
This instability is purely 
of topologic-kinetic nature, and it is {\it not} generated
by an energetic instability 
such as the Ehrlich-Schwoebel instability. It is rather
a consequence of the fact that surface diffusion currents
(in two or higher dimensions)
are vectorial quantities, and while some of their components
have a smoothening and stabilizing effect in a certain subspace, 
in other directions they may act as instabilities, as is
shown schematically in Fig. \ref{schematic}. This figure is a top view 
of a two dimensional substrate on which there are two vicinal
surfaces. Due to diffusion, the adatoms on the flat terraces
may reach the edge of the terrace, however they are assumed
to feel no energetic barriers at the step edges.
After reaching the edge of the step the adatoms will tend to
remain attached
to it since it is an energetically more favorable position
(with at least one more extra bond when compared to the position
on the terrace). 
The atoms on the step edge may relocate
through {\it line-diffusion} along the edge to energetically even
more favorable places (kink sites), thus reducing the differences
in the local chemical potential. (For simplicity we assume no
desorption from the growth front.)
For example, if there is a protuberance on (or a dent in) the step 
as shown in Fig. \ref{schematic} (a), 
the step-edge atoms will preferentially diffuse
towards the base of the hump (or towards filling up the dent),
creating a net current towards the region of larger height,
i.e. {\it uphill} as shown by the thick white arrow in 
Fig. \ref{schematic} (a).

\begin{figure}[htbp]
\epsfxsize=3.3 in   
\epsfbox{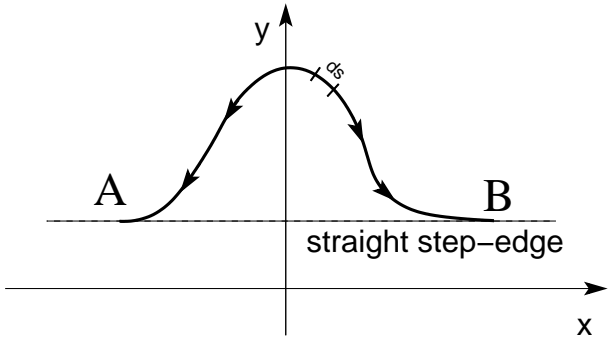}  \vspace{0.0cm}  
\caption{
A straight step-edge with a hump. The step-edge current
density generated by the local curvature gradient
along the hump creates a net uphill current.
\label{hump}} \end{figure}

Note that although this current stabilizes the step-edge, 
it {\it destabilizes} the $2+1$ dimensional surface $h(x,y)$ 
at long wavelengths,
creating mounds on the substrate
induced by the uphill current. The local current
density
${\bf J_l}$ along the step-line is proportional to the local gradient
along the edge of the chemical potential $\mu$, 
i.e., one can write 
${\bf J_l}=-\nu \mbox {\boldmath $\nabla$}_e \mu$, 
where $\nu$ is the mobility of the
atoms along the step edge \cite{K}. Assuming local thermodynamic
equilibrium, the chemical potential may be expressed as a
functional of a free energy which in turn is written as that of 
an elastic
line of line-tension $\sigma$. 
The result is that $\mu$ is 
proportional to the local curvature $\kappa$ of the line, 
$\mu=-\sigma \kappa$ \cite{K}. $\kappa$ 
is the local curvature on the line
$h(x,y)=\mbox{const}$, and thus it can be expressed in terms of
the local derivatives of $h$:
\begin{eqnarray}
{\bf J_l} = 
-\sigma \nu \mbox {\boldmath $\nabla$} \left( |\nabla h|^{-3}
\left|
\begin{array}{ccc}
 h_{xx} & h_{xy} & h_x \\ h_{yx} & h_{yy} & h_y \\
h_x & h_y & 0 
\end{array} \right| \right) \label{jl}
\end{eqnarray}
Next, we calculate the
uphill current due to the  current density (in Eq.
(\ref{jl})) created by an
arbitrary shaped hump created at the edge of a 
single straight step, and then specialize the
expression for a simple, cosine shaped hump.
We assume that one can choose an $(x,y)$ coordinate system
such that the equation for the step-profile $h(x,y)=const$
in this system is simply expressed by a single valued function: 
$y=y(x)$, see Fig. \ref{hump}.
In this simple setup the edge is parametrized
by $x$, and thus the step-edge current in Eq.
(\ref{jl}) becomes:
\begin{equation}
{\bf J_l} = \nu \sigma \;
\frac{y^{(3)} \left[ 1+(y')^2 \right]-3 y' (y'')^2}
{\left[ 1+(y')^2 \right]^{7/2}}
\left({\bf e_x} + y' {\bf e_y} \right) \label{pjl}
\end{equation}
where $y^{(3)}, y''$ and $y'$ are the third, second and first
order derivatives of the profile, and ${\bf e_x}$ and 
${\bf e_y}$ are the unit vectors along the $x$ and $y$ axes,
respectively.  The total mass
transported per unit time by curvature gradients  
between points
$A$ and $B$ on the step-edge is defined as the line-integral of the
edge current density:
\begin{equation}
{\bf I}_{AB} = \int_{A}^{B} ds\; {\bf J_l} \label{uph}
\end{equation}
where $ds$ is the integration element along the curve of the
step-edge.  In the following we illustrate
that this current points toward the base of the higher step.

Let us choose a step-profile given by 
$y(x)=a \cos{(b x)}$ between the points
$x_A = -\pi/b$ and $x_B = \pi/b$. After performing
the integrals in Eq. (\ref{uph}), one obtains:
\begin{eqnarray}
{\bf I}_{AB}
= - \frac{4 \nu\sigma b}{15}\left[ 
\frac{1+11 z + 6 z^2}{(1+z)^2}
{\bf E}(i\sqrt{z}) \right. && \nonumber \\
\left. -\frac{1+3z}{1+z}{\bf K}(i\sqrt{z}) \right]
{\bf e_y} &&
\end{eqnarray}
where $z = a^2 b^2$ is a dimensionless number, and ${\bf K}(k)$
and ${\bf E}(k)$ are complete elliptic integrals of first and
second kinds respectively.
Since $1+11z+6z^2 > 1+4z+3z^2$, and 
${\bf E}(i\sqrt{z}) > {\bf K}(i\sqrt{z})$, for all 
$z > 0$, the mass-current
has a negative $y$ component  (the $x$ component is 
zero because of symmetry),
i.e., it points towards the base of the higher step, 
uphill. In the  limit of small
humps, $z \ll 1$ (small slopes approximation),
${\bf K}(i\sqrt{z})\simeq \frac{\pi}{4}(1-\frac{z}{4})$,
${\bf E}(i\sqrt{z})\simeq \frac{\pi}{4}(1+\frac{z}{4})$, so
the current becomes proportional
to the square of the hump's height, $I_{AB} \sim - \pi \nu \sigma
b^3 a^2$ and for
peaked humps ($z \gg 1$),
${\bf K}(i\sqrt{z})\simeq \ln{(4\sqrt{z})}/\sqrt{z}$,
${\bf E}(i\sqrt{z})\simeq \sqrt{z}$
the current depends linearly on the height
of the hump, $I_{AB} \sim \frac{8}{5} \sigma \nu b^2 a$. 
The same analysis can be repeated
when there is a dent in the step profile, with similar conclusions.
Certainly, the currents expressed above are instantaneous 
local currents. 
If one is interested in the evolution of the step profile, one 
can employ simple geometrical considerations 
(see \cite{K,Mullins}) to write:
\begin{equation}
\frac{\partial y}{\partial t} = v_n \sqrt{1+(y')^2} \label{14}
\end{equation}
where $y=y(x,t)$,  $v_n$ is the normal step-edge velocity
and the prime denotes derivative with respect to $x$.
Assuming that the mass transport is solely due to the step-edge
current, one has $v_n ds =- (\partial J_l / \partial x)
dx$, with $ds=dx \sqrt{1+(y')^2}$. Thus Eq. (\ref{14})
takes the form:
\begin{eqnarray}
\frac{\partial y}{\partial t} = 
-\frac{\partial J_l}{\partial x}
= -\nu\sigma \left( \frac{y^{(4)}}{[1+(y')^2]^2}-
\frac{10y'y''y^{(3)}}{[1+(y')^2]^3}\right.&& \nonumber \\
 \left. -\frac{3[1-5(y')^2](y'')^3}
{[1+(y')^2]^4}\right).&& \label{full}
\end{eqnarray}
In the limit of small slopes, $y' << 1$, $y'' << 1$, 
$y^{(3)} <<1 $, $y^{(4)} <<1$, and so from Eq. (\ref{full})
it follows that: $\partial y / \partial t = -\nu \sigma y^{(4)}$, 
after keeping the leading term only.
This is the well known Mullins equation for the dynamic relaxation
of a step-edge,
related to the fourth order linear growth equation,
Eq. (\ref{k4neq}) with $\lambda_{22}=0$,
followed by the LC model in our growth simulations.

The continuum description presented here compellingly demonstrates 
the possibility of a destabilizing uphill current arising entirely
from step edge diffusion --- an SED instability without any
ES barriers,
as discussed in Refs. \cite{edge,ecu} recently.

\subsection{A discrete description}

In reality the deposition process (and our simulation of
limited mobility models)
takes place on the discrete and atomistic
crystalline lattice, which introduces an orientation dependence
of the line tension $\sigma$. One would naturally expect that
the high-symmetry, in-plane crystalline directions will have
the largest $\sigma$, these being the most stable. However,
there is a hierarchy even among these high-symmetry orientations,
as is illustrated in Figs. \ref{schematic} (b) and 
\ref{schematic} (c).   Fig. \ref{schematic} (b) shows a 
step-edge aligned along a diagonal of the square lattice and
in Fig. \ref{schematic} (c) 
the step-edge is oriented along one of the main
axes of the lattice. 
While the step is stable along the diagonal,
it is not as stable along the main axis, since in this latter case
(of Fig. \ref{schematic} (c))  in order for an atom to reach a higher 
coordination site along the line, it has to detach itself from the
step-edge (to become an adatom) first, which is energetically less 
favorable.

Let us now analyze in more detail the effects of SED 
in the case of the two most important MBE-motivated
limited mobility models, WV and DT models.
The diffusion rules for the mobile atoms for these two
models are only slightly different: in the case of WV model
particles
seek to maximize their coordination while in the DT model they 
only seek to increase it. Nevertheless the two models
present  completely different behaviors. As we have seen in 
Subsection II.C, the NRT version of WV model produces mounded
structures, however this does not happen for the DT model.
In the following we explain this unexpected
difference between WV and DT models based on local SED currents. 
Given the rather ``minor'' differences between DT and WV rules
(and their almost identical growth morphologies in 1+1 dimensions)
it is remarkable how different their 2+1 dimensional NRT growth
morphologies are (WV has spectacular mounded morphology as seen in
Fig. \ref{wvsur} and DT has essentially statistically scale invariant
kinetically rough growth as in Fig. \ref{dtsur}).

\begin{figure}[htbp]\noindent
\begin{minipage}{2.4 in}  
\hspace*{1.3cm}\epsfxsize=2.4 in  \epsfbox{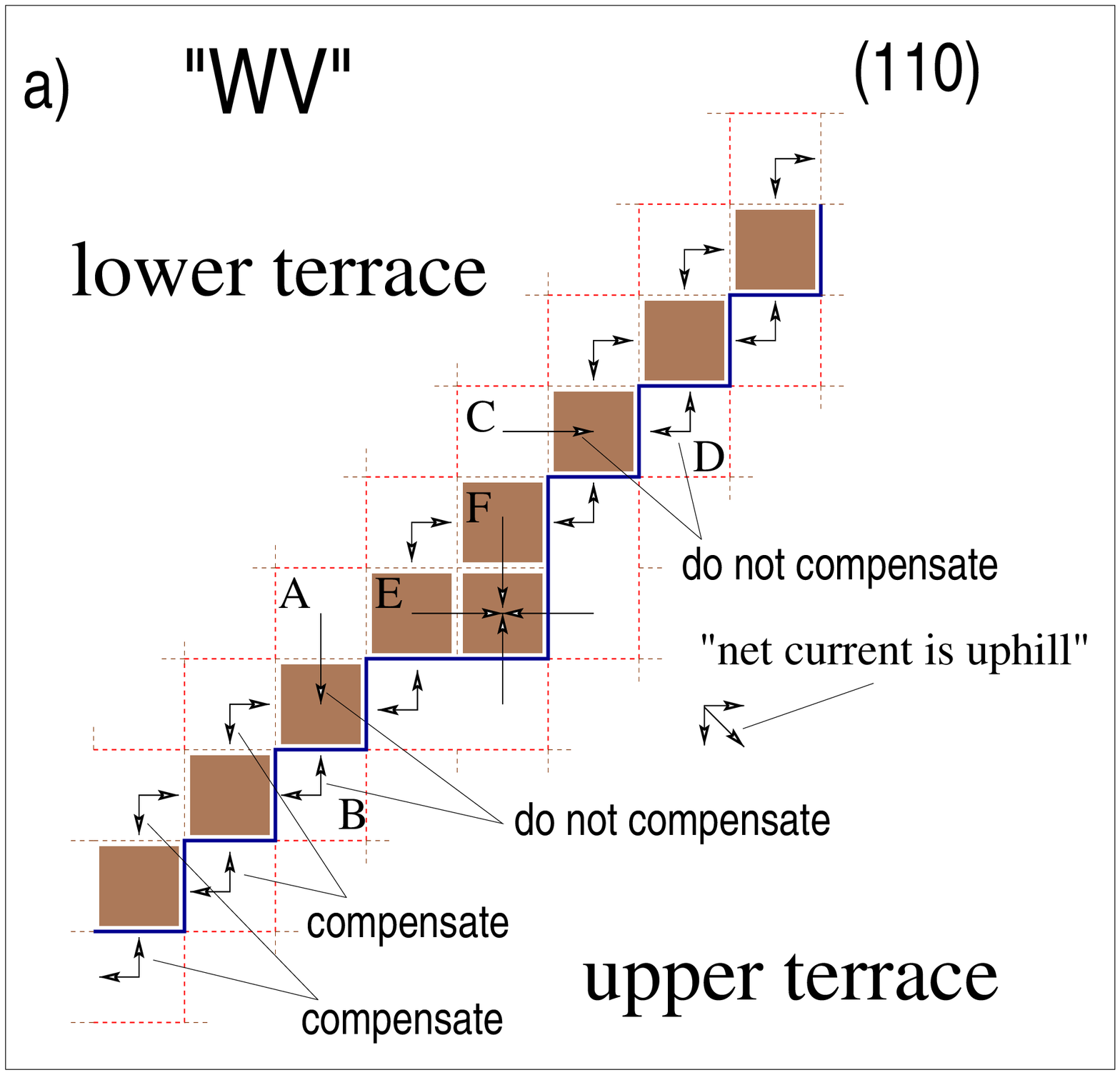} 
\end{minipage}\\ 
\noindent\begin{minipage}{2.4 in} 
\hspace*{1.3cm}\epsfxsize=2.4 in  \epsfbox{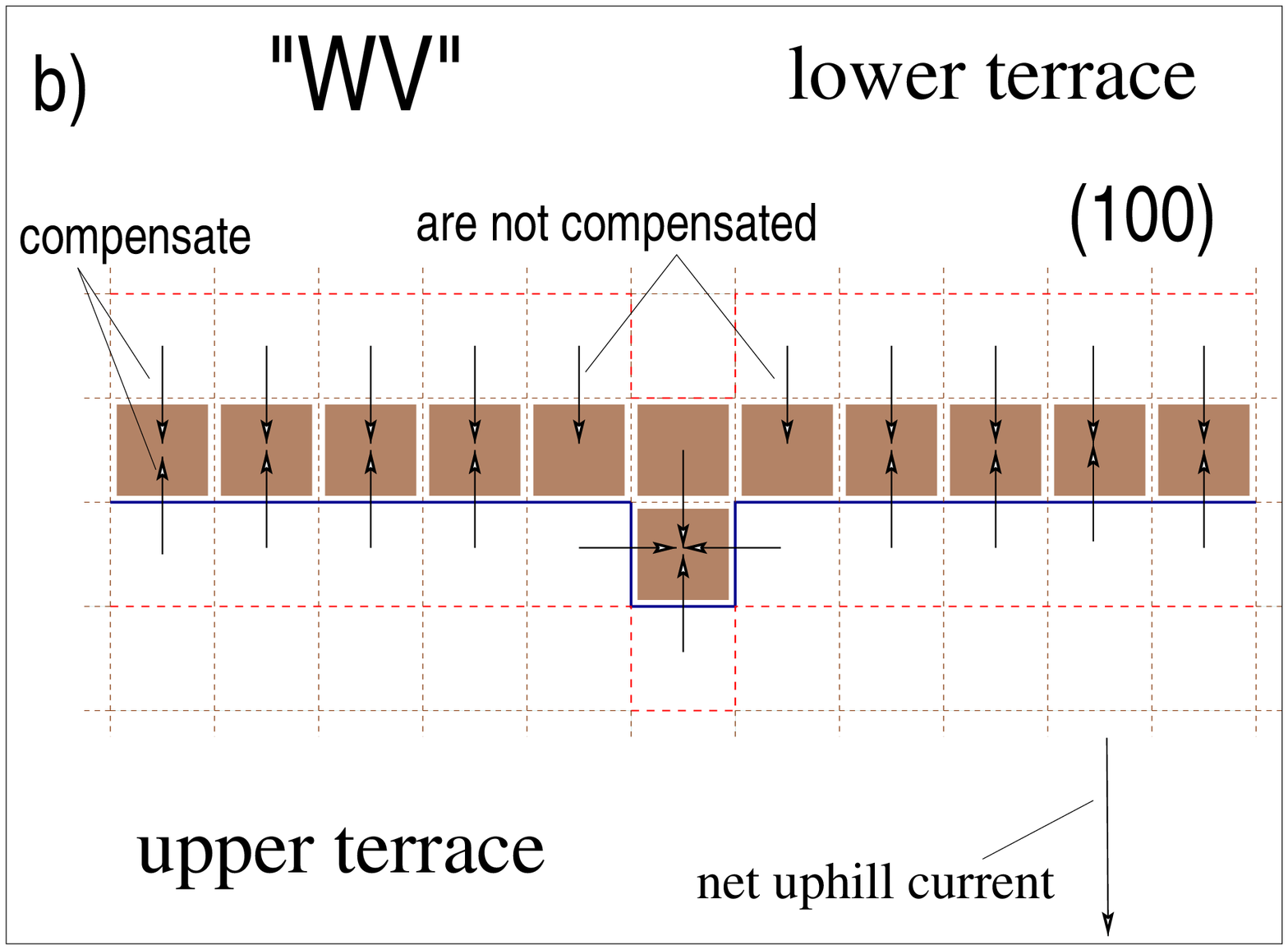} 
\end{minipage} \vspace{0.4cm}  
\caption{
Local currents along step-edges of orientations
(110) and (100) for the WV rules.
\label{currWV}} \end{figure}

Providing an understanding of this difference (between DT and WV
morphologies) in terms of the SED instability
(WV has it and DT does not) is one of the 
important accomplishments of our work.

\begin{figure}[htbp]\noindent 
\begin{minipage}{2.4 in}
\hspace*{1.3cm}\epsfxsize=2.4 in  \epsfbox{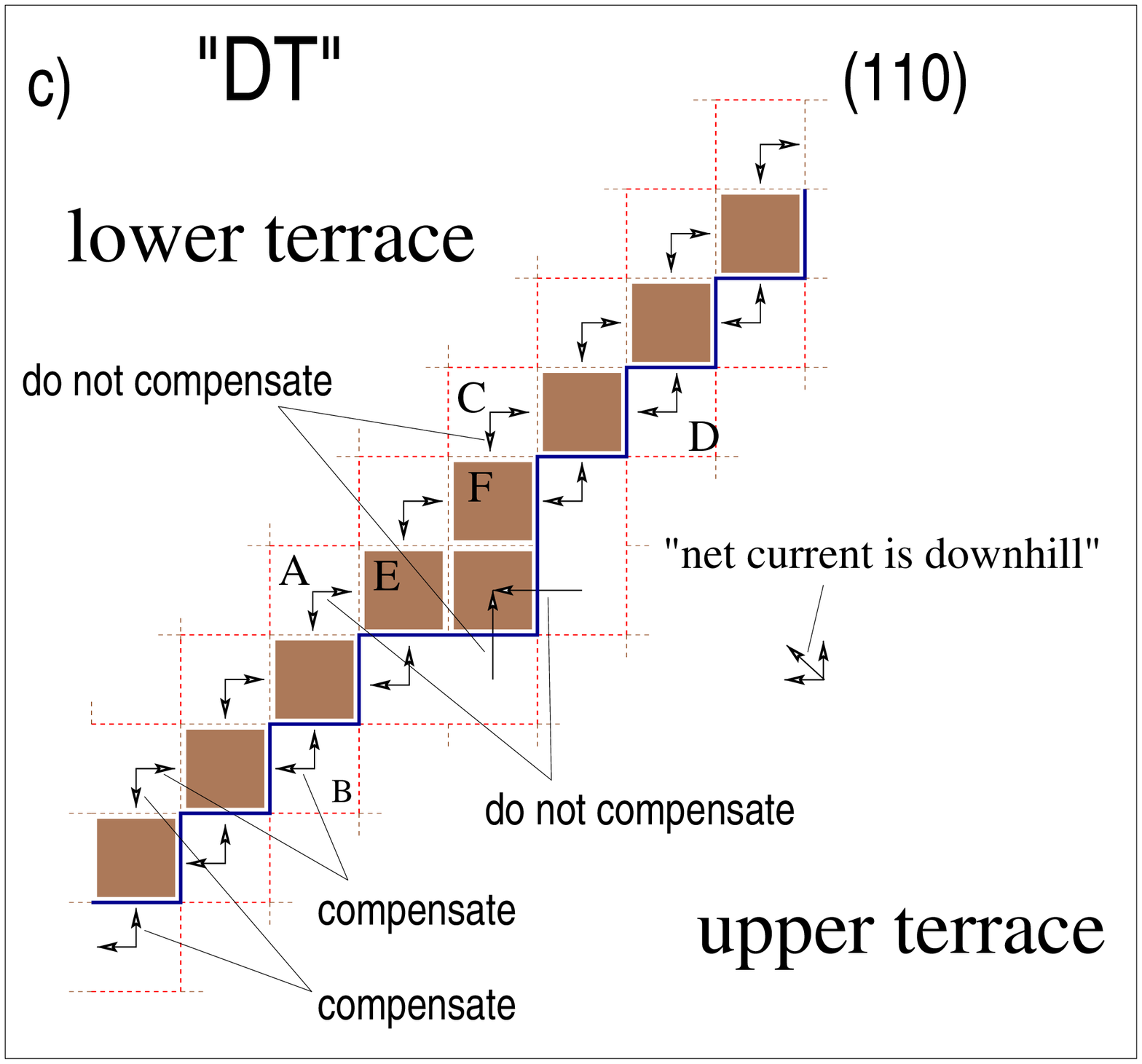}  
\end{minipage}\\  
\begin{minipage}{2.4 in}  
\hspace*{1.3cm}\epsfxsize=2.4 in  \epsfbox{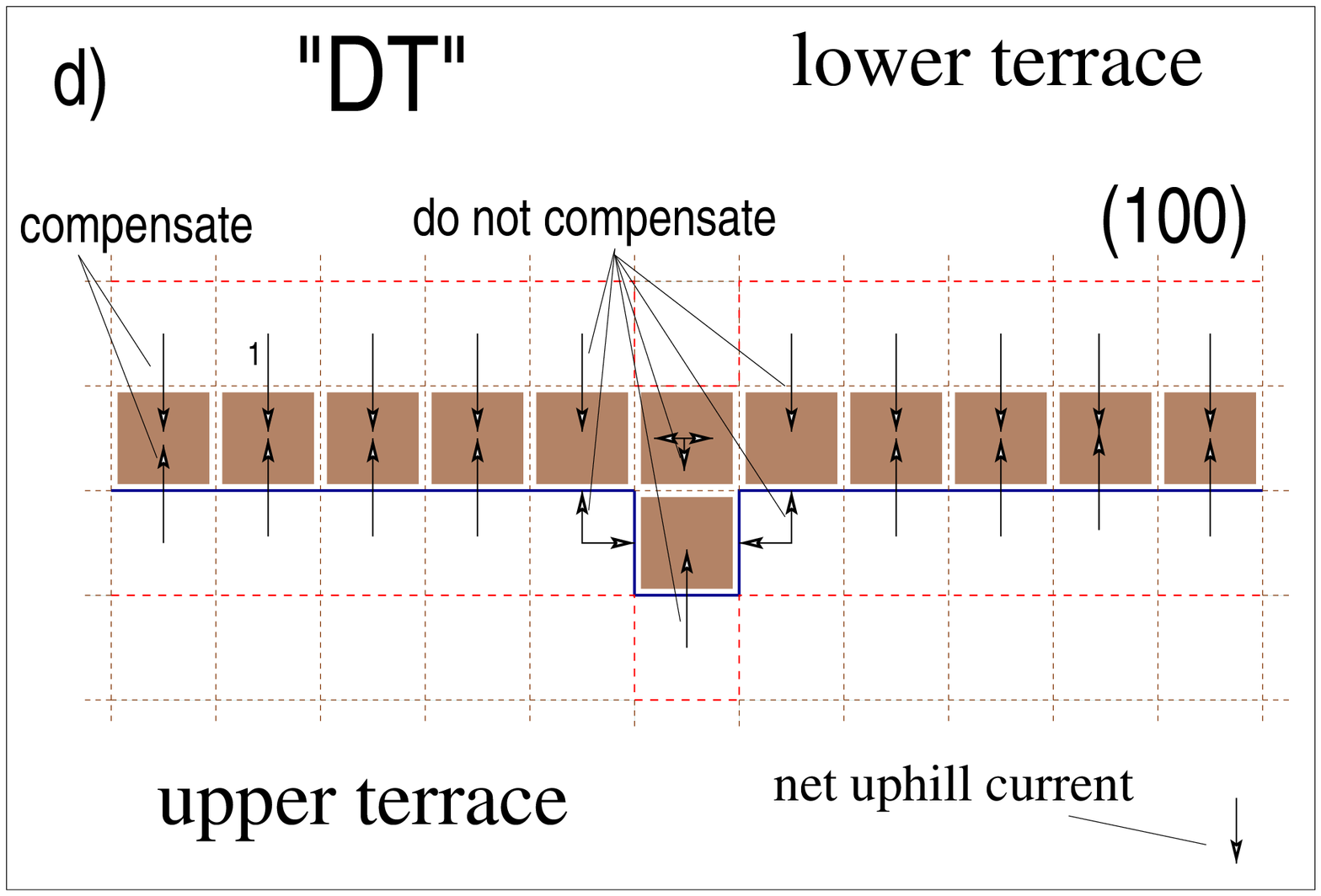}  
\end{minipage}  \vspace{0.4cm}  
\caption{
Local currents along step-edges of orientations
(110) and (100) for the  DT rules.
\label{currDT}} \end{figure}

Figs. \ref{currWV} and  \ref{currDT} show two 
high-symmetry oriented steps, (110) and
(100) respectively, with a small dent in it. The local
particle current shown by arrows are calculated using
the rules of WV and DT models. One can immediately conclude
that in the case of these configurations, the net
current is $j^{WV}_{(111)}=\sqrt{2}/2$ and $j^{WV}_{(100)}=2$
both directed {\it uphill}, $j^{DT}_{(111)}=-\sqrt{2}/2$ directed
{\it downhill} and $j^{DT}_{(100)}=1/3$ directed uphill. 
(The current is considered to be 1 across the boundary
between nearest neighbor sites X and X$^\prime$, 
if a particle once at
position X will go in one step with probability 1 
to position X$^\prime$.
In case of ties, the probability fraction corresponding to the
rules is taken).

As seen from the configurations 
above, there is a clear net uphill local current in the case
of WV model in both cases of high-symmetry orientations. 
According to the theory of SED, one should see a mounded
morphology.
Nevertheless, when we look at the morphology of Fig. 
\ref{wvsur} (a),
one only sees a rough surface not obviously a mounded landscape
in the $n_r=1$ WV growth.
The reason for this discrepancy lies at the heart
of the SED instability and 
the conditions (and constraints) for
its observability: having destabilizing
step-edge diffusion currents is {\it not sufficient} for the
creation of mounds. There is another very important necessary
ingredient (which was not analyzed in Refs. 
\cite{edge,ecu}) that is needed for the formation of
quasi-regular mounded growth patterns
in addition to the destabilizing
SED currents: {\it suppression of deposition and nucleation noise}.
In order to have a consistent uphill current build-up
(to create mounds), 
we need a sustained stability of  the step-edge, 
in other words, if we look at the step-edge as a 1+1 
dimensional `surface' in the $x-y$ plane 
(the upper terrace  being the `substrate') we need to have more or less a 
solid-on-solid type of step-edge, with perhaps some small 
overhangs eventually, but no large or frequent ones.   
The only effect which can disrupt  a stable step-edge is 
the {\it nucleation of new islands} on the terraces near the 
edge. 
Let us imagine that we start with a straight step, 
aligned along the most stable direction. If island nucleation 
takes place easily, then the islands that nucleate close to the 
stable edge will quickly grow into the step, disrupting the 
step's contour and identity. 
If the relaxation time  of the disrupted contour is larger than  the
nucleation time, small newly grown islands will disrupt the 
step even further until it ceases to exist. In this case the 
SED instability {\it will not} manifest in the mound formation
because the necessary condition for mound formation is not
satisfied globally although there may be local uphill current.
Thus, noise may prevent the SED instability from manifesting
a global mounded instability.

The importance of noise suppression for the SED instability
is strikingly apparent from comparing Figs. \ref{wvsur} (a) and 
\ref{wvsur} (b) for the WV model. 
The NRT version has mounded morphology and the $n_r=1$ 
version does not.
However for the DT model, one finds no such regular
mounded structures even when
noise and nucleation is strongly supressed (Fig. \ref{dtsur}).
The reason for this difference is that in the DT model the SED current
is essentially downhill (and thus is stabilizing).
Depending on the edge configuration (for example the (100) edge in 
Fig. \ref{currDT})
one may find a {\it small} uphill current
in the DT model, however it is
statistically not significant to destabilize the whole surface. There will
be a weak mounding tendency at early times, when the
$\nabla^4 h$ diffusion term is relevant 
and when the height-height correlation
function shows some oscillations (see
\cite{short}), indicating the presence of this
term, but then it will be quickly taken over by the stabilizing
downhill current and nucleation events.
Thus, the DT model, at best, will exhibit some weak and irregular
epitaxial mounding \cite{short}, but {\it not} a regular
patterned mounded morphology.

Another important point is that mounding due to
the SED instability {\it does not require 
any growth nonlinearity to be present}.
This is obvious from the fact that the LC model, which is
strictly linear by construction, shows spetacular epitaxial
mounding induced by the SED instability.
In the following subsection we analyze in more details the effects
of noise reduction on the SED instability.

\subsection{Noise reduction method: a numerical way to control
the step edge diffusion instability}

As emphasized in the previous subsection, 
a necessary ingredient
for creating mounds by SED is the 
suppression of noise.
In the following we restrict our analysis
to discrete limited mobility models, and 
frequently use for comparison
WV and DT models. 
Unfortunately, rigorous analytical calculations are practically 
impossible for discrete 
and nonlinear dynamical growth models of DT and WV types.
However, we make an
attempt by introducing the proper quantities to pinpoint the
effects of noise reduction on these two models
and actually show the major difference between the two models
in one and two substrate dimensions in the context of
SED instability.
First, we introduce the notion of {\it 
conditional occupation rates}. 
The conditional occupation rate $f_i$ 
associated with site $i$ in a particular
height-configuration is the {\it net gain} of transition rates 
coming from the contribution of all the neighbors within a distance
of $l$ sites on the surface and 
from the growth direction (from the deposition beam).

\begin{figure}[htbp]\noindent
\begin{minipage}{2.8 in}  
\hspace*{1cm}\epsfxsize=2.8 in  \epsfbox{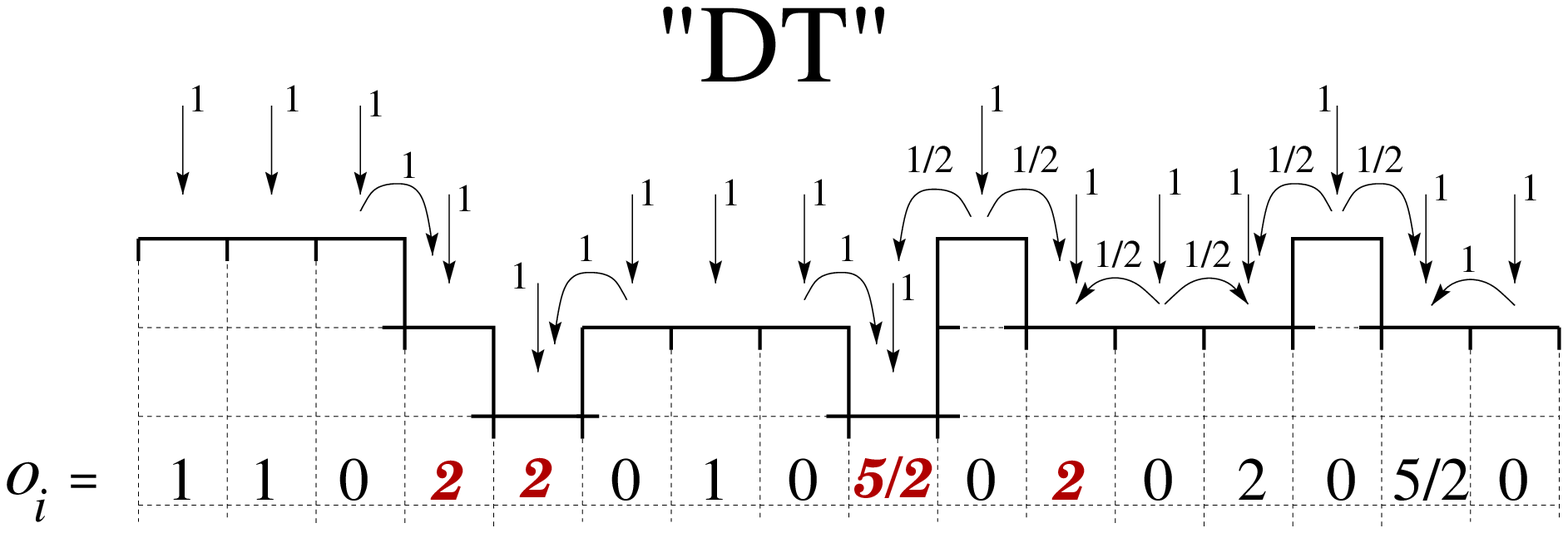} 
\end{minipage}\\  
\begin{minipage}{2.8 in} 
\hspace*{1cm}\epsfxsize=2.8 in  \epsfbox{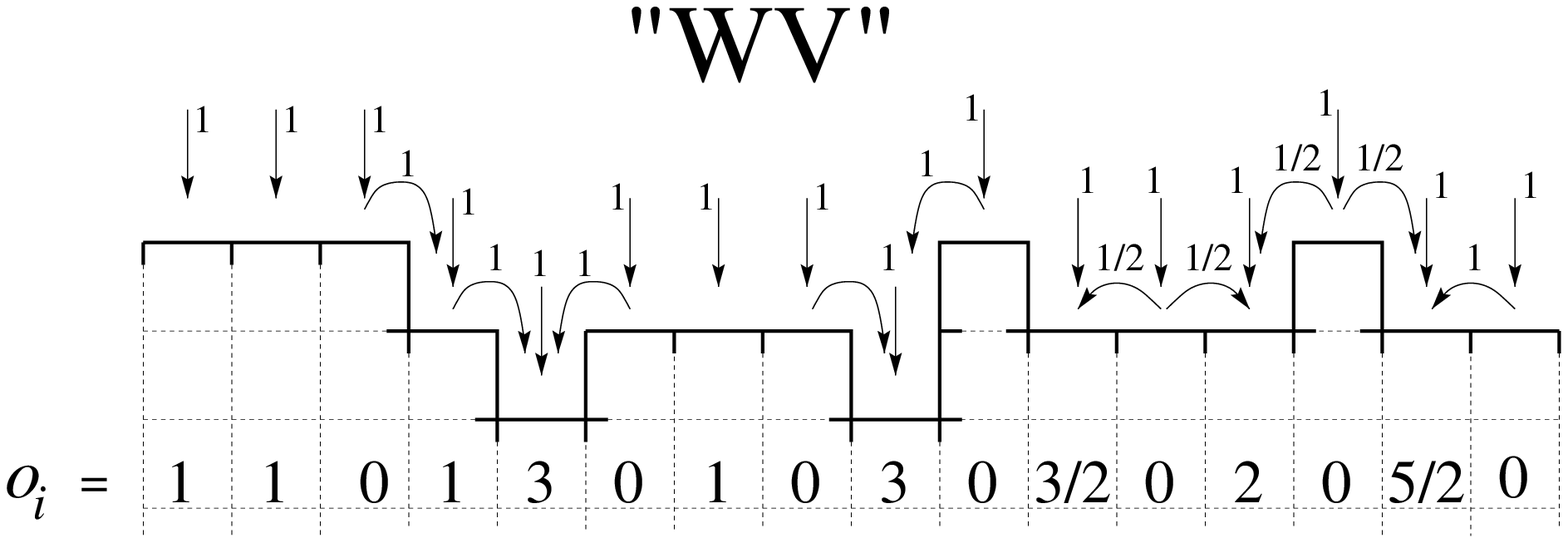}  
\end{minipage} \vspace{0.5cm}  
\caption{
Conditional occupation probabilities for two particular
configurations in  $1+1$ dimensions for DT and WV models.
The $f_i$-s that differ for the DT from WV are set in bold italics.
Here $l=1$.
\label{occs1d}} \end{figure}

The $f_i$-s can easily be calculated once the diffusion rules are known.
If ${\cal W}_{i\to i'}$ is the transition rate for an atom
landed on site $i$ to be incorporated at site $i'$ (which
is a nearest neighbor when $l=1$), then:
\begin{equation}
f_i=1+\sum_{i'}^{\{i\}_l} \left({\cal W}_{i'\to i}
- {\cal W}_{i\to i'}\right) \label{occ}
\end{equation}
where the unity in front of the sum represents the contribution
from the beam and $\{i\}_l$ means summing the contributions 
over all the neighbors of site $i$ within a distance $l$.
The $f_i$-s are obviously conditioned to the event that there
is an atom landed on site $i$ or on the neighbors $i'$. It is important
to emphasize that these quantities in 
Eq. (\ref{occ}) are introduced for the
no-desorption case only: once an atom is incorporated
in the surface it stays that way. If one wishes to add desorption
to the model, then the evaporation rates have to be included in 
Eq. (\ref{occ})
as well (the rates in Eq. (\ref{occ}) are all diffusion rates ).

\begin{figure}[htbp]\noindent
\begin{minipage}{2.4 in}  
\hspace*{1.3cm}\epsfxsize=2.4 in  \epsfbox{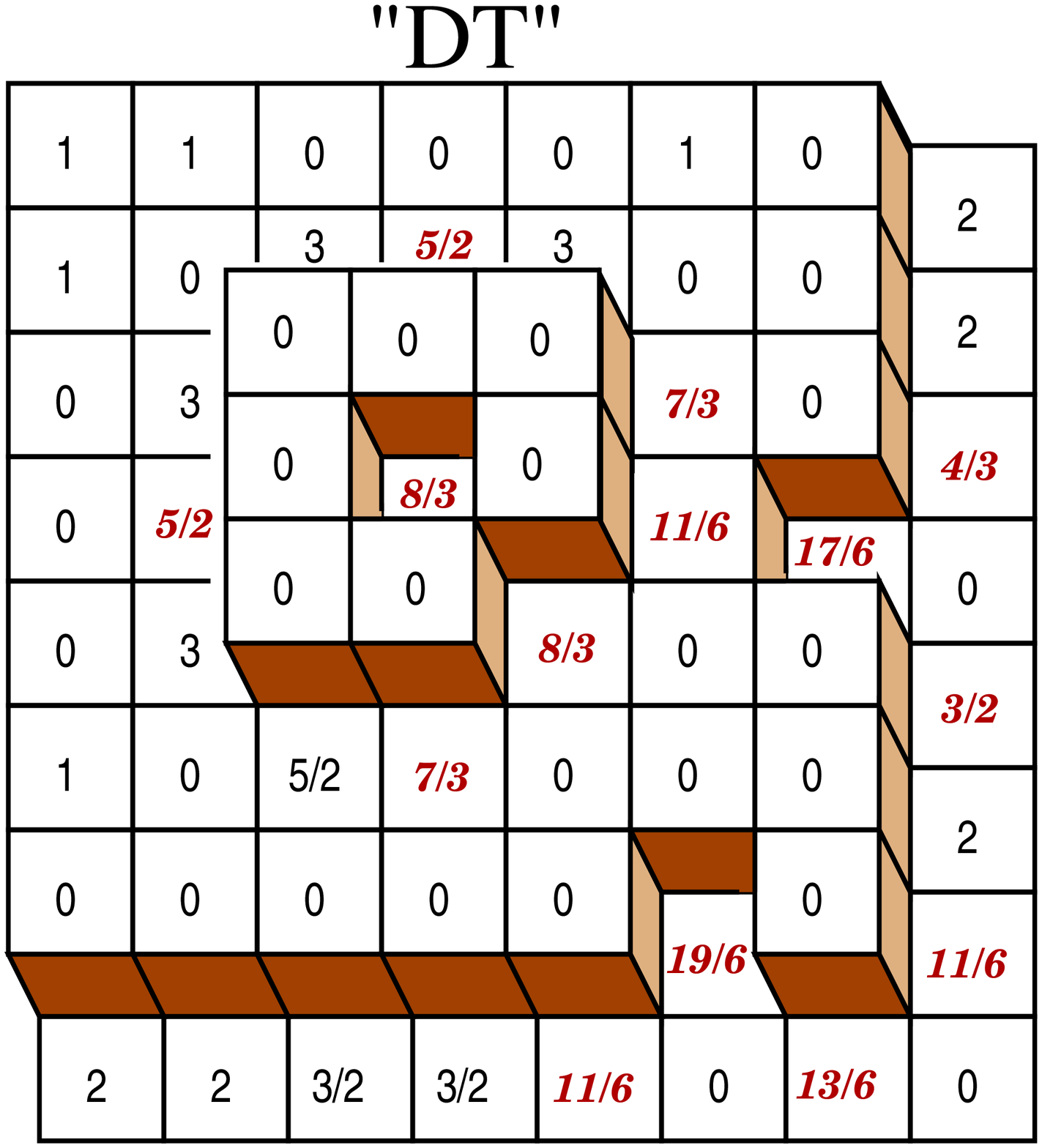}  
\end{minipage}\\   
\begin{minipage}{2.4 in}  
\hspace*{1.3cm}\epsfxsize=2.4 in  \epsfbox{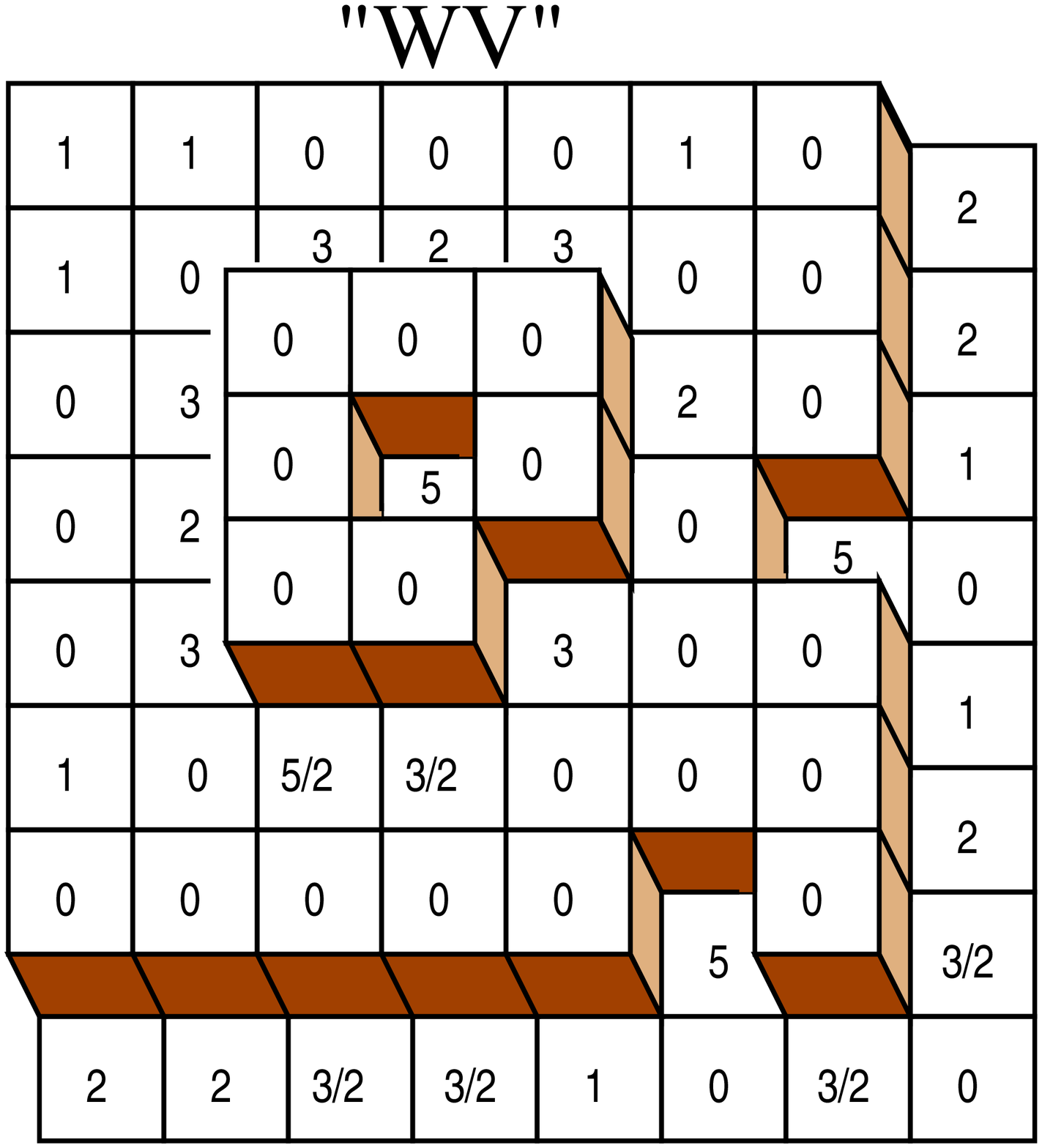}  
\end{minipage}  \vspace{0.5cm}  
\caption{
Conditional occupation probabilities for two particular
configurations in $2+1$ dimensions for DT and WV models.
The $f_i$-s that differ for the DT from WV are set in bold italics.
Here $l=1$.
\label{occs2d}} \end{figure}

Let us now illustrate in light of the two models WV and DT, the
conditional occupation rates, in 1+1 and 2+1 dimensions.
Figs. \ref{occs1d} and \ref{occs2d} show the $f_i$-s for particular 
configurations for 
one and two dimensional substrates respectively.
The fractional transition rates mean that the motion of the atom
is probabilistic: it will choose with equal probability among the
available {\it identical} states. 
This is in fact the source of the
diffusion noise $\eta_c$, arising from the stochastic 
atomistic hopping process,  
which should be added 
to the right hand side of Eq. 
(\ref{cont}), however, it is an irrelevant contribution in the 
long wavelength scaling in the
renormalization group sense.
The conditional occupation rates $f_i$ are small rational numbers
from the interval:
\begin{equation}
f_i \in [0,Z_l^{(d)}+1] \label{bounded}
\end{equation}
where $Z_l$ is the maximum $l$th order coordination
number on the $d$-dimensional substrate. 

For example, on a cubic lattice ,
$Z_1^{(d)} = 2d$. For a fixed set of deposition and diffusion
rules the number of possible occupation rates is fixed, and for
low $l$-values it is only a few. This number is determined by
how many different values can be obtained from all the possible
relative height combinations of all sites within a $d$-dimensional
ball of radius $2l+1$ (the gain coming from the neighbor at $l$-away
may depend on the configuration of radius $l$ around that neighbor).
If $l=1$ then in  $1+1$ dimensions this means that a chain of maximum 
7 sites has to be considered and in $2+1$ dimensions it is a square
domain of maximum 25 sites to exhaust all the possibilities
(the domain in this latter case is given by all $(i,j)$ sites 
that obey $|i|+|j| \leq 3$).
For example, the one dimensional WV model will have
conditional occupation rates from the set 
$\{0, 1/2, 1, 3/2, 2, 5/2, 3 \}$. For the one dimensional DT, the set is 
a bit restricted: $\{0, 1, 3/2, 2, 5/2, 3 \}$, it does not allow
for $f = 1/2$. 

\begin{figure}[htbp]
\begin{minipage}{3.4 in}  
\epsfxsize=3.4 in  \epsfbox{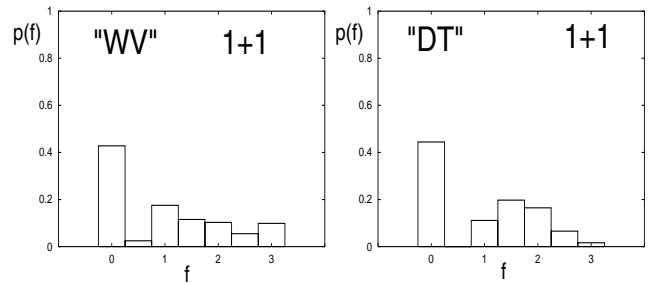} 
\end{minipage} \vspace{0.5cm} 
\caption{
Conditional occupation rates
and their occurence probability $p(f)$
for the WV and DT models.
Here $l=1$.
\label{pief}} \end{figure}

Fig. \ref{pief} shows the plot for the occurrence probabilities 
$p(f)$ versus the conditional occupation 
rates for both the WV and DT models
($n_r=1$). 
The occurrence probability of rate $f$ in one dimension 
is calculated as follows: take all the possible height configurations 
{\it that are distinguished by the deposition and diffusion rules}, 
(e.g. WV and DT) of the $2(2l+1)+1$ sites, monitor the $f$ 
that belongs to the middle site and then measure the rate of
occurrence of each value for $f$. 

It is important to count these
occurrences relative to the distinguishable number of configurations.
Both WV and DT models have their rules for the motion of the
landed atom formulated in terms of {\it coordination numbers}.
In the WV model, the landed atom will move to the site with the
maximum coordination whereas in DT the atom will move to
increase its coordination. Thus only those configurations
will be distinguished by the rules that have different coordination
numbers in at least one of the sites. A given set of coordination numbers
thus determines a class of configurations. 
For example in 1+1 dimensions both the WV and DT models on 7 sites
will distinguish a total of $3^6=729$ classes, since the
deposition and diffusion rules for these two models do not
involve actual height differences, just the fact that a site
is at higher, the same, or lower level than its neighbors.
In $2+1$ dimensions the exact enumeration of these classes
is highly non-trivial, and it constitutes the subject of
a future publication.

One can note the following property of
conditional occupation rates: in every domain of sites on the
substrate that has a zero inflow and zero outflow current
across its boundaries (`island'), the sum of the rates within
the island adds up to the total number of sites within the island:
\begin{equation} 
\sum_{{\bf i} \in island} f_{\bf i} = 
\mbox{nr. of sites within the island.}
\end{equation}
This is a consequence of conservation of probability. The span of the
island may only change if a new particle is deposited
within a distance $2l+1$ from its boundary, outward from the island. 
This is valid in any dimension.

The larger the $f_i$ the more probable it is that site $i$ will
receive an atom (it can come from more directions than for one
with a lower $f$). We believe that when using the NRT, i.e.,
$n_r > 1$, the sites
with higher $f$ will have a higher chance to be actually deposited onto.
To illustrate this, we consider the simplest setup: starting from a
flat surface with a few irregularities, we perform the counting process and
see with what probability the various sites will actually be deposited
onto when the corresponding counters reach the threshold value $n_r$.

\begin{figure}[htbp]\noindent
\begin{minipage}{3.4 in}  
\hspace*{0.06cm}\epsfxsize=3.4 in  \epsfbox{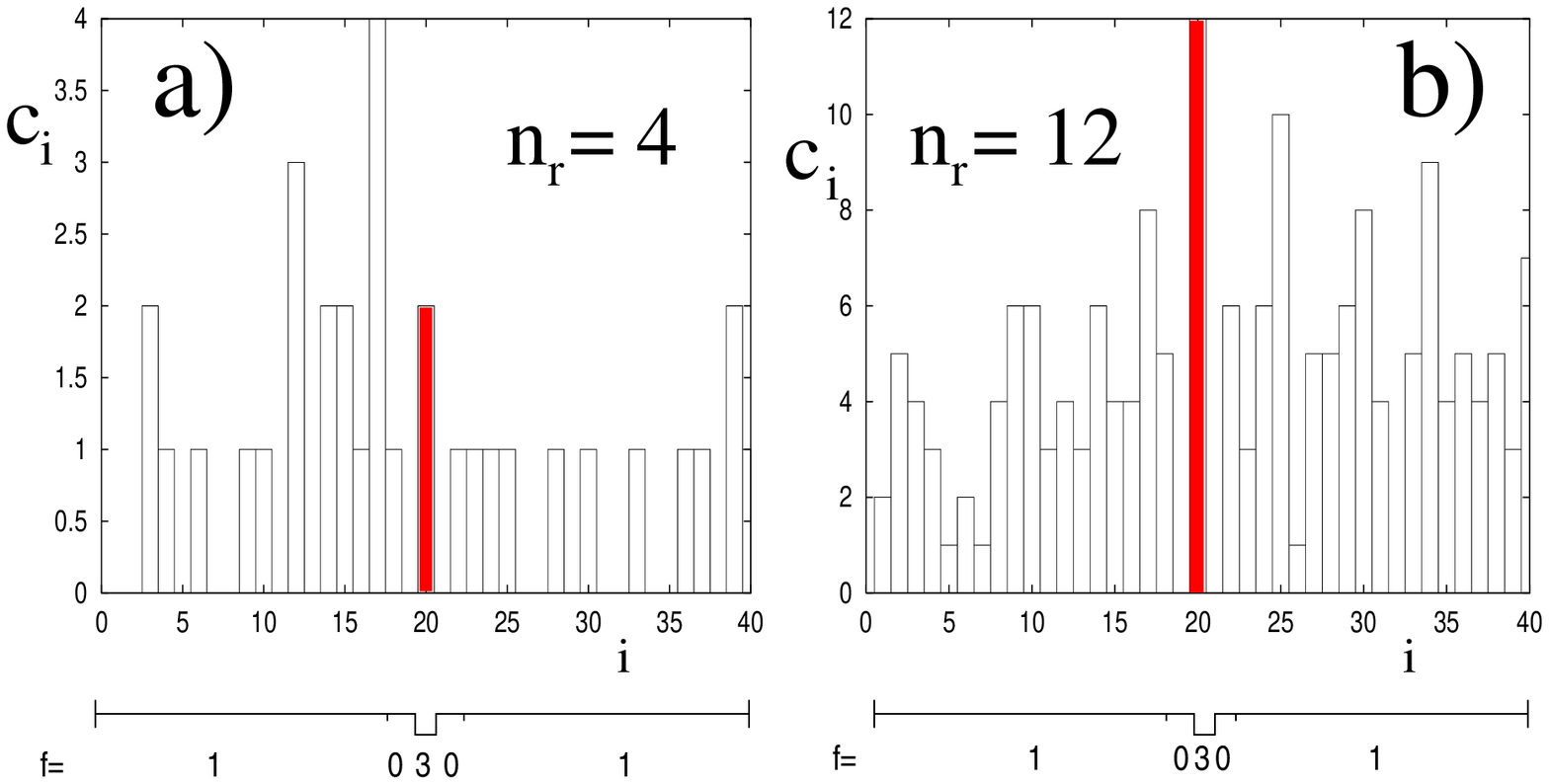} 
\end{minipage}\\ 
\begin{minipage}{3.4 in}  
\epsfxsize=3.4 in  \epsfbox{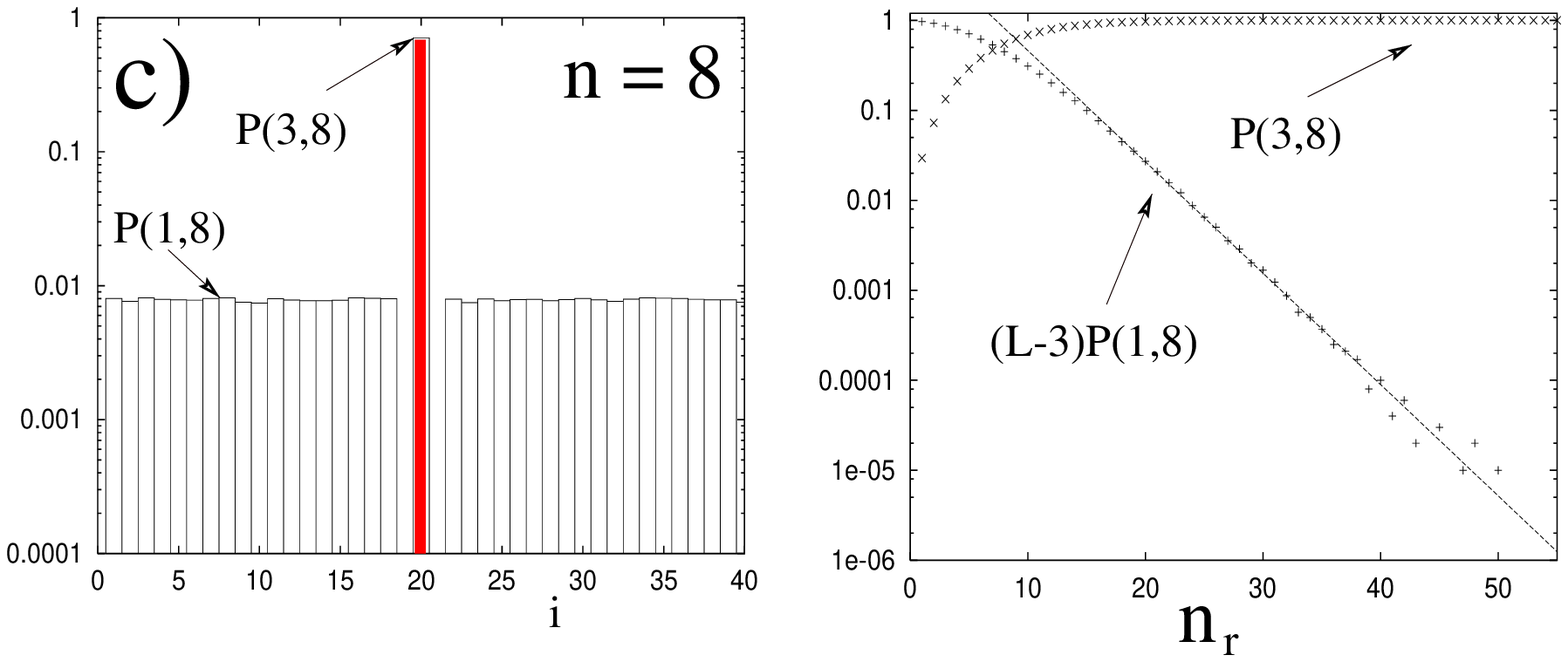} 
\end{minipage} \vspace{0.5cm} 
\caption{
Particular counter values with
different noise reduction parameters
a) $n_r=4$, and b) $n_r=12$. c) shows the
probability that site $i$ reaches
the treshold for the first time.
d) Plot of probability for nucleation on the
terrace (of $L-3$ sites) and deposition
at the dent site with $f=3$. The straight
line is an exponential fit with slope $-0.285$ and intercept $8$.
\label{cis}} \end{figure}

Unfortunately rigorous analytic derivations are difficult even in the
simplest case because of the fact that surface growth is driven by
the first-passage time probabilities of the counters (or waiting
times). 
We therefore follow numerically the evolution of counters on a simple
substrate which is flat everywhere except at one site where the
height is lower by $-1$ (see the surface
profile just inder Fig. \ref{cis} (a)). 
According to both rules (WV and DT)
the occupation rate is $1$ except for the lower site where it is
3 and its two nearest neighbors where it is $0$.
Fig. \ref{cis} (c) shows the probability 
that the first real deposition
event occurs at site $i$ for $n_r=8$, $L=40$. 
Obviously all sites on the flat region
will have the same chance of $P(f,n_r)=P(1,n_r)$ 
to reach the threshold for the
first time, while the site with $f=3$ will be the most probable
to reach it, and for its two nearest neighbors it is 
entirely unlikely: $P(0,n_r)=0$. 
In case of Fig. \ref{cis} (c), $P(1,8)=0.0077$ and $P(3,8)=0.71$. 
This means that in spite
of the uniform randomness of the shot noise, $71\%$ of the 
time the actual deposition will occur on that {\it single} 
site with higher coordination and only 
$21\%$ of the time we will actually
be able to land an atom somewhere on the 
{\it whole flat} island of 37 sites.
This clearly shows that nucleation 
of new islands on flat regions is strongly
suppressed. 
In Fig. \ref{cis} (d) we
represent the two probabilities 
$P(3,n_r)$ and $P(1,n_r)$ as functions
of $n_r$ (on a log-linear scale). For higher
noise reduction parameter values we are 
practically forced almost each time
to fill in a site with the highest 
occupation rate, and with almost zero probability
to land an atom on the flat terrace. 
Looking back on Figs \ref{currWV} (a) and \ref{currWV} (b) 
for the 2+1 WV model, it becomes
clear that the main particle current will be {\it along the step edge} 
(the shaded squares on the figures) and it 
will be greatly enhanced by NRT. 
At the same time the rate of  island nucleation 
is drastically reduced, so the relaxation of the step-edge becomes
rather fast compared to the creation of new islands.
Since the step-edge dynamics becomes fast 
compared with the nucleation dynamics
the mounds that are formed will 
grow into each other preserving their
pyramidal shapes, exhibiting the observed 
mound-coarsening behavior.
Models in 2+1 dimensions can be related to the
conditional occupation rates shown in Fig. \ref{occs2d}.
The hole in the middle of the top island, and also the dents
in the step-edges have much higher rates to be occupied in the
WV model than in the DT model. Thus the step-edge currents are a lot
stronger for the WV model than for the DT model. 
In addition the middle site at the
base of the upper island has a lower occupation rate for the WV ($f=2$)
than for the DT ($f=5/2=2.5$) model. 
This means that in the case of WV it is
harder to disrupt a straight step-edge than for the DT model.
We believe that the spectacular difference between the DT and
WV noise reduced morphologies discovered in this work arise from 
the conditional occupation differences in the two models,
which lead to an SED instability in the WV, but not in
the DT model.
While we have provided a rather compelling qualitative picture
in this subsection explaining why the SED instability produces
spectacular mounded morphologies in the NRT version
($n_r \neq 1$) of the WV model, but neither in the DT model
($n_r =1$ or $n_r \neq 1$ version) nor in the $n_r =1$ WV model,
we have not been able to come up with an analytical theory 
for the mound formation, which remains an important 
open problem for the future.

\section{Properties of the mounded morphologies}

In this section, we discuss the dynamical scaling properties
of the mounded morphologies presented in Section II.
A conventional way \cite{short,sdsppdtes}
to study mounding is to look
at the height-height correlation function 
\begin{equation}
H(r) = \langle h({\bf x}) h({\bf x} + {\bf r})
       \rangle _{\bf x},
\label{hhcor}
\end{equation}
where $r=|{\bf r}|$ is the distance between two sites
on the substrate, and the $\langle ... \rangle$ denotes
a substrate averaging.

\begin{figure}[htbp]\noindent
\begin{minipage}{2.2 in}  
\hspace*{1.3cm}\epsfxsize=2.0 in  \epsfbox{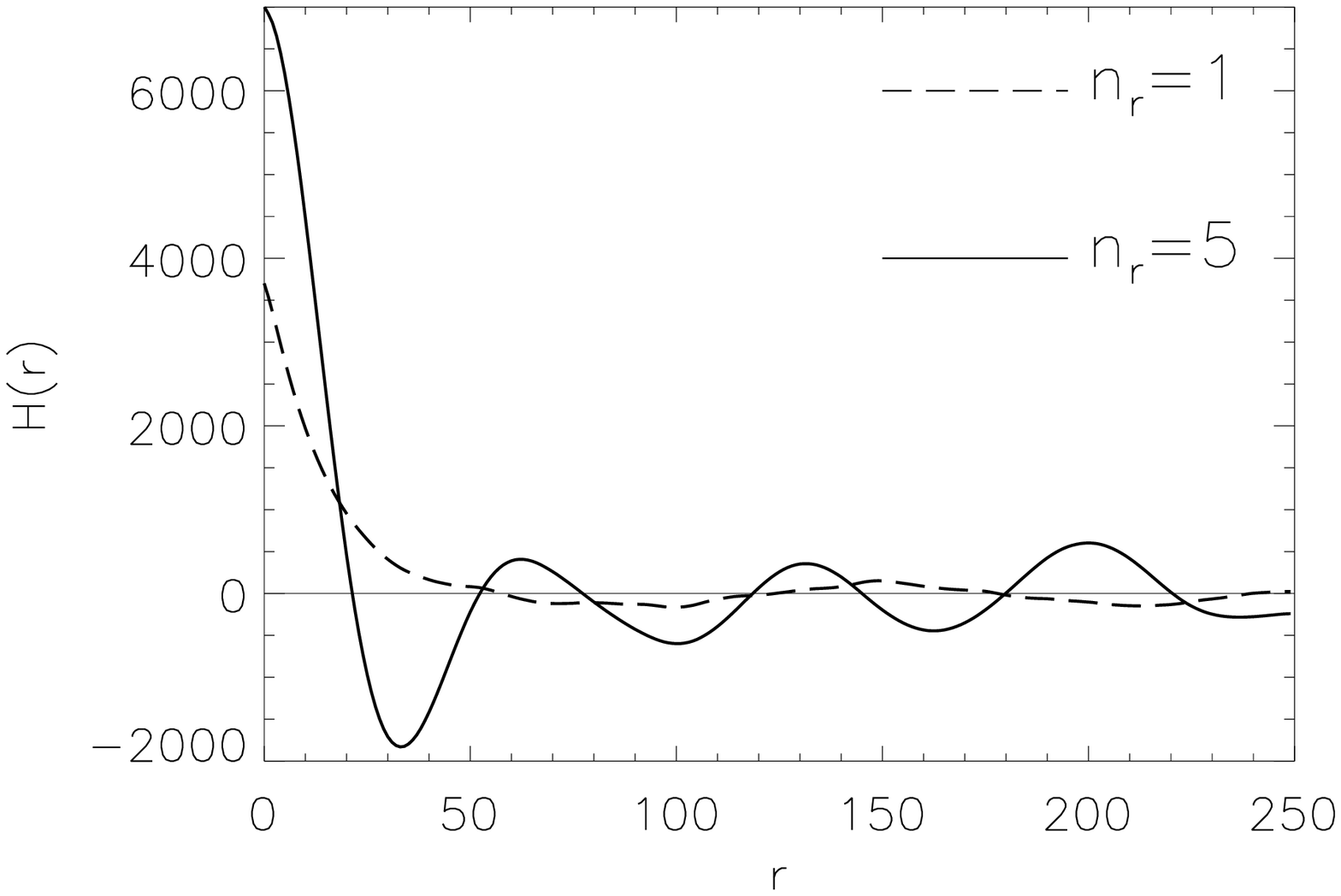}    
\end{minipage}\\  
\begin{minipage}{2.2 in}  
\hspace*{1.3cm}\epsfxsize=2.0 in  \epsfbox{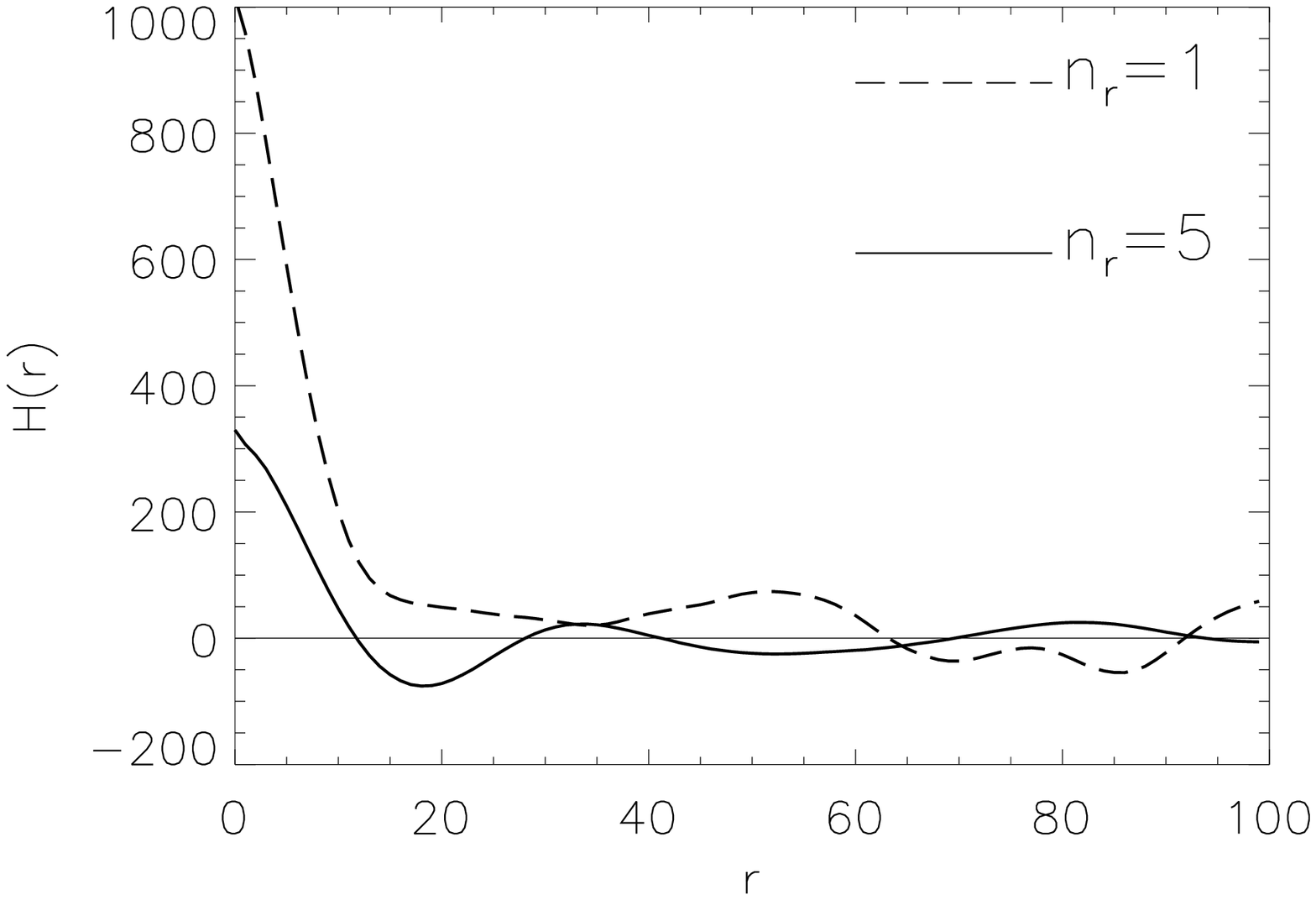}    
\end{minipage}\\
\begin{minipage}{2.2 in}  
\hspace*{1.3cm}\epsfxsize=2.0 in  \epsfbox{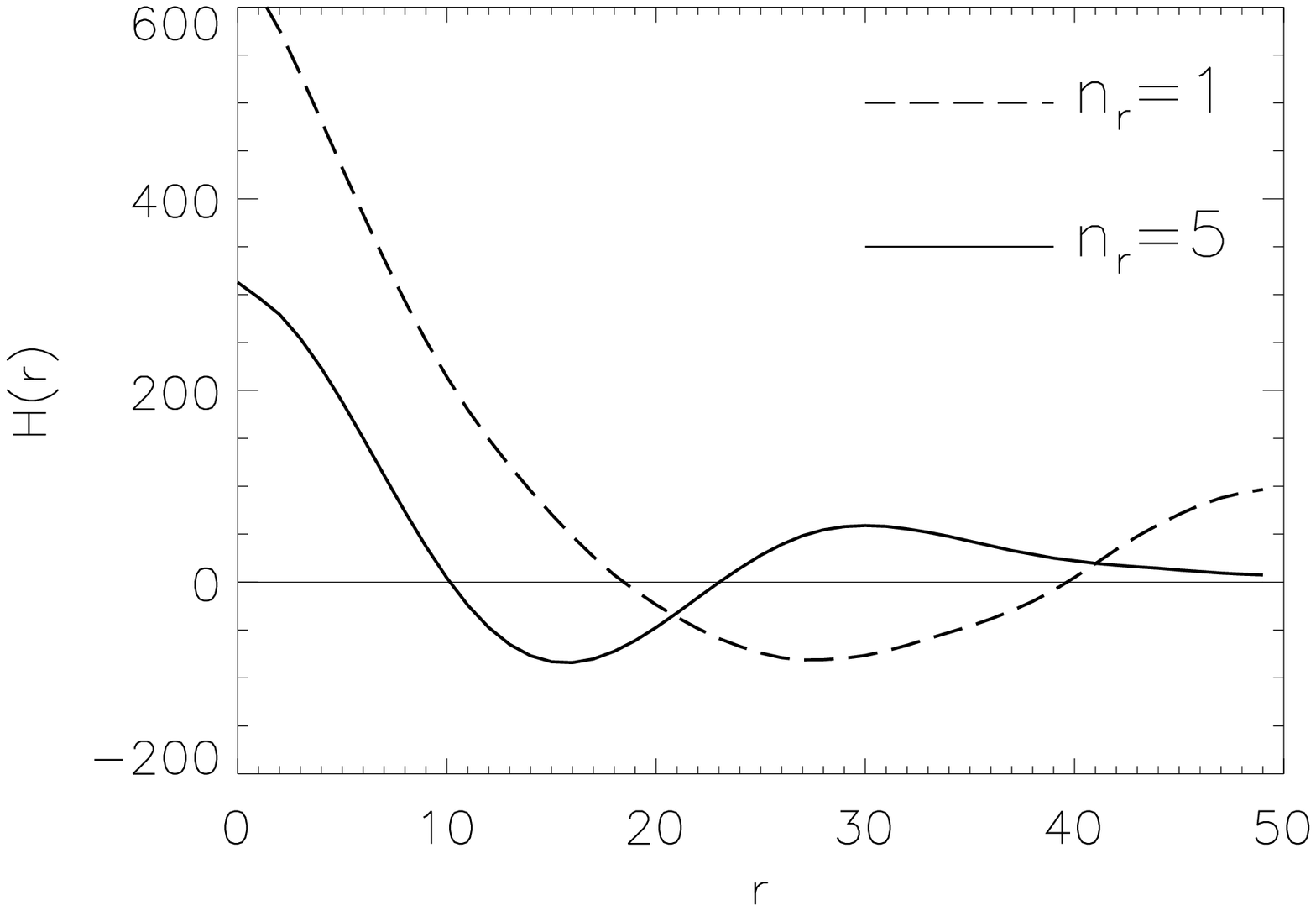}    
\end{minipage} 
\begin{minipage}{2.2 in}  
\hspace*{1.3cm}\epsfxsize=2.0 in  \epsfbox{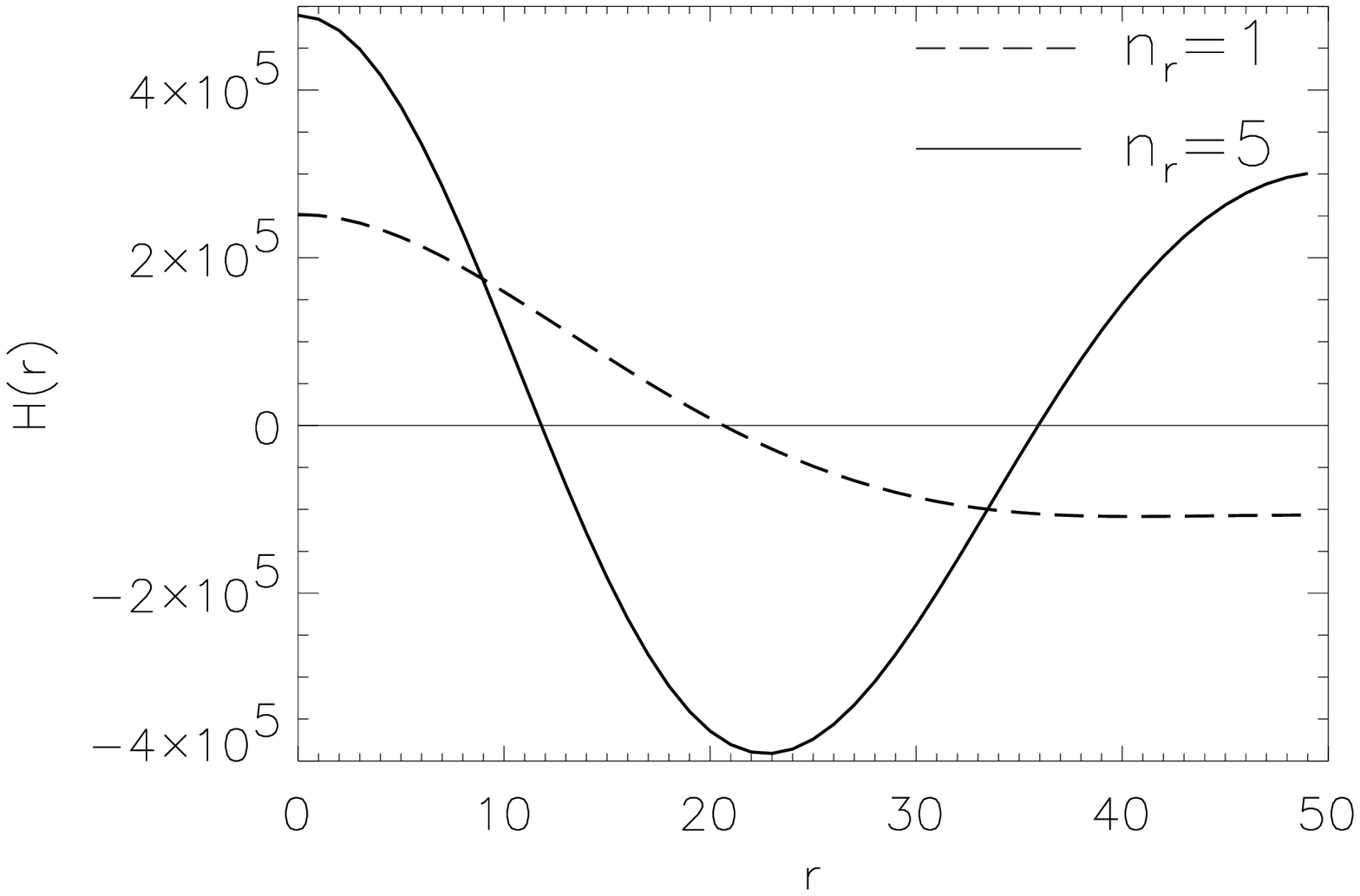}    
\end{minipage}
\caption{
The height-height correlation functions $H(r)$
corresponding to the mounded morphologies presented
in Fig. \ref{wvsur}, \ref{4lsur}, \ref{4nsur},
and \ref{6lsur}.
From top to bottom :
the WV model at $10^4$ ML,
the LC model at $10^3$ ML,
the nonlinear fourth order at $10^3$ ML,
and the linear sixth order at $10^6$ ML.
\label{hhosc}} \end{figure}

The calculated $H(r)$ oscillates as a function of $r$ in
systems with mounded patterns.
Fig. \ref{hhosc} shows the oscillating correlation
function from the WV,
the linear fourth order (LC model), the nonlinear
fourth order, and the linear sixth order models.
Note that noisy oscillations are already detectable 
in the original models with $n_r=1$, 
but the oscillations here are weak and irregular \cite{short}. 
The oscillations become spectacularly 
enhanced and rather evident when 
the noise reduction
technique is used. 
This corroborates the fact that NRT
suppresess the noise, but does not
change the growth universality class.

\begin{figure}[htbp]\noindent
\begin{minipage}{2.2 in}  
\hspace*{1.3cm}\epsfxsize=2.2 in  \epsfbox{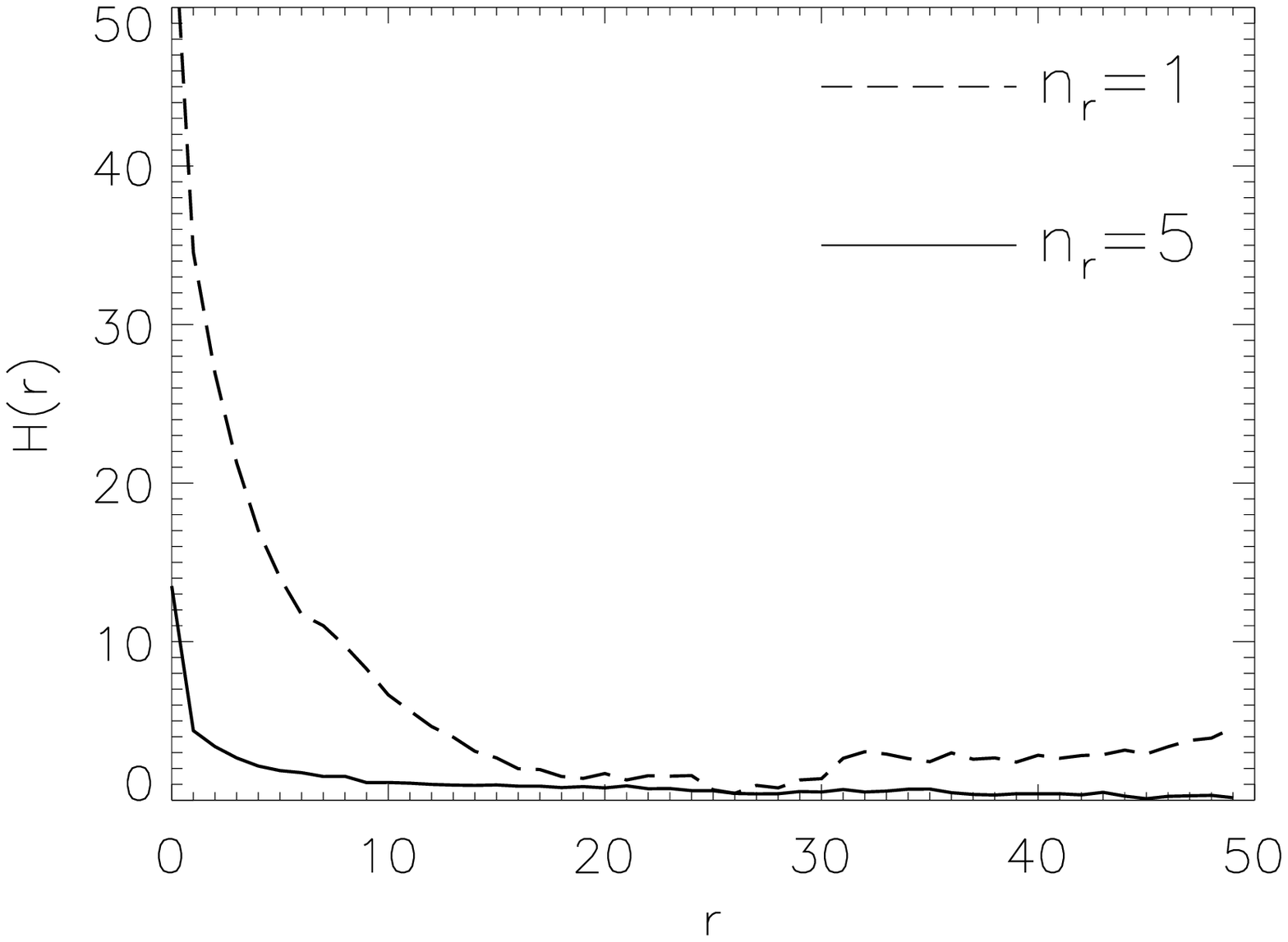}   
\end{minipage}\\  
\begin{minipage}{2.2 in}  
\hspace*{1.3cm}\epsfxsize=2.2 in  \epsfbox{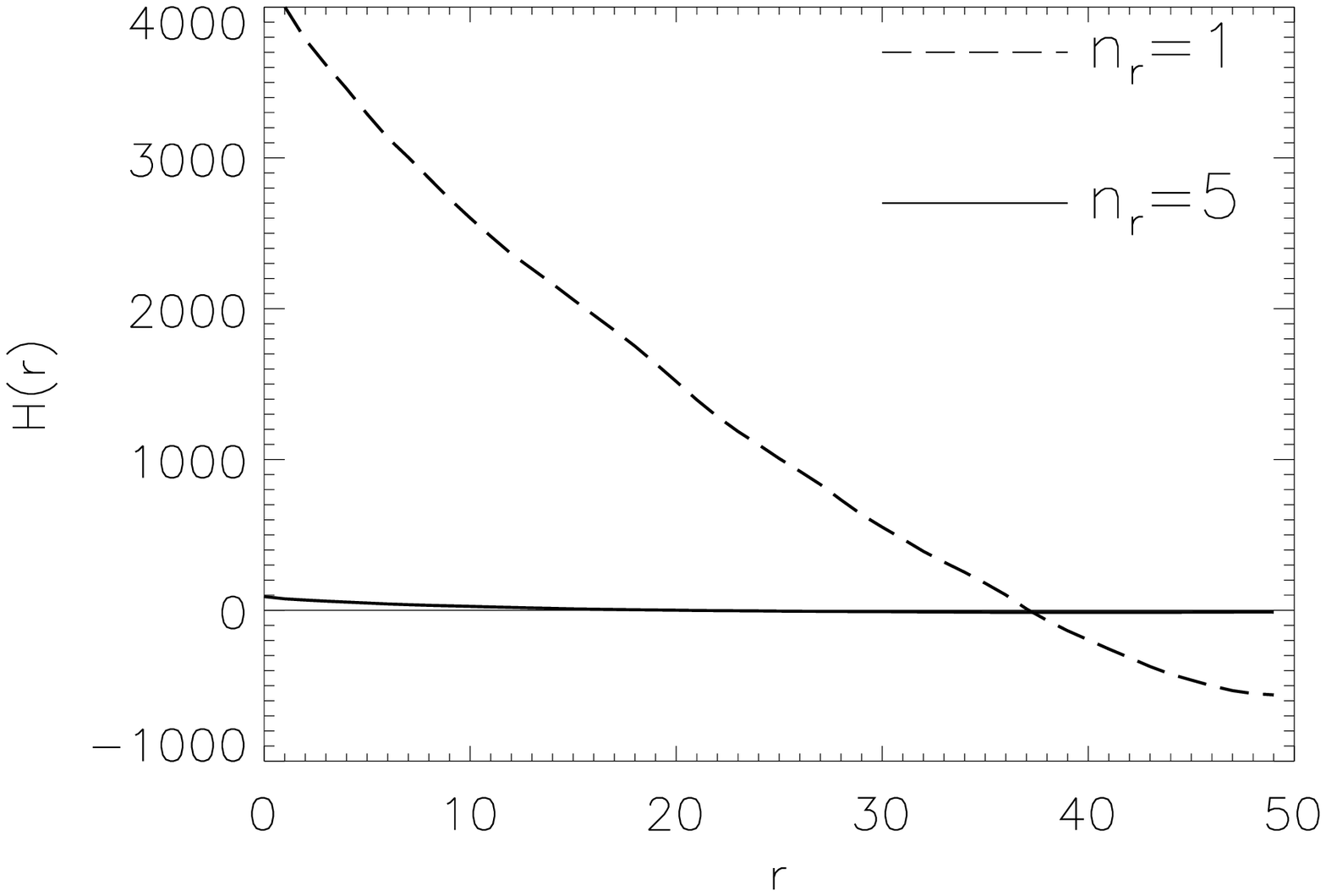}   
\end{minipage}\\  
\begin{minipage}{2.2 in}  
\hspace*{1.3cm}\epsfxsize=2.2 in  \epsfbox{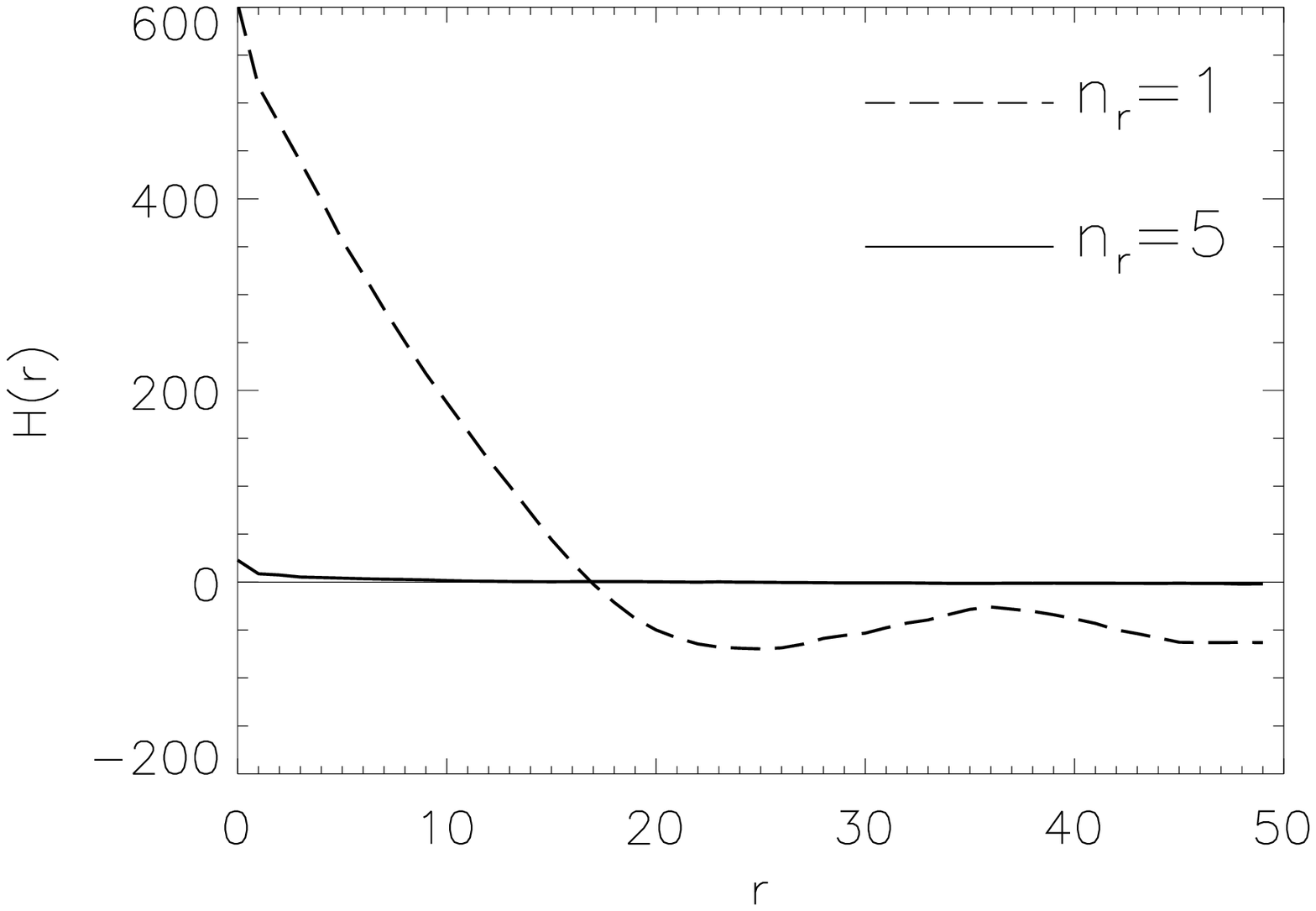}  
\end{minipage}
\caption{
The height-height correlation functions $H(r)$
corresponding to the morphologies presented
in Fig. \ref{fmsur}, \ref{dtsur}, and \ref{4ssur}.
From top to bottom :
the F model at $10^6$ ML,
the DT model at $10^6$ ML,
and the nonlinear fourth order with higher nonlinearities
at $10^3$ ML.
\label{hhflat}} \end{figure}

In Fig. \ref{hhflat} we present the non-oscillatory 
$H(r)$ corresponding to
the non-mounding interfaces of F, DT, and the nonlinear
fourth order equation with infinite order nonlinearities 
(Eq. (\ref{4seq})) models .

The morphologies from the original version ($n_r=1$) of these
models are kinetically rough. But when the noise is suppressed
with $n_r>1$, no specific oscillatory
pattern emerges although some weak and irregular $H(r)$
oscillating could sometimes be detected in the DT model
\cite{short} (particularly for $n_r=1$),
most likely caused by the presence of the $\nabla ^4 h$
Mullins term in the DT growth equation.
As a matter of fact,
these rough surfaces (F, DT, etc.)
merely become ``smoother'' and flatter for $n_r>1$.

\begin{figure}[htbp]
\begin{minipage}{3.4 in}  
\epsfxsize=3.3 in  \epsfbox{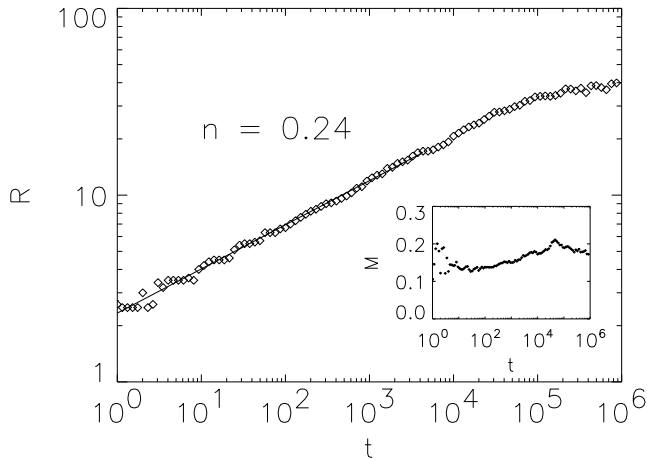}  
\end{minipage}\vspace*{0.5cm}
\caption{
The average mound radius as a function of time
from the WV model with $n_r=5$ from a substrate of size
$100 \times 100$. The average mound slope of the same
system is shown in the inset
\label{wvrm}} \end{figure}

Other important characteristics of mounded patterns 
\cite{short,sdsppdtes} are the
average mound radius and the average mound slope.
Conventionally, the distance of the first zero-crossing of
the correlation function $H(r)$ is taken to be the average
mound radius $R$, and $[H(r=0)]^{1/2}$ is the average mound
height. 

\begin{figure}[htbp] 
\begin{minipage}{3.4 in}  
\epsfxsize=3.3 in  \epsfbox{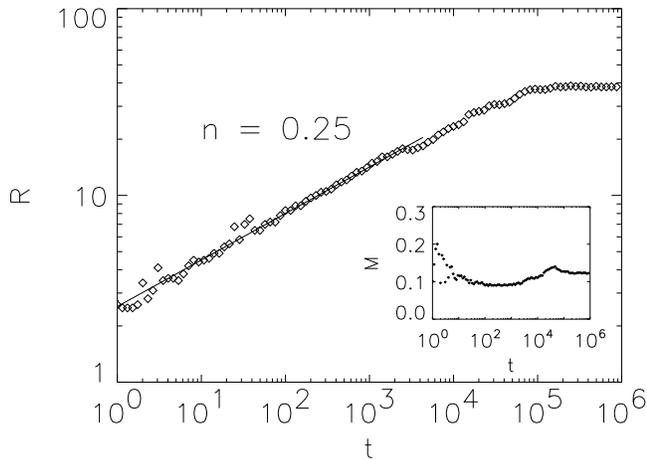}  
\end{minipage}\vspace*{0.5cm}
\caption{
The average mound radius as a function of time
from the LC model with $n_r=5$ from a substrate of size
$100 \times 100$. The average mound slope of the same
system is shown in the inset
\label{lcrm}} \end{figure}

The average mound slope ($M$) is naturally the ratio of
average mound height and average mound radius.
The scaling relation for $R$ can be written as
\begin{equation}
R(t) \sim t^n
\label{eqrt}
\end{equation}
where $n$ is the {\it coarsening} exponent.
The average mound slope $M$ increases in time as
\begin{equation}
M(t) \sim t^ \lambda
\label{eqmt}
\end{equation}
where $\lambda$ is the {\it steepening} exponent.
In Fig. \ref{wvrm} we show the evolution of the average
mound radius in the WV model with $n_r=5$. 
The average mound radius is obviously increasing in time,
showing that the coarsening process is dominant during the
growth period with the coarsening exponent $n = 0.24$. 
The inset shows the average mound slope which does not
change much in time ($\lambda \sim 0$).

\begin{figure}[htbp]
\begin{minipage}{3.4 in}  
\epsfxsize=3.3 in  \epsfbox{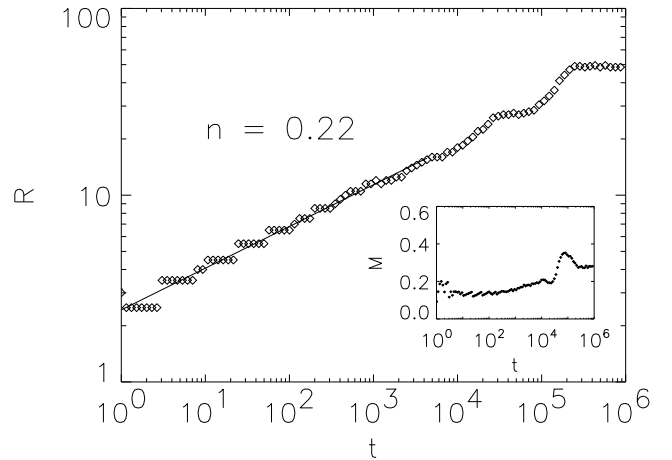}  
\end{minipage}\vspace*{0.5cm}
\caption{
The average mound radius as a function of time
from the nonlinear fourth order model 
with $n_r=5$ from a substrate of size
$100 \times 100$. The average mound slope of the same
system is shown in the inset
\label{4nrm}} \end{figure}

This shows that the mounds in 
the WV model have constant magic `slopes' or 
equivalently, the WV mounded growth exhibits slope selection.
The value of this selected slope in the WV model is approximately
$\tan \theta_0 \approx 0.1-0.2$ or
$\theta_0 \approx 6-12 ^\circ$ and it does not seem to
depend strongly on the value of $n_r$.
Similar behavior is also found in the SED-induced mounded
morphologies in the LC model with $n_r=5$ (Fig. \ref{lcrm})
and the nonlinear fourth order model with $n_r=5$ (Fig. \ref{4nrm}).
We note that the scaling properties of the ES barrier-induced
mounded morphologies are quite distinct from the SED-induced
mounding instability.
Asymptotically, there is no slope selection
in the mound formation caused by an ES barrier \cite{sdsppdtes}. 
After an initial
crossover period, the coarsening drastically slows down with
$n \rightarrow 0$ and the steepening exponent 
becomes exceptionally large with
$\lambda \rightarrow 0.5$.
Thus SED induced mounding has slope selection and ES barrier
induced mounding typically does not (asymptotically).

\subsection{Comparison with experiments}

In this subsection we discuss experimental epitaxial mound formation
in light of our findings about the limited mobility growth models.
First and foremost we caution against
the wisdom of automatically assuming 
the existence of an ES barrier
whenever a mounded morphology is seen. 
As we have argued in this paper, the ES barrier is most
certainly a sufficient condition to create 
mounded growth morphologies
(nominally without any slope selection,
unless some additional mechanisms, not intrinsic
to ES barrier, is invoked), 
but by all means, it is not the {\it only}
possible cause or a necessary condition. 
When analyzing experimental mounding data,
therefore, an
ES barrier should not be automatically assumed and more than
one possibilities should be critically considered.
We hope to make our point by using a recent experiment
as an example \cite{zuoexp}.
It is instructive to carry out a comparison between 
the observed mound
morphology in Cu(100) growth \cite{zuoexp} and the mound
morphology from the LC (the simple linear fourth order) model
shown in this paper.

\begin{figure}[htbp]
\begin{minipage}{2.4 in}  
\epsfxsize=2.4 in  \epsfbox{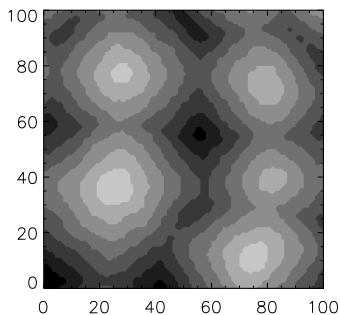}  
\end{minipage}\vspace*{0.5cm}
\caption{
The contour plot of the ``equal priority'' LC morphology
with $l=2$ and $n_r=8$ at $10^4$ ML.
\label{lcexp}} \end{figure}

In Fig. \ref{lcexp}, we show a contour plot of our mounded
growth front simulated from the simple LC model with diffusion
length $l=2$ and noise reduction factor $n_r=8$.
The plot looks very similar (in fact, almost identical)
to the STM images of the
mound morphology in reversible homoepitaxy on Cu(100)
in Ref. \cite{zuoexp};
both systems show mounds with square bases and
the mounds are lined up in a somewhat regular pattern.
Not only are the two systems strikingly comparable visually, 
but the quantitative dynamical scaling properties for the two cases
are also approximately the same.
In the experiment \cite{zuoexp}, the mounds have a magic slope
of $5.6 \pm 1.3 ^\circ$ and the coarsening exponent is
given as $n=0.23 \pm 0.01$.
In our LC simulation, $n \approx 0.25$ and the slope selection
process is clearly observed (see Fig. \ref{lcrm}) with
typical mound slope being around $10^\circ$.
The selected slopes in the two systems are not equal,
but this is not particularly crucial. The LC model is only a simple
dynamic growth model which in general should not be compared
quantitatively with real MBE growth. 
We are not making a claim that there is absolutely no ES barrier
effect in the Cu(100) epitaxial growth.
What we want to point out here is that the simple LC model
{\it without} ES barrier can produce mounded morphology that is
almost exactly the same as the Cu(100) 
mounded growth and the possibility
that the mound morphology in Cu(100) may involve mechanisms
other than the ES barrier should be explored.
Clearly the SED mechanism as a possible source for slope
selected mounding in Cu(100) growth should be taken seriously.

Finally, we recall a ``puzzling'' observation from Ref. \cite{zuoexp},
namely, that the {\it average} mound slope calculated from the 
ratio between the average
mound height ($[H(r=0)]^{1/2}$) and the average mound radius
(first zero crossing of $H(r)$ correlation function) is much
smaller than the {\it typical} mound slope measured from the 
morphology directly. 
In the Cu(100) case \cite{zuoexp}, the
average selected slope is $2.4^\circ$ and the typical mound slope
is $5.6^\circ$. In our LC study, the average slope is approximately
$6^\circ$ while the typical slope is in the range between 
$10^\circ$ to $15^\circ$.
This is so because $[H(r=0)]^{1/2}$ (taken as the average mound height)
does not equal the typical mound height. 

\begin{figure}[htbp]
\begin{minipage}{3.4 in}  
\epsfxsize=3.4 in  \epsfbox{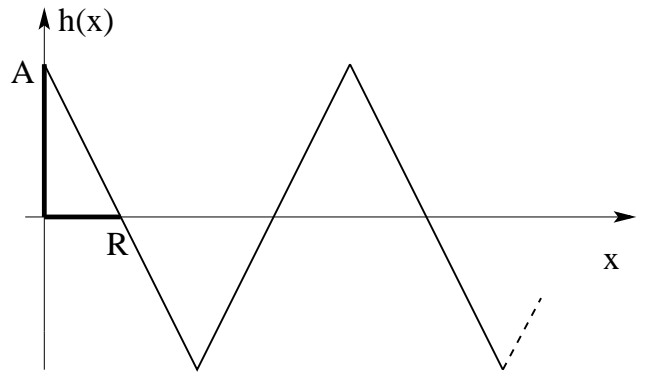}    
\end{minipage}\vspace*{0.5cm}
\caption{
A simple shematic setup to show the difference in the
definitions for the average mound slope.
\label{Schem}} \end{figure}

Let us consider
the simplest ``mounded'' pattern which mimics the height
variation in the morphology along a cut perpendicular to the
base of the pyramids (see the schematic Fig. \ref{Schem}).
In the simplest situation this is described by the periodic,
piecewise linear function:
$h(x) = 
-\frac{A}{R}(x-4kR)+A$, if $x\in (4kR, 4k+2R)$, and
$h(x)=\frac{A}{R}(x-4kR-2R)-A$, if $x\in (4kR+2R, 4k+4R)$,
$k=0,1,2,...$,
where the meaning of $A$ and $R$ is obvious from Fig. \ref{Schem}.
Thus, one has for this simple shape $[H(r=0)]^{1/2} = 
(A/R)/\sqrt{3}=0.57.. s$, where $s=A/R$ is the typical mound
slope, illustrating the difference between the two
definitions. In addition we also have a small discrepancy
introduced by the fact that slope measured from the ratio
of two averages is different from the average of the ratios. 
Thus, our simple LC model mounding explains another puzzling
aspect of experimental mounding not easily explained via
the ES barrier mechanism.
We refrain from carrying out further comparison with other
experimentally reported epitaxial mounding in the literature
because the main point we are making is qualitative:
SED may be playing an important role in observed epitaxial mounding.

\subsection{Global particle diffusion currents}

In the {\it conserved} growth models where the continuity
equation, Eq. (\ref{full4theq}), applies, global particle diffusion
currents can be measured directly from growth on
tilted substrates \cite{KPS}.
If the measured current is a {\it downhill} current (negative) 
then it corresponds to a {\it stable} surface that belongs
asymptotically to the EW universality class, and an
{\it uphill} particle current will point to an {\it unstable}
surface (i.e. mounded morphology) \cite{KPS}.
Our particle current results, however, show that
this concept may be misleading for growth in d=2+1 
dimensional physical surfaces since
the {\it global} current
may show no simple systematic behavior as a function of tilt.
In general, substrate current in 2+1 dimensional growth may
be positive or negative depending on tilt direction.
In fact, we believe that the current measurement on tilted
substrates is a useful technique for determining the 
universality class of growth models only in 1+1 dimensions.
In 2+1 dimensions, where the surface current could be
stabilizing in one direction (e.g. 100) and destabilizing
in another (e.g. 111), the technique fails.
This was not appreciated in Ref. \cite{KPS} and has not been
earlier discussed in the literature.

An important example is the particle current in
the WV model in various dimensions. 
It was reported \cite{KPS} that the WV model
has downhill current in both one and two dimensions. 
This implies that the WV model belongs to the
EW universality class, which was the conclusion in
Ref. \cite{KPS}.
However, our Fig. \ref{wvsur} shows 
that the two dimensional version of the WV model
has mounded morphology induced by an SED instability.
This seems confusing as mound morphology should have
uphill current. We repeated the particle current study on the two
dimensional WV model and found the same downhill current
as in the literature \cite{KPS}, both in the original WV 
model with $n_r=1$ and in the $n_r=5$ WV model.
But all the global particle current measurements in the literature
were obtained \cite{KPS,RP}
by tilting the substrate in the (100)
direction!
We have also studied the diffusion current 
by tilting the substrate
along the diagonal, i.e. in the (111) direction
,
and found a completely different result.
The WV current from the (111) tilted substrate is
{\it uphill} (positive). In the $n_r=1$ model, 
the (111) current starts
out as uphill and then crosses over to a negative value, i.e.,
downhill with 
crossover time increasing with increasing substrate size $L$.
This is already a weak indication that there are some
instabilities present.
In the $n_r>1$ WV model, however, we measure strong (111) uphill
current which stays positive up to $10^6$ ML which is our 
longest simulation time, confirming the suspicion that in fact
the WV model in two dimensions has the SED mounding instability
induced by (111) uphill current, and a simple measurement of
(100) downhill current gives a qualitatively incorrect result.

Our result shows that the global particle current strongly 
depends on how the substrate is tilted. This 
tilt direction (e.g. 100 versus 111) dependence may 
be very complicated in systems with two or higher dimensional
substrates since surface current is a vectorial quantity in
2 (or higher) substrate dimensions.
The particle current from one particular tilt may
be highly misleading with respect 
to the overall behavior of the system.
The downhill (100)
current in the WV model had incorrectly suggested
that the model belongs to the EW universality class
and should exhibit relatively flat morphology in two
dimensions. 
However, by properly suppressing the noise 
our study shows that the two dimensional WV
model actually becomes unstable due to a strong SED driven
uphill (111) current.
As a matter of fact, there has been at least
one prior report of mounding in the WV model in 
unphysical higher dimensions
(three and four dimensional substrates) \cite{kot}.
In higher substrate dimensions the noise effects 
are intrinsically a lot weaker than in one or two substrate
dimensions, because diffusion in lower dimensional substrates
is topologicaly more constrained, and therefore
the SED instability is easier to observe (even without 
noise reduction) in higher dimensions.
We have explicitly verified through direct numerical simulations
that higher (3+1 and 4+1) dimensional WV growth has spectacular
SED instability-induced mound formation even in the absence
of any noise reduction, thus verifying the old and unexplained
results of Ref. \cite{kot}.
The higher dimensional DT growth 
is smooth and flat, and exhibits no such mounding.
This arises from the much stronger edge diffusion effects in 
higher dimensions which effectively introduces very strong 
intrinsic suppression of noise.
Our work therefore resolves an old puzzle,
explaining why the higher dimensional simulations
\cite{kot} 
of the WV model produced mounded morphologies.

\section{Conclusions}
 
We have shown that one has to carefully account for 
subtle topological-kinetic
instabilities when trying to understand mounding in
the surface growth process.
These instabilities are simply the result of the fact that in
higher (higher than 1+1)
dimensions a local current may act as a stabilizing one
in a certain projection on the substrate and as a destabilizing
one in another projection. The local anisotropy of current 
effects averaged over the whole substrate may generate mounded
growth even in situations (e.g. WV model) where mounding 
is completely unanticipated.
The situations where SED instabilities may be present are
rather subtle to analyze and each case has to be investigated
rather carefully using the noise reduction technique.
The fact that SED mounding instability exists in the noise
reduced WV, LC, and the nonlinear fourth order continuum model,
but not in F, DT, and the infinite order nonlinear model
shows that the kinetic-topological nature of SED mounding
may not be obvious at all in a given situation. 
We note here that the DT model also exhibits very weak mound-like
morphologies \cite{short}. However, the weak mound-like structures
in the DT model arise from the fact that the roughness exponent
of the model is exceptionally large
(arising from the presence of the $\nabla ^4 h$ term
in the growth equation). 
The mounds (and the
oscillation in the height correlation $H(r)$) in the 2+1 DT
model are random and irregular
while the SED-induced mound formations
(in e.g. noise reduced WV model) exhibit regular patterns
with strong periodic oscillations in $H(r)$.

One of our important findings is that SED generically 
leads to slope-selected mounded morphology whereas
the ES instability generically leads to slope steepening 
mounding. We believe that many of the slope selected
mounding morphologies observed in MBE experiments
(with typical selected slopes being small $\sim 10^\circ$
or less) may actually arise from the kinetic-topological mounding
we find generically in the LC or the WV model in this paper.
Further experimental and theoretical work is required
to settle the competing roles of various instabilities
contributing to epitaxial mounding in real surface growth.

\begin{acknowledgments}

This work is supported by the NSF-DMR-MRSEC at the University 
of Maryland and by the US-ONR.
ZT also acknowledges partial support from the DOE under
contract W-7405-ENG-36
during the final stage of the manuscript preparation.

\end{acknowledgments}

\end{document}